\documentclass[11pt]{article}
\newcounter{ourcount}
\usepackage{amsmath,amsfonts,amssymb,graphicx,epsfig,pdflscape,multirow,theorem}
\usepackage{arydshln,cite}
\usepackage[matrix,arrow,curve]{xy} 
\numberwithin{equation}{section}
\graphicspath{{./eps/}}
\usepackage{cite}
\usepackage[usenames]{color}
\usepackage{pstricks}
\usepackage[backref=false]{hyperref}
\hypersetup{
colorlinks=true,
citecolor=red,
linkcolor=darkblue
}
\definecolor{darkblue}{rgb}{0,0,.8}
\definecolor{red}{rgb}{1,0,0}

\theorembodyfont{\itshape} 
\theoremheaderfont{\scshape}
\theoremstyle{plain}  
\newtheorem{Lemme}{Lemma}[section]

\newtheorem{Proposition}[Lemme]{Proposition}

\newtheorem{Corollary}[Lemme]{Corollary}

\numberwithin{equation}{section}

\newcommand{\nc}{\newcommand}
\nc{\fh}{\hat{f}}
\nc{\muh}{\hat{\mu}}
\nc{\nuh}{\hat{\nu}}
\nc{\disp}{\displaystyle}
\nc{\cosec}{\mathop{\mbox{cosec}}}
\nc{\ir}{\mathrm{i}}
\def\Re{\mathop{\mbox{Re}}}
\def\Im{\mathop{\mbox{Im}}}
\nc{\bib}{\bibitem}
\nc{\al}{\alpha}
\nc{\g}{\gamma}
\nc{\G}{\Gamma}
\nc{\D}{\Delta}
\nc{\eps}{\epsilon}
\nc{\la}{\lambda}
\nc{\La}{\Lambda}
\nc{\var}{\varphi}
\nc{\pa}{\partial}
\nc{\nn}{\nonumber \\ }
\nc{\hf}{\frac{1}{2}}
\nc{\dz}{\frac{dz}{2\pi i}}
\nc{\bin}[2]{\left(\!\!\!\begin{array}{c} {#1}\\ {#2} \end{array}\!\!\!\right)}
\nc{\be}{\begin{equation}}
\nc{\ee}{\end{equation}}
\nc{\bea}{\begin{eqnarray}}
\nc{\eea}{\end{eqnarray}}
\nc{\bra}[1]{\langle {#1}|}
\nc{\ket}[1]{|{#1}\rangle}
\nc{\chit}{\raisebox{0.25ex}{$\chi$}}
\nc{\Dbh}{\mbox{\boldmath $\hat D$}}
\nc{\Dbb}{\mbox{\boldmath $\bar D$}}
\nc{\Dbm}{\mbox{\boldmath $\mathcal D$}}
\nc{\db}{\mbox{\boldmath $d$}}
\nc{\Ab}{\mbox{\boldmath $A$}}
\nc{\Bb}{\mbox{\boldmath $B$}}
\nc{\Cb}{\mbox{\boldmath $C$}}
\nc{\Db}{\mbox{\boldmath $D$}}
\nc{\eb}{\mbox{\boldmath $e$}}
\nc{\Fb}{\mbox{\boldmath $F$}}
\nc{\Fbt}{\mbox{\boldmath $\tilde{F}$}}
\nc{\fb}{\mbox{\boldmath $f$}}
\nc{\fbt}{\mbox{\boldmath $\tilde{f}$}}
\nc{\Gb}{\mbox{\boldmath $G$}}
\nc{\Hb}{\mbox{\boldmath $H$}}
\nc{\Jb}{\mbox{\boldmath $J$}}
\nc{\Kb}{\mbox{\boldmath $K$}}
\nc{\Mb}{\mbox{\boldmath $M$}}
\nc{\Pb}{\mbox{\boldmath $P$}}
\nc{\Qb}{\mbox{\boldmath $Q$}}
\nc{\Tb}{\mbox{\boldmath $T$}}
\nc{\Tbb}{\mbox{\boldmath $\bar T$}}
\nc{\Tbm}{\mbox{\boldmath $\mathcal T$}}
\nc{\tb}{\mbox{\boldmath $t$}}
\nc{\Ub}{\mbox{\boldmath $U$}}
\nc{\Vb}{\mbox{\boldmath $V$}}
\nc{\Wb}{\mbox{\boldmath $W$}}
\nc{\Xb}{\mbox{\boldmath $X$}}
\nc{\yb}{\mbox{\boldmath $y$}}
\nc{\Zb}{\mbox{\boldmath $Z$}}
\nc{\Hc}{{\cal H}}
\nc{\Rc}{{\cal R}}
\nc{\Lc}{{\cal L}}
\nc{\Vc}{{\cal V}}
\nc{\Ib}{\mbox{\boldmath $I$}}
\nc{\qb}{\bar{q}}
\nc{\oN}{\mathbb{N}}
\nc{\oZ}{\mathbb{Z}}
\nc{\oR}{\mathbb{R}}
\newrgbcolor{darkgreen}{0., 0.733333, 0.0621042}
\def\vvdots{\mathinner{\mkern1mu\raise1pt\vbox{\kern7pt\hbox{.}}\mkern2mu
  \raise4pt\hbox{.}\mkern2mu\raise7pt\hbox{.}\mkern1mu}}
\nc{\gauss}[2]{\left[\!\!\begin{array}{c} {#1}\\ {#2} \end{array}\!\!\right]}
\nc{\sbin}[2]{\left\{\!\!\!\begin{array}{c} {#1}\\ {#2} 
\end{array}\!\!\!\right\}}
\nc{\sbinlr}[2]{\Big\langle\!\!\begin{array}{c} {#1}\\ {#2} 
\end{array}\!\!\Big\rangle}
\nc{\bino}[2]{\left(\!\!\begin{array}{c} {#1}\\ {#2} \end{array}\!\!\right)}
\def\half {\mbox{$\textstyle \frac{1}{2}$}}

\definecolor{lightblue}{rgb}{.7,.7,1}
\definecolor{lightlightblue}{rgb}{.85,.85,1}
\definecolor{midblue}{rgb}{.7,.7,1}

\def\loopa{
\psframe[linewidth=.25pt](0,0)(1,1)
\psarc[linewidth=1.5pt,linecolor=blue](1,0){.5}{90}{180}
\psarc[linewidth=1.5pt,linecolor=blue](0,1){.5}{-90}{0}
}
\def\loopb{
\psframe[linewidth=.25pt](0,0)(1,1)
\psarc[linewidth=1.5pt,linecolor=blue](0,0){.5}{0}{90}
\psarc[linewidth=1.5pt,linecolor=blue](1,1){.5}{180}{270}
}

\def\facegrid#1#2{
\psframe[fillstyle=solid,fillcolor=lightlightblue,linewidth=0pt]#1#2
\psgrid[gridlabels=0pt,subgriddiv=1]#1#2}

\def\floor#1{{\lfloor #1\rfloor}}
\def\looopa{
\psarc[linewidth=1.5pt,linecolor=blue](1,0){.5}{90}{180}
\psarc[linewidth=1.5pt,linecolor=blue](0,1){.5}{-90}{0}
}
\def\looopb{
\psarc[linewidth=1.5pt,linecolor=blue](0,0){.5}{0}{90}
\psarc[linewidth=1.5pt,linecolor=blue](1,1){.5}{180}{270}
}

\newcommand{\statei}{{\ket{-\frac{3}{32}}}\ar@{};[0,0];}
\newcommand{\stateii}{{\ket{\frac{5}{32}}}\ar@{};[0,0];}
\newcommand{\stateiii}{{\ket{\frac{21}{32}}}\ar@{};[0,0];}
\newcommand{\stateiv}{{\ket{\frac{45}{32}}}\ar@{};[0,0];}
\newcommand{\statev}{{\ket{\frac{77}{32}}}\ar@{};[0,0];}
\newcommand{\statevi}{{\ket{\frac{117}{32}}}\ar@{};[0,0];}
\newcommand{\stated}{{\dots}\ar@{};[0,0];}
\newcommand{\statee}{{}\ar@{};[0,0];}

\begin{document}

\topmargin -5mm
\oddsidemargin 5mm

\begin{titlepage}

\vspace{8mm}
\begin{center}
{\LARGE \bf Critical dense polymers with Robin boundary conditions,}\\[7pt]
{\LARGE \bf half-integer Kac labels and $\oZ_4$ fermions}
\end{center}

\vspace{8mm}
\begin{center}
{\vspace{-5mm}\Large Paul A. Pearce$^\ast$,\, J{\o}rgen Rasmussen$^\dagger$,\, Ilya Yu.\! Tipunin$^\ddagger$}
\\[.4cm]
{\em {}$^\ast$Department of Mathematics and Statistics, University of Melbourne}\\
{\em Parkville, Victoria 3010, Australia}
\\[.25cm]
{\em {}$^\dagger$School of Mathematics and Physics, University of Queensland}\\
{\em St Lucia, Brisbane, Queensland 4072, Australia}
\\[.25cm]
{\em {}$^\ddagger$TAMM Theory Division, Lebedev Physics Institute}\\
{\em Leninski Pr., 53, Moscow 119991, Russia} 
\\[.4cm]
{\tt p.pearce\,@\,ms.unimelb.edu.au}
\quad
{\tt j.rasmussen\,@\,uq.edu.au}
\quad
{\tt tipunin\,@\,gmail.com}
\end{center}

\vspace{10mm}
\centerline{{\bf{Abstract}}}
\vskip.4cm
\noindent
For general Temperley-Lieb loop models, including the logarithmic minimal models ${\cal LM}(p,p')$ with $p,p'$ coprime integers, 
we construct an infinite family of Robin boundary conditions on the strip as linear combinations of Neumann and Dirichlet boundary 
conditions. These boundary conditions are Yang-Baxter integrable and allow loop segments to terminate on the boundary. 
Algebraically, the Robin boundary conditions are described by the one-boundary Temperley-Lieb algebra. 
Solvable critical dense polymers is the first member ${\cal LM}(1,2)$ of the family of logarithmic minimal models and
has loop fugacity $\beta=0$ and central charge $c=-2$.
Specializing to ${\cal LM}(1,2)$ with our Robin boundary conditions, we solve the model exactly on strips of arbitrary finite 
size $N$ and extract the finite-size conformal corrections using an Euler-Maclaurin formula. 
The key to the solution is an inversion identity satisfied by the commuting 
double row transfer matrices. This inversion identity is established directly in the Temperley-Lieb algebra. 
We classify the eigenvalues of the double row transfer matrices using the physical combinatorics 
of the patterns of zeros in the complex spectral parameter plane and obtain finitized characters related to spaces of
coinvariants of $\oZ_4$ fermions.
In the continuum scaling limit, the Robin boundary conditions are associated with irreducible Virasoro Verma modules with 
conformal weights $\Delta_{r,s-\frac{1}{2}}=\frac{1}{32}(L^2-4)$ where $L=2s-1-4r$, $r\in\oZ$, $s\in\oN$. 
These conformal weights populate a Kac table with half-integer Kac labels. 
Fusion of the corresponding modules with the generators of the Kac fusion algebra is examined and general fusion rules are proposed.
\end{titlepage}

\newpage

\setcounter{page}{2}

\tableofcontents
\clearpage

\section{Introduction}

The exactly solvable model ${\cal LM}(1,2)$~\cite{PR2007,PRV1210} of critical dense 
polymers\cite{Flory,Gennes,Cloizeaux,Saleur87a,Duplantier86,Saleur87b,SaleurHalfInt87} 
is the first member of the family of logarithmic minimal models ${\cal LM}(p,p')$~\cite{PRZ} 
where $1\le p<p'$ and $p,p'$ are coprime integers. These models are Yang-Baxter integrable~\cite{BaxBook} Temperley-Lieb (TL) loop 
models~\cite{TL,Nienhuis,BN}
on the square lattice. The TL loop models are distinguished, one from the other, by the value of the crossing parameter
$\lambda\in\oR$ in terms of which the loop fugacity is given by $\beta=2\cos\lambda$.
In the case of ${\cal LM}(p,p')$, the crossing parameter is a rational multiple of $\pi$, parameterised as
$\lambda=\frac{(p'-p)\pi}{p'}$. For critical dense polymers, with $\lambda=\frac{\pi}{2}$, 
the loop fugacity $\beta=0$ vanishes so closed loops are not allowed. The next member of the series 
${\cal LM}(2,3)$ is critical (bond) percolation~\cite{BroadHamm57} with $\lambda=\frac{\pi}{3}$ and $\beta=1$. 
These models are important prototypical examples of a large class of geometrical critical systems with nonlocal degrees of freedom in 
the form of extended polymers or connectivities. 

The study of the conformal properties of such systems started with Saleur and Duplantier 
in the late eighties. Remarkably, they found that certain conformal weights are given by the Kac formula
\be
 \Delta_{r,s}^{p,p'}=\frac{(rp'-sp)^2-(p'-p)^2}{4pp'},\qquad r,s\in\tfrac{1}{2}\,\oN
\label{confWts}
\ee
but where the Kac labels $r,s$ can (i)~take integer values that are outside~\cite{Saleur87a,Saleur87b} of the known Kac 
tables (\mbox{$1\le r\le p-1$}, \mbox{$1\le s\le p'-1$}) for the unitary minimal models ${\cal M}(p,p')$ with $p'=p+1$, or (ii)~be 
half-integers~\cite{SaleurHalfInt87,Duplantier86}. These differences in operator content are allowed because the minimal models 
${\cal M}(p,p\!+\!1)$ are rational and unitary whereas the geometrical theories ${\cal LM}(p,p+1)$ are nonunitary and not rational. 
The existence of a family of spin fields with
conformal weights $\Delta^{p,p+1}_{k+\frac{1}{2},0}$ for $k\in\oN$ has recently been posited~\cite{Delfino}.
It has also been suggested that fields with half-integer Kac labels play a role in the description of critical percolation~\cite{Rid0808}.

More fundamentally, it is now known that, when the crossing parameter $\lambda$ of the nonlocal loop model is a rational multiple of 
$\pi$, the continuum scaling limit of the   loop model ${\cal LM}(p,p')$~\cite{PRZ} is described by a logarithmic CFT~\cite{Gurarie93,JPA} 
with central charge 
\be
 c=1-\frac{6(p'-p)^2}{pp'}
\ee
The central charge is thus $c=-2$ for critical dense polymers and $c=0$ for critical percolation. Compared to the local degrees of 
freedom of the minimal models ${\cal M}(p,p')$, the nonlocal nature of the degrees of freedom of the logarithmic minimal models 
${\cal LM}(p,p')$ has profound implications for the associated CFT. For example, a rational CFT is 
described~\cite{MooreSeiberg} by a finite 
number of irreducible representations which close under fusion. In contrast, the representation content of a logarithmic CFT is very rich 
and has not been completely classified even for the simplest theories. In the context of the 
logarithmic minimal models ${\cal LM}(p,p')$ in the so-called Virasoro picture, it is known that there is an infinite family of 
reducible yet indecomposable Kac representations~\cite{PRZ,RP0707,Ras1012,BGT1102} labeled by the integer Kac labels $r,s\in\oN$. 
The conformal weights of the Kac representations coincide with the conformal weights (\ref{confWts}) with integer labels 
$r,s\in\oN$ lying in infinitely extended Kac tables, thus giving rise to Kac labels outside of the rational Kac tables. 
By allowing a ${\cal W}$-extended conformal symmetry 
algebra~\cite{Kausch91,Flohr9509,Kausch9510,GK9606,FHST0306,CF0508,Walgebra1,Walgebra2,Walgebra3,AM0707,KS0901,NT0902}, 
the infinity of Virasoro-Kac representations can be reorganized into a finite family of reducible yet indecomposable ${\cal W}$-Kac 
representations~\cite{Ras1106}. 
In addition, minimal-irreducible, ${\cal W}$-irreducible and projective representations also exist as discussed 
in~\cite{FK0705,GR0707,PRR0803,RP0804,Ras0805,GRW0905,Ras0906,Wood0907}, 
for example, but all of these together still do not exhaust the possible representations of ${\cal LM}(p,p')$. 
In the principal series ${\cal LM}(p,p+1)$, in particular, there should exist representations associated to half-integer Kac labels. 

Many representations, such as the Kac representations, admit conjugate boundary conditions on the lattice. In this paper, we introduce 
a family of Yang-Baxter integrable Robin boundary conditions~\cite{Robin} for general TL loop models, including
the logarithmic minimal models ${\cal LM}(p,p')$. These 
boundary conditions allow loop segments to terminate on the boundary. They satisfy the boundary Yang-Baxter equation and are 
constructed as linear combinations of Neumann and Dirichlet boundary conditions. A motivation to study these boundary conditions is 
that, for ${\cal LM}(p,p')$, they are generally expected to be conjugate to representations with noninteger Kac labels. 

The introduction of so-called $r$-type seams has been a very successful way of constructing new Yang-Baxter integrable boundary 
conditions, initially in rational lattice models~\cite{BP96,BP01}, 
but more recently also in logarithmic minimal models~\cite{PRZ,Annecy,PRV1210}.
Motivated by this, it is natural to look for similar constructions of boundary conditions in TL loop models with loop segments allowed
to terminate on the boundary. 
The Robin boundary conditions are thus labelled by two nonnegative integers $w$ and $d$ where $w$ measures the width of a 
boundary seam while $d$ denotes the number of defects or through-lines. 
Within such a seam, a projection operation is applied to project onto a specific vector space of link states yielding 
a family of well-defined and commuting transfer matrices. In the case of
${\cal LM}(1,2)$, we find that the parameters $w$ and $d$ are related to a pair of Kac labels $r,s-\tfrac{1}{2}$ with $r\in\oZ$ 
and $s\in\mathbb{N}$. 

Algebraically, the Robin boundary conditions are constructed using the generators of the
one-boundary TL or blob algebra~\cite{MaSa93,MaWood2000,NRG2005,Nichols2006a,Nichols2006b}.
That is, the loop configurations can be expressed and examined by means of a diagrammatic realisation of this algebra, 
and most subsequent manipulations and calculations are accordingly done algebraically. This does not imply, however, that the Robin
boundary conditions and the associated Robin link states give rise to new representations of the one-boundary TL algebra. Rather, 
the one-boundary TL algebra provides the framework and machinery for the construction of a new family of commuting transfer matrices.

To study in detail the properties of the Robin boundary conditions, we specialize to the case of critical dense polymers ${\cal LM}(1,2)$. 
In this case, the model can be solved exactly on an arbitrary finite lattice allowing the conformal spectra to be extracted analytically using 
an Euler-Maclaurin formula. In this way, we obtain conformal weights with half-integer Kac labels
\be
 \Delta^{1,2}_{r,s-\frac{1}{2}}=\Delta^{1,2}_{0,\frac{L}{2}}=\frac{1}{32}(L^2-4)=-\frac{3}{32}, \frac{5}{32}, \frac{21}{32}, \frac{45}{32}, 
  \frac{77}{32}, \ldots \qquad L=2s-1-4r,\qquad r\in\oZ, s\in\oN
\label{D12}
\ee
We recall that, for the principal series ${\cal LM}(p,p+1)$, the conformal weight of the twist operator
(changing boundary conditions from Neumann to Dirichlet~\cite{JSbdy}) is
\be
 \Delta^{p,p+1}_{0,\frac{1}{2}}=\Delta^{p,p+1}_{\frac{p}{2},\frac{p}{2}}=\frac{p^2-4}{16p(p+1)}=-\frac{3}{32}, 0, \frac{5}{192}, \frac{3}{80}, 
   \ldots  \qquad p=1,2,3,4,\ldots
\label{RobinConfWts}
\ee
with $\Delta^{1,2}_{0,\frac{1}{2}}=-\frac{3}{32}$ for critical dense polymers, in accordance with (\ref{D12}). 

It is stressed that, while this paper has some overlap with the paper of Jacobsen and Saleur~\cite{JSbdy}, 
our general Robin boundary conditions are new. The loop configurations in~\cite{JSbdy} are defined on a tilted square lattice
with periodic boundary conditions forming an annulus, whereas we consider the regular square lattice on the strip. Jacobsen and Saleur
study their model at the isotropic point allowing them to consider the situation with all boundary loops blobbed along the outer rim of 
the annulus.
By contrast, this is not possible in our scenario as the Dirichlet boundary condition alone does not provide a solution to the 
spectral parameter
dependent boundary Yang-Baxter equations. Instead, our Robin boundary conditions are functions of the spectral parameter. 
This has the advantage that finite-size corrections, to the eigenvalues of the transfer matrix, can be described by means of 
physical combinatorics associated with the patterns of zeros in the complex $u$-plane of the spectral parameter.

In a logarithmic CFT setting, 
critical dense polymers ${\cal LM}(1,2)$ with Robin boundary conditions is described by the $\oZ_4$ sector of symplectic 
fermions~\cite{SaleurSUSY,Kausch2000,CQS2006,PRV2010}. In particular, the characters of the representations with conformal weights 
(\ref{D12}) and half-integer Kac labels are irreducible and associated with Virasoro Verma modules. Even stronger, the finitized
characters obtained from the lattice implementation of the Robin boundary conditions
are found to match the characters over certain spaces of coinvariants of $\oZ_4$ fermions.

The layout of the paper is as follows. The $su(2)$ loop model on the square lattice is introduced in Section~\ref{Sec:Loop} in terms of 
bulk face operators and Neumann and Dirichlet boundary triangles. The local properties of the face operators are described in the planar 
TL algebra. Robin boundary conditions, given as linear combinations of Neumann and Dirichlet boundary conditions, are 
constructed as solutions to the boundary Yang-Baxter equation in Section~\ref{Sec:Robin}. The construction uses the one-boundary 
TL algebra. The relation between the one-boundary TL and blob algebra is described in 
Appendix~\ref{Sec:Blob}. A simple Robin twist boundary condition is constructed first and then general Robin boundary conditions 
are constructed by allowing defects and incorporating a boundary seam. 
To study the spectra in the continuum scaling limit of the model, it is necessary to specify the vector space of link states on which
the transfer tangles, as elements of the one-boundary TL algebra, act. 
In Section~\ref{Sec:LinkStates}, the Robin link states are thus defined and their relation to so-called standard modules is explained. 
The commuting double row transfer matrices, with Neumann boundary conditions on the left and Robin boundary conditions on the right, 
are set up in Section~\ref{Sec:Transfer}. The associated quantum Hamiltonians are also derived in this section. Specializing to critical 
dense polymers, the double row transfer matrices satisfy functional equations in the form of inversion identities. The inversion identity 
solved in this paper is presented in Section~\ref{Sec:CritDensePol}. Inversion identities for general Robin boundary conditions 
in critical dense polymers are 
derived in Appendix~\ref{App:GenInv}. Section~\ref{Sec:CritDensePol} also contains the derivation of the exact finite-size spectra by 
using empirical physical combinatorics, an Euler-Maclaurin formula and finitized characters. The conformal data is summarized in 
Section~\ref{Sec:CFT} where we relate critical dense polymers with Robin boundary conditions to $\oZ_4$ fermions.
Using results obtained in Appendix~\ref{App:NGK}, we also give the fusion rules between Robin modules and the generators of the 
Kac fusion algebra. Section~\ref{Sec:Discussion} contains a concluding discussion.

\section{Lattice loop model}
\label{Sec:Loop}

\subsection{Statistical lattice model on a strip}
\label{Sec:StatLatMod}

We consider a square lattice model of densely packed, non-oriented and non-intersecting loops defined on a rectangular strip
of width $N$ and even height $M$.
In a given lattice configuration, a single bulk face contains a pair of loop segments linking the edges pairwise as
\be
\psset{unit=.77cm}
\begin{pspicture}[shift=-.42](1,1)
\facegrid{(0,0)}{(1,1)}
\rput[bl](0,0){\loopa}
\end{pspicture}
\qquad \mathrm{or}\qquad
\psset{unit=.77cm}
\begin{pspicture}[shift=-.42](1,1)
\facegrid{(0,0)}{(1,1)}
\rput[bl](0,0){\loopb}
\end{pspicture}
\label{bulk}
\ee
As indicated in the configuration to the left
in Figure~\ref{conf}, the left boundary is closed off with half-arcs linking adjacent faces pairwise,
while loops can terminate at the right boundary or reflect back into the bulk via half-arcs linking adjacent faces at heights
$2j-1$ and $2j$, $j\in\oN$.
\begin{figure}[t]
\begin{center}
\psset{unit=0.75}
\begin{pspicture}(-0.7,-0.5)(5,8)
\facegrid{(0,0)}{(4,8)}
\psline{<->}(0.1,-0.3)(3.9,-0.3)
\psline{<->}(-0.8, 0.1)(-0.8, 7.9)
\rput(2,-0.7){$N$}
\rput(-1.25,4){$M$}
\rput(4.5,7.5){$\bullet$}
\rput(4.5,6.5){$\bullet$}
\rput(4.5,5.5){$\bullet$}
\rput(4.5,4.5){$\bullet$}
\rput(4.5,1.5){$\bullet$}
\rput(4.5,0.5){$\bullet$}
\psline[linewidth=1pt,linestyle=dashed, dash=1pt 1pt](4, 8)(4.5,8)
\psline[linewidth=1pt,linestyle=dashed, dash=1pt 1pt](4, 6)(4.5,6)
\psline[linewidth=1pt,linestyle=dashed, dash=1pt 1pt](4, 4)(4.5,4)
\psline[linewidth=1pt,linestyle=dashed, dash=1pt 1pt](4, 2)(4.5,2)
\psline[linewidth=1pt,linestyle=dashed, dash=1pt 1pt](4, 0)(4.5,0)
\psset{linecolor=blue}
\psset{linewidth=1.5pt}
\psarc(4,3){0.5}{-90}{90}
\psarc(0,1){0.5}{90}{270}
\psarc(0,3){0.5}{90}{270}
\psarc(0,5){0.5}{90}{270}
\psarc(0,7){0.5}{90}{270}
\put(0,7){$\loopa$}\put(1,7){$\loopa$}\put(2,7){$\loopb$}\put(3,7){$\loopa$}
\put(0,6){$\loopa$}\put(1,6){$\loopa$}\put(2,6){$\loopb$}\put(3,6){$\loopa$}
\put(0,5){$\loopb$}\put(1,5){$\loopb$}\put(2,5){$\loopa$}\put(3,5){$\loopa$}
\put(0,4){$\loopb$}\put(1,4){$\loopb$}\put(2,4){$\loopb$}\put(3,4){$\loopb$}
\put(0,3){$\loopa$}\put(1,3){$\loopa$}\put(2,3){$\loopa$}\put(3,3){$\loopa$}
\put(0,2){$\loopa$}\put(1,2){$\loopa$}\put(2,2){$\loopa$}\put(3,2){$\loopb$}
\put(0,1){$\loopb$}\put(1,1){$\loopb$}\put(2,1){$\loopa$}\put(3,1){$\loopb$}
\put(0,0){$\loopa$}\put(1,0){$\loopb$}\put(2,0){$\loopa$}\put(3,0){$\loopb$}
\psline(4,7.5)(4.5,7.5)
\psline(4,6.5)(4.5,6.5)
\psline(4,5.5)(4.5,5.5)
\psline(4,4.5)(4.5,4.5)
\psline(4,1.5)(4.5,1.5)
\psline(4,0.5)(4.5,0.5)
\end{pspicture}
\qquad\qquad\qquad
%
\begin{pspicture}(-0.7,-0.5)(5,8)
\facegrid{(0,0)}{(4,8)}
\multirput(0,0)(0,2){4}{\pspolygon[fillstyle=solid,fillcolor=lightlightblue](-1,0)(0,1)(-1,2)}
\psline{<->}(0.1,-0.3)(3.9,-0.3)
\psline{<->}(-1.55, 0.1)(-1.55, 7.9)
\rput(2,-0.7){$N$}
\rput(-2,4){$M$}
\pspolygon[fillstyle=solid,fillcolor=lightlightblue](4,7)(5,8)(5,6)(4,7)
\pspolygon[fillstyle=solid,fillcolor=lightlightblue](4,5)(5,6)(5,4)(4,5)
\pspolygon[fillstyle=solid,fillcolor=lightlightblue](4,3)(5,4)(5,2)(4,3)
\pspolygon[fillstyle=solid,fillcolor=lightlightblue](4,1)(5,2)(5,0)(4,1)
\psset{linecolor=blue}
\psset{linewidth=1.5pt}
\psarc(0,1){0.5}{90}{270}
\psarc(0,3){0.5}{90}{270}
\psarc(0,5){0.5}{90}{270}
\psarc(0,7){0.5}{90}{270}
\put(0,7){$\loopa$}\put(1,7){$\loopa$}\put(2,7){$\loopb$}\put(3,7){$\loopa$}
\put(0,6){$\loopa$}\put(1,6){$\loopa$}\put(2,6){$\loopb$}\put(3,6){$\loopa$}
\put(0,5){$\loopb$}\put(1,5){$\loopb$}\put(2,5){$\loopa$}\put(3,5){$\loopa$}
\put(0,4){$\loopb$}\put(1,4){$\loopb$}\put(2,4){$\loopb$}\put(3,4){$\loopb$}
\put(0,3){$\loopa$}\put(1,3){$\loopa$}\put(2,3){$\loopa$}\put(3,3){$\loopa$}
\put(0,2){$\loopa$}\put(1,2){$\loopa$}\put(2,2){$\loopa$}\put(3,2){$\loopb$}
\put(0,1){$\loopb$}\put(1,1){$\loopb$}\put(2,1){$\loopa$}\put(3,1){$\loopb$}
\put(0,0){$\loopa$}\put(1,0){$\loopb$}\put(2,0){$\loopa$}\put(3,0){$\loopb$}
\psline(4,7.5)(5,7.5)
\psline(4,6.5)(5,6.5)
\psline(4,5.5)(5,5.5)
\psline(4,4.5)(5,4.5)
\psbezier(4,2.5)(5,2.4)(5,3.6)(4,3.5)
\psline(4,1.5)(5,1.5)
\psline(4,0.5)(5,0.5)
\end{pspicture}
\caption{Lattice configuration $\sigma$ with weight $W_\sigma=w_1^{18}w_2^{14}a_1a_2^3\,\beta^4\beta_1^2\,\beta_2$. 
Closed bulk loops are not allowed for critical dense polymers with $\beta=0$.}
\label{conf}
\end{center}
\end{figure}
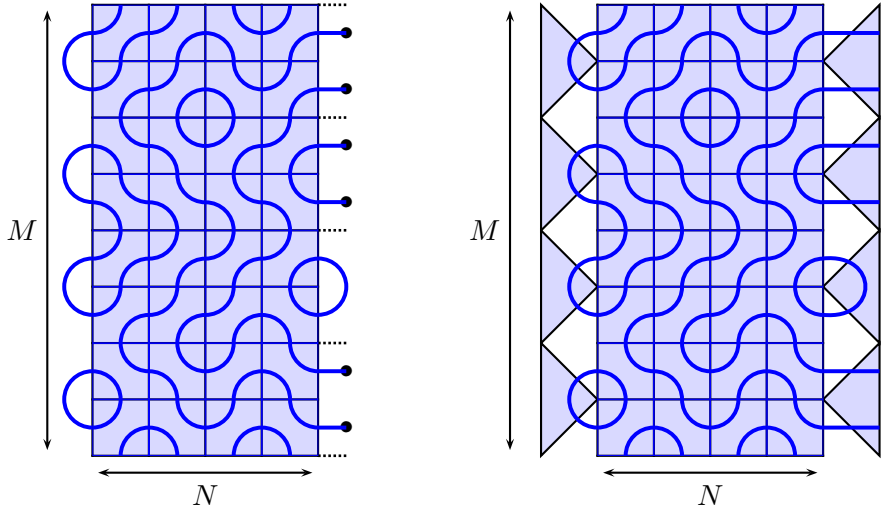
We distinguish between bulk and boundary loops.
First, loops not terminating at the boundary are assigned the bulk loop fugacity $\beta$. 
Second, loops terminating at the right boundary are classified
according to whether the lower attachment point is located at an odd or even height. 
The corresponding boundary loop fugacity is denoted by $\beta_1$ or $\beta_2$, respectively.

It is convenient to introduce boundary triangles to describe the possible boundary conditions on the right of the strip.
A boundary triangle thus comes with one of the two possible configurations
\be
\psset{unit=.6cm}
\begin{pspicture}[shift=-0.89](0,0)(1,2)
\pspolygon[fillstyle=solid,fillcolor=lightlightblue](0,1)(1,2)(1,0)(0,1)
\psarc[linewidth=1.5pt,linecolor=blue](0,1){.7}{-45}{45}
\end{pspicture}
\qquad \mathrm{and}\qquad\!
\begin{pspicture}[shift=-0.89](0,0)(1,2)
\pspolygon[fillstyle=solid,fillcolor=lightlightblue](0,1)(1,2)(1,0)(0,1)
\psline[linecolor=blue,linewidth=1.5pt]{-}(0.4,0.6)(1,0.6)
\psline[linecolor=blue,linewidth=1.5pt]{-}(0.4,1.4)(1,1.4)
\end{pspicture}
\label{boundary}
\ee
where the horizontal line segments indicate that the corresponding loop segments terminate at the boundary.
We refer to the boundary conditions in (\ref{boundary}) as {\em Neumann and Dirichlet boundary conditions}, respectively.
The configuration on the right in Figure~\ref{conf} is equivalent to the configuration on the left,
and the two types of boundary loops are assigned fugacities as indicated here
\be
\psset{unit=.6cm}
\beta_1:\qquad
\begin{pspicture}[shift=-2.39](0,0)(1,5)
\pspolygon[fillstyle=solid,fillcolor=lightlightblue](0,1)(1,2)(1,0)(0,1)
\pspolygon[fillstyle=solid,fillcolor=lightlightblue](0,4)(1,5)(1,3)(0,4)
\rput(1,2.7){$\vdots$}
\psline[linecolor=blue,linewidth=1.5pt](0.4,0.6)(1,0.6)
\psline[linecolor=blue,linewidth=1.5pt](0.4,4.4)(1,4.4)
\psbezier[linewidth=1.5pt,linecolor=blue](0.4,0.6)(-1.4,0.45)(-1.4,4.55)(0.4,4.4)
\rput(1.8,0.65){\small odd}
\rput(1.8,4.4){\small even}
\end{pspicture}
\hspace{2cm}
\beta_2:\quad\
\begin{pspicture}[shift=-2.39](0,0)(1,5)
\pspolygon[fillstyle=solid,fillcolor=lightlightblue](0,1)(1,2)(1,0)(0,1)
\pspolygon[fillstyle=solid,fillcolor=lightlightblue](0,4)(1,5)(1,3)(0,4)
\rput(1,2.7){$\vdots$}
\psline[linecolor=blue,linewidth=1.5pt](0.4,3.6)(1,3.6)
\psline[linecolor=blue,linewidth=1.5pt](0.4,1.4)(1,1.4)
\psbezier[linewidth=1.5pt,linecolor=blue](0.4,1.4)(-0.8,1.5)(-0.8,3.5)(0.4,3.6)
\rput(1.8,1.4){\small even}
\rput(1.8,3.65){\small odd}
\end{pspicture}
\qquad\qquad 
\ee

To obtain a statistical model, local Boltzmann weights are assigned to the bulk faces and boundary triangles. 
The weights $w_1$ and $w_2$ are thus assigned to the bulk faces 
\,$
\begin{pspicture}[shift=-0.05](0.4,0.5)
\psset{unit=0.4}
\facegrid{(0,0)}{(1,1)}
\rput[bl](0,0){\loopa}
\end{pspicture}
$\, 
and
\,$
\begin{pspicture}[shift=-0.05](0.4,0.5)
\psset{unit=0.4}
\facegrid{(0,0)}{(1,1)}
\rput[bl](0,0){\loopb}
\end{pspicture}
$\,,
respectively, while the weights $a_1$ and $a_2$ are assigned to the boundary triangles 
$
\begin{pspicture}[shift=-0.12](0.45,0.6)
\psset{unit=0.25}
\pspolygon[fillstyle=solid,fillcolor=lightlightblue](0,1)(1,2)(1,0)(0,1)
\psarc[linewidth=1.35pt,linecolor=blue](0,1){.7}{-45}{45}
\end{pspicture}
$\!\!
and
$
\begin{pspicture}[shift=-0.12](0.45,0.6)
\psset{unit=0.25}
\pspolygon[fillstyle=solid,fillcolor=lightlightblue](0,1)(1,2)(1,0)(0,1)
\psline[linecolor=blue,linewidth=1.2pt]{-}(0.4,0.6)(1,0.6)
\psline[linecolor=blue,linewidth=1.2pt]{-}(0.4,1.4)(1,1.4)
\end{pspicture}
$\!\!,
respectively. Viewing the loop fugacities as non-local Boltzmann weights,
the weight of a lattice configuration $\sigma$ is thus given by
\be
 W_\sigma=w_1^{n_1}w_2^{n_2}a_1^{m_1}a_2^{m_2}\beta^\ell\beta_1^{\ell_1}\beta_2^{\ell_2}
\ee
where $n_1$, $n_2$, $m_1$ and $m_2$ indicate the numbers of the various faces and triangles in $\sigma$, while $\ell$, $\ell_1$
and $\ell_2$ indicate the numbers of loops.
For example, the weight of the configuration in Figure~\ref{conf} is given by
\be
 W_\sigma=w_1^{18}w_2^{14}a_1a_2^3\,\beta^4\beta_1^2\,\beta_2
\ee
As usual, the partition function is obtained by summing over all possible lattice configurations
\be
 Z=\sum_\sigma W_\sigma
\ee

\subsection{Face operators and local relations}
\label{Sec:Bulk}

To obtain a Yang-Baxter integrable lattice model, we parameterise the bulk loop fugacity as
\be
 \beta=2\cos\lambda,\qquad 0<\lambda<\pi
\label{beta}
\ee 
where $\lambda$ is the crossing parameter of the model.
Letting $u$ denote the spectral parameter, the bulk of the lattice is then described by the elementary bulk face operators
\be
\psset{unit=.77cm}
\begin{pspicture}[shift=-.42](1,1)
\facegrid{(0,0)}{(1,1)}
\psarc[linewidth=0.025]{-}(0,0){0.16}{0}{90}
\rput(.5,.5){$_u$}
\end{pspicture}
\, :=\ 
s_1(-u)\ \begin{pspicture}[shift=-.45](1,1)
\facegrid{(0,0)}{(1,1)}
\rput[bl](0,0){\loopa}
\end{pspicture}
\;+\,s_0(u)\
\begin{pspicture}[shift=-.42](1,1)
\facegrid{(0,0)}{(1,1)}
\rput[bl](0,0){\loopb}
\end{pspicture}
\label{1x1}
\ee
where
\be 
 s_k(u):=\frac{\sin (u+k\lambda)}{\sin\lambda},\qquad k\in\oZ
\label{sk}
\ee
These face operators are crossing symmetric
\be
\psset{unit=.77cm}
\begin{pspicture}[shift=-.42](1,1)
\facegrid{(0,0)}{(1,1)}
\psarc[linewidth=0.025]{-}(0,0){0.16}{0}{90}
\rput(.5,.5){$_u$}
\end{pspicture}
\ =\
\begin{pspicture}[shift=-.42](1,1)
\facegrid{(0,0)}{(1,1)}
\psarc[linewidth=0.025]{-}(1,0){0.16}{90}{180}
\rput(.5,.5){$_{\lambda-u}$}
\end{pspicture}
\ =\
\begin{pspicture}[shift=-.42](1,1)
\facegrid{(0,0)}{(1,1)}
\psarc[linewidth=0.025]{-}(1,1){0.16}{180}{270}
\rput(.5,.5){$_u$}
\end{pspicture}
\ =\ 
\begin{pspicture}[shift=-.42](1,1)
\facegrid{(0,0)}{(1,1)}
\psarc[linewidth=0.025]{-}(0,1){0.16}{-90}{0}
\rput(.5,.5){$_{\lambda-u}$}
\end{pspicture}
\label{crossing11}
\ee
and satisfy the Yang-Baxter equation (YBE)
\be
\psset{unit=.77cm}
\begin{pspicture}[shift=-0.89](0,0)(3,2)
\facegrid{(2,0)}{(3,2)}
\pspolygon[fillstyle=solid,fillcolor=lightlightblue](0,1)(1,2)(2,1)(1,0)(0,1)
\psarc[linewidth=0.025]{-}(2,0){0.16}{0}{90}
\psarc[linewidth=0.025]{-}(2,1){0.16}{0}{90}
\psarc[linewidth=0.025]{-}(0,1){0.16}{-45}{45}
\rput(2.5,.5){$_u$}
\rput(2.5,1.5){$_v$}
\rput(1,1){$_{u-v}$}
\psline[linecolor=blue,linewidth=1.5pt]{-}(2,0.5)(1.5,0.5)
\psline[linecolor=blue,linewidth=1.5pt]{-}(2,1.5)(1.5,1.5)
\end{pspicture}
\ \ =\ \ 
\begin{pspicture}[shift=-0.89](0,0)(3,2)
\facegrid{(0,0)}{(1,2)}
\pspolygon[fillstyle=solid,fillcolor=lightlightblue](1,1)(2,2)(3,1)(2,0)(1,1)
\psarc[linewidth=0.025]{-}(0,0){0.16}{0}{90}
\psarc[linewidth=0.025]{-}(0,1){0.16}{0}{90}
\psarc[linewidth=0.025]{-}(1,1){0.16}{-45}{45}
\rput(.5,.5){$_v$}
\rput(.5,1.5){$_u$}
\rput(2,1){$_{u-v}$}
\psline[linecolor=blue,linewidth=1.5pt]{-}(1,0.5)(1.5,0.5)
\psline[linecolor=blue,linewidth=1.5pt]{-}(1,1.5)(1.5,1.5)
\end{pspicture}\ \
\label{YBE}
\ee
The face operators also commute in the sense that
\be
\psset{unit=.6cm}
\begin{pspicture}[shift=-0.89](0,0)(4,2)
\pspolygon[fillstyle=solid,fillcolor=lightlightblue](0,1)(1,2)(2,1)(1,0)(0,1)
\psarc[linewidth=1.5pt,linecolor=blue](2,1){.7}{45}{135}
\psarc[linewidth=1.5pt,linecolor=blue](2,1){.7}{-135}{-45}
\psarc[linewidth=0.025]{-}(0,1){0.16}{-45}{45}
\pspolygon[fillstyle=solid,fillcolor=lightlightblue](2,1)(3,2)(4,1)(3,0)(2,1)
\psarc[linewidth=0.025]{-}(2,1){0.16}{-45}{45}
\rput(1,1){$_{u}$}
\rput(3,1){$_{v}$}
\end{pspicture}
\ = \
\begin{pspicture}[shift=-0.89](0,0)(4,2)
\pspolygon[fillstyle=solid,fillcolor=lightlightblue](0,1)(1,2)(2,1)(1,0)(0,1)
\psarc[linewidth=1.5pt,linecolor=blue](2,1){.7}{45}{135}
\psarc[linewidth=1.5pt,linecolor=blue](2,1){.7}{-135}{-45}
\psarc[linewidth=0.025]{-}(0,1){0.16}{-45}{45}
\pspolygon[fillstyle=solid,fillcolor=lightlightblue](2,1)(3,2)(4,1)(3,0)(2,1)
\psarc[linewidth=0.025]{-}(2,1){0.16}{-45}{45}
\rput(1,1){$_{v}$}
\rput(3,1){$_{u}$}
\end{pspicture}
\ee
and satisfy the local inversion relation
\be
\psset{unit=.6cm}
\begin{pspicture}[shift=-0.89](0,0)(4,2)
\pspolygon[fillstyle=solid,fillcolor=lightlightblue](0,1)(1,2)(2,1)(1,0)(0,1)
\psarc[linewidth=1.5pt,linecolor=blue](2,1){.7}{45}{135}
\psarc[linewidth=1.5pt,linecolor=blue](2,1){.7}{-135}{-45}
\psarc[linewidth=0.025]{-}(0,1){0.16}{-45}{45}
\pspolygon[fillstyle=solid,fillcolor=lightlightblue](2,1)(3,2)(4,1)(3,0)(2,1)
\psarc[linewidth=0.025]{-}(2,1){0.16}{-45}{45}
\rput(1,1){$_{u}$}
\rput(3,1){$_{-u}$}
\end{pspicture}
\ =\,s_1(u)s_1(-u) \
\begin{pspicture}[shift=-0.89](0,0)(2,2)
\pspolygon[fillstyle=solid,fillcolor=lightlightblue](0,1)(1,2)(2,1)(1,0)(0,1)
\psarc[linewidth=1.5pt,linecolor=blue](1,0){.7}{45}{135}
\psarc[linewidth=1.5pt,linecolor=blue](1,2){.7}{-135}{-45}
\end{pspicture}
\label{inv}
\ee

The bulk face operators generate a planar TL algebra~\cite{TL,Jones} 
where multiplication is performed by gluing or linking diagrams together.
Here we are interested in the model defined on a strip, so most products are formed by simply stacking face operators together to
form parts of the rectangular lattice. As indicated in Figure~\ref{conf}, we furthermore choose vertical as the direction of transfer
in which case the ensuing diagram algebra in the bulk is generated by the $N$-tangles
\be
 I:=\!
\begin{pspicture}[shift=-0.55](-0.1,-0.65)(2.0,0.45)
\rput(1.4,0.0){\small$...$}
\psline[linecolor=blue,linewidth=1.5pt]{-}(0.2,0.35)(0.2,-0.35)\rput(0.2,-0.55){$_1$}
\psline[linecolor=blue,linewidth=1.5pt]{-}(0.6,0.35)(0.6,-0.35)
\psline[linecolor=blue,linewidth=1.5pt]{-}(1.0,0.35)(1.0,-0.35)
\psline[linecolor=blue,linewidth=1.5pt]{-}(1.8,0.35)(1.8,-0.35)\rput(1.8,-0.55){$_N$}
\end{pspicture} 
 ,\qquad
 e_j:=\!
 \begin{pspicture}[shift=-0.55](-0.1,-0.65)(3.2,0.45)
\rput(0.6,0.0){\small$...$}
\rput(2.6,0.0){\small$...$}
\psline[linecolor=blue,linewidth=1.5pt]{-}(0.2,0.35)(0.2,-0.35)\rput(0.2,-0.55){$_1$}
\psline[linecolor=blue,linewidth=1.5pt]{-}(1.0,0.35)(1.0,-0.35)
\psline[linecolor=blue,linewidth=1.5pt]{-}(2.2,0.35)(2.2,-0.35)
\psline[linecolor=blue,linewidth=1.5pt]{-}(3.0,0.35)(3.0,-0.35)\rput(3.0,-0.55){$_{N}$}
\psarc[linecolor=blue,linewidth=1.5pt]{-}(1.6,0.35){0.2}{180}{0}\rput(1.4,-0.55){$_j$}
\psarc[linecolor=blue,linewidth=1.5pt]{-}(1.6,-0.35){0.2}{0}{180}
\end{pspicture} 
,\qquad j=1,\ldots,N-1
\ee
In this setting, 
multiplication is by vertical concatenation of diagrams placing the $N$-tangle $c_2$ atop the $N$-tangle $c_1$ to form the product $c_1c_2$,
and the algebra is recognised as the usual loop representation of the ordinary TL algebra $TL_N(\beta)$ on $N$ nodes or 
strands. This bulk algebra is extended in Section~\ref{Sec:OneTL} to handle the Robin boundary conditions along the right edge of the 
strip. As an element of $TL_N(\beta)$, the face operator in (\ref{1x1}), turned $45^\circ$ in the counterclockwise direction, reads
\be
 X_j(u)=s_1(-u)I+s_0(u)e_j
\ee
and satisfies the YBE
\be
 X_{j+1}(u)X_{j}(u+v)X_{j+1}(v)=X_j(v)X_{j+1}(u+v)X_j(u)
\ee

\section{Robin boundary conditions}
\label{Sec:Robin}

\subsection{Twist boundary condition}

Similar to the construction of the bulk face operator (\ref{1x1}), a boundary triangle is defined as a linear combination of the
Neumann and Dirichlet boundary configurations (\ref{boundary}) where the coefficients are chosen such that the triangles satisfy the
boundary Yang-Baxter equation (BYBE)
\be
\psset{unit=.6cm}
\begin{pspicture}[shift=-1.89](0,0)(3,4)
\pspolygon[fillstyle=solid,fillcolor=lightlightblue](0,1)(1,2)(2,1)(1,0)(0,1)
\pspolygon[fillstyle=solid,fillcolor=lightlightblue](1,2)(2,3)(3,2)(2,1)(1,2)
\pspolygon[fillstyle=solid,fillcolor=lightlightblue](2,1)(3,2)(3,0)(2,1)
\pspolygon[fillstyle=solid,fillcolor=lightlightblue](2,3)(3,4)(3,2)(2,3)
\psarc[linewidth=1.5pt,linecolor=blue](2,1){.7}{-135}{-45}
\psarc[linewidth=0.025]{-}(0,1){0.16}{-45}{45}
\psarc[linewidth=0.025]{-}(1,2){0.16}{-45}{45}
\rput(1,1){$_{u-v}$}
\rput(2.05,2){$_{\lambda-u-v}$}
\rput(2.65,1){$_{u}$}
\rput(2.65,3){$_{v}$}
\end{pspicture}
\quad =\ \
\begin{pspicture}[shift=-1.89](0,0)(3,4)
\pspolygon[fillstyle=solid,fillcolor=lightlightblue](0,3)(1,4)(2,3)(1,2)(0,3)
\pspolygon[fillstyle=solid,fillcolor=lightlightblue](1,2)(2,3)(3,2)(2,1)(1,2)
\pspolygon[fillstyle=solid,fillcolor=lightlightblue](2,1)(3,2)(3,0)(2,1)
\pspolygon[fillstyle=solid,fillcolor=lightlightblue](2,3)(3,4)(3,2)(2,3)
\psarc[linewidth=1.5pt,linecolor=blue](2,3){.7}{45}{135}
\psarc[linewidth=0.025]{-}(0,3){0.16}{-45}{45}
\psarc[linewidth=0.025]{-}(1,2){0.16}{-45}{45}
\rput(1,3){$_{u-v}$}
\rput(2.05,2){$_{\lambda-u-v}$}
\rput(2.65,1){$_{v}$}
\rput(2.65,3){$_{u}$}
\end{pspicture}
\label{BYBE1}
\ee
Using the crossing symmetry (\ref{crossing11}), this is readily seen to be equivalent to
\be
\psset{unit=.6cm}
\begin{pspicture}[shift=-1.89](0,-1)(2,4)
\pspolygon[fillstyle=solid,fillcolor=lightlightblue](0,0)(1,1)(2,0)(1,-1)(0,0)
\pspolygon[fillstyle=solid,fillcolor=lightlightblue](0,2)(1,3)(2,2)(1,1)(0,2)
\pspolygon[fillstyle=solid,fillcolor=lightlightblue](1,1)(2,2)(2,0)(1,1)
\pspolygon[fillstyle=solid,fillcolor=lightlightblue](1,3)(2,4)(2,2)(1,3)
\psarc[linewidth=1.5pt,linecolor=blue](1,1){.7}{135}{225}
\psarc[linewidth=0.025](1,-1){0.16}{45}{135}
\psarc[linewidth=0.025](1,1){0.16}{45}{135}
\rput(1,0){$_{u-v}$}
\rput(1,2){$_{u+v}$}
\rput(1.65,1){$_{u}$}
\rput(1.65,3){$_{v}$}
\end{pspicture}
\quad =\ \ 
\begin{pspicture}[shift=-1.89](0,0)(2,4)
\pspolygon[fillstyle=solid,fillcolor=lightlightblue](0,4)(1,5)(2,4)(1,3)(0,4)
\pspolygon[fillstyle=solid,fillcolor=lightlightblue](0,2)(1,3)(2,2)(1,1)(0,2)
\pspolygon[fillstyle=solid,fillcolor=lightlightblue](1,1)(2,2)(2,0)(1,1)
\pspolygon[fillstyle=solid,fillcolor=lightlightblue](1,3)(2,4)(2,2)(1,3)
\psarc[linewidth=1.5pt,linecolor=blue](1,3){.7}{135}{225}
\psarc[linewidth=0.025](1,3){0.16}{45}{135}
\psarc[linewidth=0.025](1,1){0.16}{45}{135}
\rput(1,4){$_{u-v}$}
\rput(1,2){$_{u+v}$}
\rput(1.65,1){$_{v}$}
\rput(1.65,3){$_{u}$}
\end{pspicture}
\label{BYBE2}
\ee
A simple solution is provided by the Neumann boundary conditions.

Here we are interested in solutions of the form
\be
\psset{unit=.6cm}
\begin{pspicture}[shift=-0.89](0,0)(1,2)
\pspolygon[fillstyle=solid,fillcolor=lightlightblue](0,1)(1,2)(1,0)(0,1)
\rput(0.65,1){$_u$}
\end{pspicture}
\ =\Gamma(u)\
\begin{pspicture}[shift=-0.89](0,0)(1,2)
\pspolygon[fillstyle=solid,fillcolor=lightlightblue](0,1)(1,2)(1,0)(0,1)
\psarc[linewidth=1.5pt,linecolor=blue](0,1){.7}{-45}{45}
\end{pspicture}
\ +s_0(2u)\
\begin{pspicture}[shift=-0.89](0,0)(1,2)
\pspolygon[fillstyle=solid,fillcolor=lightlightblue](0,1)(1,2)(1,0)(0,1)
\psline[linecolor=blue,linewidth=1.5pt]{-}(0.4,0.6)(1,0.6)
\psline[linecolor=blue,linewidth=1.5pt]{-}(0.4,1.4)(1,1.4)
\end{pspicture}
\label{twist}
\ee
where $\Gamma(u)$ is analytic and 
where the coefficient in front of the Dirichlet term has been chosen to ensure that the triangle (\ref{twist})
reduces to a Neumann term as $u\to0$,
\be
\psset{unit=.6cm}
\lim_{u\to0}\,
\begin{pspicture}[shift=-0.89](0,0)(1,2)
\pspolygon[fillstyle=solid,fillcolor=lightlightblue](0,1)(1,2)(1,0)(0,1)
\rput(0.65,1){$_u$}
\end{pspicture}
\ =\Gamma(0)\ 
\begin{pspicture}[shift=-0.89](0,0)(1,2)
\pspolygon[fillstyle=solid,fillcolor=lightlightblue](0,1)(1,2)(1,0)(0,1)
\psarc[linewidth=1.5pt,linecolor=blue](0,1){.7}{-45}{45}
\end{pspicture}
\label{uG0}
\ee 
More general solutions will be discussed in Section~\ref{Sec:GenRobin}.
\begin{Proposition}
The general solution of the form (\ref{twist}) to the BYBE (\ref{BYBE2}) is given by
\be
 \Gamma(u)=\Gamma_{\gamma}(u):=\gamma-\beta_1[s_0(u)]^2+\beta_2[s_0(u-\tfrac{\lambda}{2})]^2,\qquad \gamma\in\mathbb{C}
\label{GGg}
\ee
\label{Prop:BYBE}
\end{Proposition}
\noindent{\scshape Proof:} 
Each side of the BYBE (\ref{BYBE2}) decomposes into the six connectivity diagrams
\be
\begin{pspicture}[shift=-0.4](0,0)(1,1)
\psline[linecolor=blue,linewidth=1.5pt](0,0)(0,1)
\psline[linecolor=blue,linewidth=1.5pt](0.5,0)(0.5,1)
\end{pspicture} 
\qquad
\begin{pspicture}[shift=-0.4](0,0)(1,1)
\psarc[linecolor=blue,linewidth=1.5pt](0.25,0){0.25}{0}{180}
\psarc[linecolor=blue,linewidth=1.5pt](0.25,1){0.25}{180}{360}
\end{pspicture} 
\qquad
\begin{pspicture}[shift=-0.4](0,0)(1,1)
\psline[linecolor=blue,linewidth=1.5pt](0,0)(0,1)
\psbezier[linecolor=blue,linewidth=1.5pt](0.5,0)(0.5,0.25)(0.65,0.25)(0.75,0.25)
\psbezier[linecolor=blue,linewidth=1.5pt](0.5,1)(0.5,0.75)(0.65,0.75)(0.75,0.75)
\rput(0.75,0.25){$_\bullet$}
\rput(0.75,0.75){$_\bullet$}
\end{pspicture} 
\qquad\ \
\begin{pspicture}[shift=-0.4](0,0)(1,1)
\psbezier[linecolor=blue,linewidth=1.5pt](0,0)(0,0.4)(0.4,0.4)(0.75,0.4)
\psbezier[linecolor=blue,linewidth=1.5pt](0.5,0)(0.5,0.2)(0.65,0.2)(0.75,0.2)
\psbezier[linecolor=blue,linewidth=1.5pt](0,1)(0,0.6)(0.4,0.6)(0.75,0.6)
\psbezier[linecolor=blue,linewidth=1.5pt](0.5,1)(0.5,0.8)(0.65,0.8)(0.75,0.8)
\rput(0.75,0.2){$_\bullet$}
\rput(0.75,0.4){$_\bullet$}
\rput(0.75,0.6){$_\bullet$}
\rput(0.75,0.8){$_\bullet$}
\end{pspicture} 
\qquad\ \
\begin{pspicture}[shift=-0.4](0,0)(1,1)
\psarc[linecolor=blue,linewidth=1.5pt](0.25,1){0.25}{180}{360}
\psbezier[linecolor=blue,linewidth=1.5pt](0,0)(0,0.4)(0.4,0.4)(0.75,0.4)
\psbezier[linecolor=blue,linewidth=1.5pt](0.5,0)(0.5,0.2)(0.65,0.2)(0.75,0.2)
\rput(0.75,0.2){$_\bullet$}
\rput(0.75,0.4){$_\bullet$}
\end{pspicture} 
\qquad\ \
\begin{pspicture}[shift=-0.4](0,0)(1,1)
\psarc[linecolor=blue,linewidth=1.5pt](0.25,0){0.25}{0}{180}
\psbezier[linecolor=blue,linewidth=1.5pt](0,1)(0,0.6)(0.4,0.6)(0.75,0.6)
\psbezier[linecolor=blue,linewidth=1.5pt](0.5,1)(0.5,0.8)(0.65,0.8)(0.75,0.8)
\rput(0.75,0.6){$_\bullet$}
\rput(0.75,0.8){$_\bullet$}
\end{pspicture} 
\ee
The four first are reflection symmetric with respect to a horizontal line and the corresponding decomposition coefficients
on the two sides of the equation match for all $u,v$. Requiring that the coefficients to the fifth connectivity also match yields
the relation
\bea
 &&\hspace{-2cm}s_1(-u+v)s_0(2u)s_0(u+v)\Gamma(v)\nn
 &&=s_0(2v)s_0(u+v)\Gamma(u)s_1(-u+v)+\Gamma(v)s_1(-u-v)s_0(2u)s_0(u-v)\nn
  &&+\;s_0(2v)s_1(-u-v)\Gamma(u)s_0(u-v)+\beta s_0(2v)s_0(u+v)\Gamma(u)s_0(u-v)\nn
 &&+\;\beta_1s_0(2v)s_0(u+v)s_0(2u)s_0(u-v)+\beta_2s_0(2v)s_1(-u-v)s_0(2u)s_0(u-v)
\label{trigrel}
\eea
Since the fifth and sixth diagrams are mapped into one another under the reflection above, the exact same relation
follows from matching up the coefficients of the sixth connectivity, as is readily verified.
Manipulations of the trigonometric functions now allow us to write the relation (\ref{trigrel}) as
\be
 0=s_0(2u)s_0(2v)\Big(\big(\Gamma(u)+\beta_1[s_0(u)]^2-\beta_2[s_0(u-\tfrac{\lambda}{2})]^2\big)-
  \big(\Gamma(v)+\beta_1[s_0(v)]^2-\beta_2[s_0(v-\tfrac{\lambda}{2})]^2\big)\Big)
\ee
As this is required to hold for all $u,v$, the general solution for analytic $\Gamma$ is given by (\ref{GGg}).
\hfill $\square$
\medskip
%

Adopting terminology of~\cite{KPS2004,JSbdy}, we shall refer to (\ref{twist}) as the {\em twist boundary condition}. 
As is easily verified, the corresponding boundary tangles satisfy the local boundary crossing relation
\be
\psset{unit=.77cm}
\begin{pspicture}[shift=-0.89](0,0)(3,2)
\pspolygon[fillstyle=solid,fillcolor=lightlightblue](0,1)(1,2)(2,1)(1,0)(0,1)
\psarc[linewidth=1.5pt,linecolor=blue](2,1){.7}{45}{135}
\psarc[linewidth=1.5pt,linecolor=blue](2,1){.7}{-135}{-45}
\psarc[linewidth=0.025]{-}(0,1){0.16}{-45}{45}
\pspolygon[fillstyle=solid,fillcolor=lightlightblue](2,1)(3,2)(3,0)(2,1)
\rput(1,1){$_{2u-\lambda}$}
\rput(2.65,1){$_{u}$}
\end{pspicture}
\ =s_0(2u)\
\begin{pspicture}[shift=-0.89](0,0)(1,2)
\pspolygon[fillstyle=solid,fillcolor=lightlightblue](0,1)(1,2)(1,0)(0,1)
\rput(0.58,1){$_{\lambda-u}$}
\end{pspicture}
\label{btt}
\ee
It is also noted that the Neumann boundary conditions along the left edge of the strip lattice in Figure~\ref{conf} satisfy
\be
\psset{unit=.77cm}
\begin{pspicture}[shift=-0.89](0,0)(3,2)
\pspolygon[fillstyle=solid,fillcolor=lightlightblue](0,0)(1,1)(0,2)(0,0)
\pspolygon[fillstyle=solid,fillcolor=lightlightblue](1,1)(2,0)(3,1)(2,2)(1,1)
\psarc[linewidth=1.5pt,linecolor=blue](1,1){.7}{45}{-45}
\psarc[linewidth=0.025](1,1){0.16}{-45}{45}
\rput(2,1){$_{\lambda-2u}$}
\end{pspicture}
\ =s_2(-2u)\
\begin{pspicture}[shift=-0.89](0,0)(1,2)
\pspolygon[fillstyle=solid,fillcolor=lightlightblue](0,0)(1,1)(0,2)(0,0)
\psarc[linewidth=1.5pt,linecolor=blue](1,1){.7}{135}{-135}
\end{pspicture}
\label{Nb}
\vspace{.1cm}
\ee
It is furthermore stressed that, unlike the Neumann boundary condition, the Dirichlet boundary condition alone does not in general
provide a solution to the BYBE (\ref{BYBE2}).

\subsection{One-boundary Temperley-Lieb algebra}
\label{Sec:OneTL}

Writing 
\be
\psset{unit=.6cm}
K_N(u):=\!
\begin{pspicture}[shift=-0.89](0,0)(5,2)
\psline[linecolor=blue,linewidth=1.5pt](0.5,0)(0.5,2)
\psline[linecolor=blue,linewidth=1.5pt](1.5,0)(1.5,2)
\rput(2.5,1){$\ldots$}
\psline[linecolor=blue,linewidth=1.5pt](3.5,0)(3.5,2)
\pspolygon[fillstyle=solid,fillcolor=lightlightblue](4,1)(5,2)(5,0)(4,1)
\psline[linecolor=blue,linewidth=1.5pt](4.5,1.5)(4.5,2)
\psline[linecolor=blue,linewidth=1.5pt](4.5,0)(4.5,0.5)
\rput(4.65,1){$_u$}
\end{pspicture}
\ =\Gamma(u)I+s_0(2u)f_N
\ee
the diagram algebra discussed at the end of Section~\ref{Sec:Bulk} is extended by an element $f_N$ taking the
Dirichlet boundary condition into account. The ensuing diagram algebra is thus generated by
\be
 I:=\!
\begin{pspicture}[shift=-0.55](-0.1,-0.65)(2.0,0.45)
\rput(1.4,0.0){\small$...$}
\psline[linecolor=blue,linewidth=1.5pt](0.2,0.35)(0.2,-0.35)\rput(0.2,-0.55){$_1$}
\psline[linecolor=blue,linewidth=1.5pt](0.6,0.35)(0.6,-0.35)
\psline[linecolor=blue,linewidth=1.5pt](1.0,0.35)(1.0,-0.35)
\psline[linecolor=blue,linewidth=1.5pt](1.8,0.35)(1.8,-0.35)\rput(1.8,-0.55){$_N$}
\end{pspicture} 
 ,\qquad
 e_j:=\!
 \begin{pspicture}[shift=-0.55](-0.1,-0.65)(3.2,0.45)
\rput(0.6,0.0){\small$...$}
\rput(2.6,0.0){\small$...$}
\psline[linecolor=blue,linewidth=1.5pt](0.2,0.35)(0.2,-0.35)\rput(0.2,-0.55){$_1$}
\psline[linecolor=blue,linewidth=1.5pt](1.0,0.35)(1.0,-0.35)
\psline[linecolor=blue,linewidth=1.5pt](2.2,0.35)(2.2,-0.35)
\psline[linecolor=blue,linewidth=1.5pt](3.0,0.35)(3.0,-0.35)\rput(3.0,-0.55){$_{N}$}
\psarc[linecolor=blue,linewidth=1.5pt](1.6,0.35){0.2}{180}{0}\rput(1.4,-0.55){$_j$}
\psarc[linecolor=blue,linewidth=1.5pt](1.6,-0.35){0.2}{0}{180}
\end{pspicture} 
 ,\qquad
 f_N:=\!
\begin{pspicture}[shift=-0.55](-0.1,-0.65)(2.0,0.45)
\rput(1.4,0.0){\small$...$}
\psline[linecolor=blue,linewidth=1.5pt](0.2,0.35)(0.2,-0.35)\rput(0.2,-0.55){$_1$}
\psline[linecolor=blue,linewidth=1.5pt](0.6,0.35)(0.6,-0.35)
\psline[linecolor=blue,linewidth=1.5pt](1.0,0.35)(1.0,-0.35)
\psline[linecolor=blue,linewidth=1.5pt](1.8,0.35)(1.8,-0.35)
\psarc[linecolor=blue,linewidth=1.5pt](2.4,-0.35){0.2}{90}{180}
\psarc[linecolor=blue,linewidth=1.5pt](2.4,0.35){0.2}{180}{-90}\rput(2.2,-0.55){$_{N}$}
\end{pspicture} 
\label{Ief}
\ee
where $j=1,\ldots,N-1$,
and where multiplication is by vertical concatenation of diagrams. 
This algebra is a loop representation of the one-boundary TL algebra 
\be
 TL_N(\beta;\beta_1,\beta_2):=\big\langle I,\, e_j,\, f_N;\ j=1,\ldots, N-1\big\rangle
\ee
which is a unital algebra, with identity $I$, defined by the relations
\be
 \begin{array}{rcll}
 \left[e_i,e_j\right]&\!\!\!=\!\!\!&0,\qquad\quad &|i-j|>1\\[.1cm]
 e_ie_je_i&\!\!\!=\!\!\!&e_i,\qquad\quad &|i-j|=1\\[.1cm]
 e_j^2&\!\!\!=\!\!\!&\beta e_j,\qquad\quad &j=1,\ldots, N-1\\[.1cm]
 \left[e_j,f_N\right]&\!\!\!=\!\!\!&0,\qquad\quad &j=1,\ldots, N-2\\[.1cm]
 e_{N-1}f_Ne_{N-1}&\!\!\!=\!\!\!&\beta_1 e_{N-1}&\\[.1cm]
 f_N^2&\!\!\!=\!\!\!&\beta_2 f_N&
 \end{array}
\label{1TLN}
\ee
The relation between this one-boundary TL algebra and the so-called blob algebra~\cite{MaSa93}
is discussed in Appendix~\ref{Sec:Blob}.
Diagrammatically, the fundamental nontrivial relations involving $f_N$ are given by
\be
 e_{N-1}f_Ne_{N-1}=
\begin{pspicture}[shift=-1](-0.1,-1.1)(2.9,1.15)
\rput(1.4,-0.7){\small$...$}
\rput(1.4,0){\small$...$}
\rput(1.4,0.7){\small$...$}
\psline[linecolor=blue,linewidth=1.5pt](0.2,1.05)(0.2,-1.05)
\psline[linecolor=blue,linewidth=1.5pt](0.6,1.05)(0.6,-1.05)
\psline[linecolor=blue,linewidth=1.5pt](1.0,1.05)(1.0,-1.05)
\psline[linecolor=blue,linewidth=1.5pt](1.8,1.05)(1.8,-1.05)
\psline[linecolor=blue,linewidth=1.5pt](2.2,0.35)(2.2,-0.35)
\psarc[linecolor=blue,linewidth=1.5pt](2.4,-1.05){0.2}{0}{180}
\psarc[linecolor=blue,linewidth=1.5pt](2.4,0.35){0.2}{0}{180}
\psarc[linecolor=blue,linewidth=1.5pt](2.4,1.05){0.2}{180}{0}
\psarc[linecolor=blue,linewidth=1.5pt](2.4,-0.35){0.2}{180}{0}
\psarc[linecolor=blue,linewidth=1.5pt](2.8,-0.35){0.2}{90}{180}
\psarc[linecolor=blue,linewidth=1.5pt](2.8,0.35){0.2}{180}{-90}
\psline[linewidth=0.5pt,linestyle=dashed, dash=1pt 1pt](0,0.35)(2.8,0.35)
\psline[linewidth=0.5pt,linestyle=dashed, dash=1pt 1pt](0,-0.35)(2.8,-0.35)
\rput(0.2,-1.25){$_1$}
\rput(2.6,-1.25){$_{N}$}
\end{pspicture} 
\;=
\beta_1\times\!\!
\begin{pspicture}[shift=-1](-0.1,-1.1)(2.9,0.45)
\rput(1.4,0){\small$...$}
\psline[linecolor=blue,linewidth=1.5pt](0.2,0.35)(0.2,-0.35)
\psline[linecolor=blue,linewidth=1.5pt](0.6,0.35)(0.6,-0.35)
\psline[linecolor=blue,linewidth=1.5pt](1.0,0.35)(1.0,-0.35)
\psline[linecolor=blue,linewidth=1.5pt](1.8,0.35)(1.8,-0.35)
\psarc[linecolor=blue,linewidth=1.5pt](2.4,-0.35){0.2}{0}{180}
\psarc[linecolor=blue,linewidth=1.5pt](2.4,0.35){0.2}{180}{0}
\rput(0.2,-0.55){$_1$}
\rput(2.6,-0.55){$_{N}$}
\end{pspicture} 
=\beta_1e_{N-1}
\ee
and
\be
 f_N^2=
\begin{pspicture}[shift=-0.55](-0.1,-0.65)(2.5,1)
\rput(1.4,-0.35){\small$...$}
\rput(1.4,0.35){\small$...$}
\psline[linecolor=blue,linewidth=1.5pt](0.2,0.7)(0.2,-0.7)\rput(0.2,-0.9){$_1$}
\psline[linecolor=blue,linewidth=1.5pt](0.6,0.7)(0.6,-0.7)
\psline[linecolor=blue,linewidth=1.5pt](1.0,0.7)(1.0,-0.7)
\psline[linecolor=blue,linewidth=1.5pt](1.8,0.7)(1.8,-0.7)
\psarc[linecolor=blue,linewidth=1.5pt](2.4,-0.7){0.2}{90}{180}
\psarc[linecolor=blue,linewidth=1.5pt](2.4,0){0.2}{90}{270}
\psarc[linecolor=blue,linewidth=1.5pt](2.4,0.7){0.2}{180}{-90}
\psline[linewidth=0.5pt,linestyle=dashed, dash=1pt 1pt](0,0)(2.4,0)
\rput(2.2,-0.9){$_{N}$}
\end{pspicture} 
\;=
\beta_2\times\!\!
\begin{pspicture}[shift=-0.55](-0.1,-0.65)(2.5,0.45)
\rput(1.4,0.0){\small$...$}
\psline[linecolor=blue,linewidth=1.5pt](0.2,0.35)(0.2,-0.35)\rput(0.2,-0.55){$_1$}
\psline[linecolor=blue,linewidth=1.5pt](0.6,0.35)(0.6,-0.35)
\psline[linecolor=blue,linewidth=1.5pt](1.0,0.35)(1.0,-0.35)
\psline[linecolor=blue,linewidth=1.5pt](1.8,0.35)(1.8,-0.35)
\psarc[linecolor=blue,linewidth=1.5pt](2.4,-0.35){0.2}{90}{180}
\psarc[linecolor=blue,linewidth=1.5pt](2.4,0.35){0.2}{180}{-90}\rput(2.2,-0.55){$_{N}$}
\end{pspicture} 
\;=\beta_2f_N
\\[.4cm]
\ee
As a relation in $TL_N(\beta;\beta_1,\beta_2)$, the BYBE (\ref{BYBE2}) reads
\be
 X_{N-1}(u-v)K_N(u)X_{N-1}(u+v)K_N(v)=K_N(v)X_{N-1}(u+v)K_N(u)X_{N-1}(u-v)
\ee
while the boundary crossing relation (\ref{btt}) can be expressed as
\be
 X_{N-1}(2u-\la)K_N(u)e_{N-1}=s_0(2u) K_N(\la-u)e_{N-1}
\ee
or equivalently as
\be
 e_{N-1}K_N(u)X_{N-1}(2u-\la)=s_0(2u)e_{N-1}K_N(\la-u)
\ee

\subsection{General Robin boundary conditions}
\label{Sec:GenRobin}

We now generalise the twist boundary condition (\ref{twist}) by adding a boundary seam of width $w\in\oN_0$.
For vanishing seam width, $w=0$, the construction is meant to reduce to the twist boundary condition (\ref{twist}).
\begin{Proposition}
For every $w\in\oN_0$ and $\xi\in\mathbb{C}$, the Robin boundary condition  
\be
\psset{unit=1cm}
\begin{pspicture}[shift=-0.89](0,0)(1,2)
\pspolygon[fillstyle=solid,fillcolor=lightlightblue](0,1)(1,2)(1,0)(0,1)
\psline[linecolor=blue,linewidth=1.5pt](0,0.5)(0.5,0.5)
\psline[linecolor=blue,linewidth=1.5pt](0,1.5)(0.5,1.5)
\rput(0.6,1){$_{u,\xi}$}
\end{pspicture}
\ \ \ =\ 
\begin{pspicture}[shift=-0.89](0.4,0)(6.5,2)
\facegrid{(1,0)}{(5,2)}
\pspolygon[fillstyle=solid,fillcolor=lightlightblue](5,1)(6,2)(6,0)(5,1)
\psarc[linewidth=0.025](1,0){0.16}{0}{90}
\psarc[linewidth=0.025](1,1){0.16}{0}{90}
\psarc[linewidth=0.025](3,0){0.16}{0}{90}
\psarc[linewidth=0.025](3,1){0.16}{0}{90}
\psarc[linewidth=0.025](4,0){0.16}{0}{90}
\psarc[linewidth=0.025](4,1){0.16}{0}{90}
\rput(1.55,0.5){$_{u-\xi_{w}}$}
\rput(3.55,0.5){$_{u-\xi_2}$}
\rput(4.55,0.5){$_{u-\xi_1}$}
\rput(1.5,1.5){$_{-\!u\!-\!\xi_{w\!-\!1}}$}
\rput(3.5,1.5){$_{-u-\xi_1}$}
\rput(4.5,1.5){$_{-u-\xi_0}$}
\rput(2.5,0.5){$\ldots$}
\rput(2.5,1.5){$\ldots$}
\rput(5.65,1){$_{u}$}
\psline[linecolor=blue,linewidth=1.5pt](0.5,0.5)(1,0.5)
\psline[linecolor=blue,linewidth=1.5pt](0.5,1.5)(1,1.5)
\psline[linecolor=blue,linewidth=1.5pt](5,0.5)(5.5,0.5)
\psline[linecolor=blue,linewidth=1.5pt](5,1.5)(5.5,1.5)
\end{pspicture}
\qquad
\xi_k=\xi+k\lambda
\label{uxi}
\ee
is a solution to the BYBE (\ref{BYBE1}). 
\label{Prop:GenRobin}
\end{Proposition}
\noindent{\scshape Proof:} 
Following~\cite{BP96,BP01}, this is proven diagrammatically by
\be
\psset{unit=.6cm}
\begin{pspicture}[shift=-1.89](0,0)(3,4)
\pspolygon[fillstyle=solid,fillcolor=lightlightblue](0,1)(1,2)(2,1)(1,0)(0,1)
\pspolygon[fillstyle=solid,fillcolor=lightlightblue](1,2)(2,3)(3,2)(2,1)(1,2)
\pspolygon[fillstyle=solid,fillcolor=lightlightblue](2,1)(3,2)(3,0)(2,1)
\pspolygon[fillstyle=solid,fillcolor=lightlightblue](2,3)(3,4)(3,2)(2,3)
\psarc[linewidth=1.5pt,linecolor=blue](2,1){.7}{-135}{-45}
\psarc[linewidth=0.025]{-}(0,1){0.16}{-45}{45}
\psarc[linewidth=0.025]{-}(1,2){0.16}{-45}{45}
\rput(1,1){$_{_{u-v}}$}
\rput(2.05,2){$_{_{\lambda-u-v}}$}
\rput(2.65,1){$_{_{u,\xi}}$}
\rput(2.65,3){$_{_{v,\xi}}$}
\end{pspicture}
\quad =\
\begin{pspicture}[shift=-1.89](0,0)(8,4)
\pspolygon[fillstyle=solid,fillcolor=lightlightblue](0,1)(1,2)(2,1)(1,0)(0,1)
\pspolygon[fillstyle=solid,fillcolor=lightlightblue](1,2)(2,3)(3,2)(2,1)(1,2)
\psarc[linewidth=0.025](0,1){0.16}{-45}{45}
\psarc[linewidth=0.025](1,2){0.16}{-45}{45}
\rput(1,1){$_{_{u-v}}$}
\rput(2.05,2){$_{_{\lambda-u-v}}$}
\psline[linecolor=blue,linewidth=1.5pt](1.5,0.5)(3,0.5)
\psline[linecolor=blue,linewidth=1.5pt](2.5,1.5)(3,1.5)
\psline[linecolor=blue,linewidth=1.5pt](2.5,2.5)(3,2.5)
\facegrid{(3,0)}{(4,4)}
\rput(3.5,0.55){$_{\ldots}$}
\rput(3.5,1.55){$_{\ldots}$}
\rput(3.5,2.55){$_{\ldots}$}
\rput(3.5,3.55){$_{\ldots}$}
\pspolygon[fillstyle=solid,fillcolor=lightlightblue](4,0)(6,0)(6,1)(4,1)(4,0)
\pspolygon[fillstyle=solid,fillcolor=lightlightblue](4,1)(6,1)(6,2)(4,2)(4,1)
\pspolygon[fillstyle=solid,fillcolor=lightlightblue](4,2)(6,2)(6,3)(4,3)(4,2)
\pspolygon[fillstyle=solid,fillcolor=lightlightblue](4,3)(6,3)(6,4)(4,4)(4,3)
\psarc[linewidth=0.025](4,0){0.16}{0}{90}
\psarc[linewidth=0.025](4,1){0.16}{0}{90}
\psarc[linewidth=0.025](4,2){0.16}{0}{90}
\psarc[linewidth=0.025](4,3){0.16}{0}{90}
\rput(5.05,0.5){$_{_{u-\xi_k}}$}
\rput(5.05,1.5){$_{_{-u-\xi_{k-1}}}$}
\rput(5.05,2.5){$_{_{v-\xi_k}}$}
\rput(5.05,3.5){$_{_{-v-\xi_{k-1}}}$}
\facegrid{(6,0)}{(7,4)}
\rput(6.5,0.55){$_{\ldots}$}
\rput(6.5,1.55){$_{\ldots}$}
\rput(6.5,2.55){$_{\ldots}$}
\rput(6.5,3.55){$_{\ldots}$}
\pspolygon[fillstyle=solid,fillcolor=lightlightblue](7,1)(8,2)(8,0)(7,1)
\pspolygon[fillstyle=solid,fillcolor=lightlightblue](7,3)(8,4)(8,2)(7,3)
\psline[linecolor=blue,linewidth=1.5pt](7,0.5)(7.5,0.5)
\psline[linecolor=blue,linewidth=1.5pt](7,1.5)(7.5,1.5)
\psline[linecolor=blue,linewidth=1.5pt](7,2.5)(7.5,2.5)
\psline[linecolor=blue,linewidth=1.5pt](7,3.5)(7.5,3.5)
\rput(7.65,1){$_{_{u}}$}
\rput(7.65,3){$_{_{v}}$}
\end{pspicture}
\quad =\ \ \,
\begin{pspicture}[shift=-1.89](0,0)(7,4)
\facegrid{(0,0)}{(1,4)}
\rput(0.5,0.55){$_{\ldots}$}
\rput(0.5,1.55){$_{\ldots}$}
\rput(0.5,2.55){$_{\ldots}$}
\rput(0.5,3.55){$_{\ldots}$}
\pspolygon[fillstyle=solid,fillcolor=lightlightblue](1,0)(3,0)(3,1)(1,1)(1,0)
\pspolygon[fillstyle=solid,fillcolor=lightlightblue](1,1)(3,1)(3,2)(1,2)(1,1)
\pspolygon[fillstyle=solid,fillcolor=lightlightblue](1,2)(3,2)(3,3)(1,3)(1,2)
\pspolygon[fillstyle=solid,fillcolor=lightlightblue](1,3)(3,3)(3,4)(1,4)(1,3)
\psarc[linewidth=0.025](1,0){0.16}{0}{90}
\psarc[linewidth=0.025](1,1){0.16}{0}{90}
\psarc[linewidth=0.025](1,2){0.16}{0}{90}
\psarc[linewidth=0.025](1,3){0.16}{0}{90}
\rput(2.05,0.5){$_{_{v-\xi_k}}$}
\rput(2.05,1.5){$_{_{u-\xi_k}}$}
\rput(2.05,2.5){$_{_{-u-\xi_{k-1}}}$}
\rput(2.05,3.5){$_{_{-v-\xi_{k-1}}}$}
\facegrid{(3,0)}{(4,4)}
\rput(3.5,0.55){$_{\ldots}$}
\rput(3.5,1.55){$_{\ldots}$}
\rput(3.5,2.55){$_{\ldots}$}
\rput(3.5,3.55){$_{\ldots}$}
\pspolygon[fillstyle=solid,fillcolor=lightlightblue](4,1)(5,2)(6,1)(5,0)(4,1)
\pspolygon[fillstyle=solid,fillcolor=lightlightblue](5,2)(6,3)(7,2)(6,1)(5,2)
\psarc[linewidth=0.025](4,1){0.16}{-45}{45}
\psarc[linewidth=0.025](5,2){0.16}{-45}{45}
\rput(5,1){$_{_{u-v}}$}
\rput(6.05,2){$_{_{\lambda-u-v}}$}
\psline[linecolor=blue,linewidth=1.5pt](4,0.5)(4.5,0.5)
\psline[linecolor=blue,linewidth=1.5pt](5.5,0.5)(6.5,0.5)
\psline[linecolor=blue,linewidth=1.5pt](4,1.5)(4.5,1.5)
\psline[linecolor=blue,linewidth=1.5pt](4,2.5)(5.5,2.5)
\psline[linecolor=blue,linewidth=1.5pt](4,3.5)(6.5,3.5)
\pspolygon[fillstyle=solid,fillcolor=lightlightblue](6,1)(7,2)(7,0)(6,1)
\pspolygon[fillstyle=solid,fillcolor=lightlightblue](6,3)(7,4)(7,2)(6,3)
\rput(6.65,1){$_{_{u}}$}
\rput(6.65,3){$_{_{v}}$}
\end{pspicture}
\nonumber
\vspace{.4cm}
\ee
\be
\psset{unit=.6cm}
\quad =\ \ \,
\begin{pspicture}[shift=-1.89](0,0)(7,4)
\facegrid{(0,0)}{(1,4)}
\rput(0.5,0.55){$_{\ldots}$}
\rput(0.5,1.55){$_{\ldots}$}
\rput(0.5,2.55){$_{\ldots}$}
\rput(0.5,3.55){$_{\ldots}$}
\pspolygon[fillstyle=solid,fillcolor=lightlightblue](1,0)(3,0)(3,1)(1,1)(1,0)
\pspolygon[fillstyle=solid,fillcolor=lightlightblue](1,1)(3,1)(3,2)(1,2)(1,1)
\pspolygon[fillstyle=solid,fillcolor=lightlightblue](1,2)(3,2)(3,3)(1,3)(1,2)
\pspolygon[fillstyle=solid,fillcolor=lightlightblue](1,3)(3,3)(3,4)(1,4)(1,3)
\psarc[linewidth=0.025](1,0){0.16}{0}{90}
\psarc[linewidth=0.025](1,1){0.16}{0}{90}
\psarc[linewidth=0.025](1,2){0.16}{0}{90}
\psarc[linewidth=0.025](1,3){0.16}{0}{90}
\rput(2.05,0.5){$_{_{v-\xi_k}}$}
\rput(2.05,1.5){$_{_{u-\xi_k}}$}
\rput(2.05,2.5){$_{_{-u-\xi_{k-1}}}$}
\rput(2.05,3.5){$_{_{-v-\xi_{k-1}}}$}
\facegrid{(3,0)}{(4,4)}
\rput(3.5,0.55){$_{\ldots}$}
\rput(3.5,1.55){$_{\ldots}$}
\rput(3.5,2.55){$_{\ldots}$}
\rput(3.5,3.55){$_{\ldots}$}
\pspolygon[fillstyle=solid,fillcolor=lightlightblue](4,3)(5,4)(6,3)(5,2)(4,3)
\pspolygon[fillstyle=solid,fillcolor=lightlightblue](5,2)(6,3)(7,2)(6,1)(5,2)
\psarc[linewidth=0.025](4,3){0.16}{-45}{45}
\psarc[linewidth=0.025](5,2){0.16}{-45}{45}
\rput(5,3){$_{_{u-v}}$}
\rput(6.05,2){$_{_{\lambda-u-v}}$}
\psline[linecolor=blue,linewidth=1.5pt](4,3.5)(4.5,3.5)
\psline[linecolor=blue,linewidth=1.5pt](5.5,3.5)(6.5,3.5)
\psline[linecolor=blue,linewidth=1.5pt](4,2.5)(4.5,2.5)
\psline[linecolor=blue,linewidth=1.5pt](4,1.5)(5.5,1.5)
\psline[linecolor=blue,linewidth=1.5pt](4,0.5)(6.5,0.5)
\pspolygon[fillstyle=solid,fillcolor=lightlightblue](6,1)(7,2)(7,0)(6,1)
\pspolygon[fillstyle=solid,fillcolor=lightlightblue](6,3)(7,4)(7,2)(6,3)
\rput(6.65,1){$_{_{v}}$}
\rput(6.65,3){$_{_{u}}$}
\end{pspicture}
\quad =\
\begin{pspicture}[shift=-1.89](0,0)(8,4)
\pspolygon[fillstyle=solid,fillcolor=lightlightblue](0,3)(1,4)(2,3)(1,2)(0,3)
\pspolygon[fillstyle=solid,fillcolor=lightlightblue](1,2)(2,3)(3,2)(2,1)(1,2)
\psarc[linewidth=0.025](0,3){0.16}{-45}{45}
\psarc[linewidth=0.025](1,2){0.16}{-45}{45}
\rput(1,3){$_{_{u-v}}$}
\rput(2.05,2){$_{_{\lambda-u-v}}$}
\psline[linecolor=blue,linewidth=1.5pt](1.5,3.5)(3,3.5)
\psline[linecolor=blue,linewidth=1.5pt](2.5,1.5)(3,1.5)
\psline[linecolor=blue,linewidth=1.5pt](2.5,2.5)(3,2.5)
\facegrid{(3,0)}{(4,4)}
\rput(3.5,0.55){$_{\ldots}$}
\rput(3.5,1.55){$_{\ldots}$}
\rput(3.5,2.55){$_{\ldots}$}
\rput(3.5,3.55){$_{\ldots}$}
\pspolygon[fillstyle=solid,fillcolor=lightlightblue](4,0)(6,0)(6,1)(4,1)(4,0)
\pspolygon[fillstyle=solid,fillcolor=lightlightblue](4,1)(6,1)(6,2)(4,2)(4,1)
\pspolygon[fillstyle=solid,fillcolor=lightlightblue](4,2)(6,2)(6,3)(4,3)(4,2)
\pspolygon[fillstyle=solid,fillcolor=lightlightblue](4,3)(6,3)(6,4)(4,4)(4,3)
\psarc[linewidth=0.025](4,0){0.16}{0}{90}
\psarc[linewidth=0.025](4,1){0.16}{0}{90}
\psarc[linewidth=0.025](4,2){0.16}{0}{90}
\psarc[linewidth=0.025](4,3){0.16}{0}{90}
\rput(5.05,0.5){$_{_{v-\xi_k}}$}
\rput(5.05,1.5){$_{_{-v-\xi_{k-1}}}$}
\rput(5.05,2.5){$_{_{u-\xi_k}}$}
\rput(5.05,3.5){$_{_{-u-\xi_{k-1}}}$}
\facegrid{(6,0)}{(7,4)}
\rput(6.5,0.55){$_{\ldots}$}
\rput(6.5,1.55){$_{\ldots}$}
\rput(6.5,2.55){$_{\ldots}$}
\rput(6.5,3.55){$_{\ldots}$}
\pspolygon[fillstyle=solid,fillcolor=lightlightblue](7,1)(8,2)(8,0)(7,1)
\pspolygon[fillstyle=solid,fillcolor=lightlightblue](7,3)(8,4)(8,2)(7,3)
\psline[linecolor=blue,linewidth=1.5pt](7,0.5)(7.5,0.5)
\psline[linecolor=blue,linewidth=1.5pt](7,1.5)(7.5,1.5)
\psline[linecolor=blue,linewidth=1.5pt](7,2.5)(7.5,2.5)
\psline[linecolor=blue,linewidth=1.5pt](7,3.5)(7.5,3.5)
\rput(7.65,1){$_{_{v}}$}
\rput(7.65,3){$_{_{u}}$}
\end{pspicture}
\quad
=\ 
\begin{pspicture}[shift=-1.89](0,0)(3,4)
\pspolygon[fillstyle=solid,fillcolor=lightlightblue](0,3)(1,4)(2,3)(1,2)(0,3)
\pspolygon[fillstyle=solid,fillcolor=lightlightblue](1,2)(2,3)(3,2)(2,1)(1,2)
\pspolygon[fillstyle=solid,fillcolor=lightlightblue](2,1)(3,2)(3,0)(2,1)
\pspolygon[fillstyle=solid,fillcolor=lightlightblue](2,3)(3,4)(3,2)(2,3)
\psarc[linewidth=1.5pt,linecolor=blue](2,3){.7}{45}{135}
\psarc[linewidth=0.025]{-}(0,3){0.16}{-45}{45}
\psarc[linewidth=0.025]{-}(1,2){0.16}{-45}{45}
\rput(1,3){$_{_{u-v}}$}
\rput(2.05,2){$_{_{\lambda-u-v}}$}
\rput(2.65,1){$_{_{v,\xi}}$}
\rput(2.65,3){$_{_{u,\xi}}$}
\end{pspicture}
\ee
where the second and fourth equalities follow from repeated applications of the YBE (\ref{YBE}), 
while the third equality is an immediate consequence of the BYBE (\ref{BYBE2}).
\hfill $\square$
\medskip

Aside from being a solution to the BYBE, the construction of the Robin boundary condition (\ref{uxi}) is motivated as follows.
First, by repeated applications of the YBE (\ref{YBE}), the boundary crossing property (\ref{btt}) readily extends to
\be
\psset{unit=0.9cm}
\begin{pspicture}[shift=-0.89](0,0)(3,2)
\pspolygon[fillstyle=solid,fillcolor=lightlightblue](0,1)(1,2)(2,1)(1,0)(0,1)
\psarc[linewidth=1.5pt,linecolor=blue](2,1){.7}{45}{135}
\psarc[linewidth=1.5pt,linecolor=blue](2,1){.7}{-135}{-45}
\psarc[linewidth=0.025]{-}(0,1){0.16}{-45}{45}
\pspolygon[fillstyle=solid,fillcolor=lightlightblue](2,1)(3,2)(3,0)(2,1)
\rput(1,1){$_{2u-\lambda}$}
\rput(2.6,1){$_{u,\xi}$}
\end{pspicture}
\ =s_0(2u)\
\begin{pspicture}[shift=-0.89](0,0)(1,2)
\pspolygon[fillstyle=solid,fillcolor=lightlightblue](0,1)(1,2)(1,0)(0,1)
\rput(0.57,1){$_{_{\lambda-u,\xi}}$}
\end{pspicture}
\label{bttxi}
\ee
This ensures that the transfer tangles to be discussed in Section~\ref{Sec:D} are crossing symmetric (\ref{Dcross}).
Second, the drop-down property
\be
\begin{pspicture}[shift=-0.4](0,0)(2,1.5)
\facegrid{(0,0)}{(2,1)}
\psarc[linewidth=0.025](0,0){0.16}{0}{90}
\psarc[linewidth=0.025](1,0){0.16}{0}{90}
\psarc[linewidth=1.5pt,linecolor=blue](1,1){.5}{0}{180}
\rput(0.5,0.5){$_{v-\lambda}$}
\rput(1.5,0.5){$_{v}$}
\end{pspicture}
\ \  =
s_2(-v)s_0(v)\ \ 
\begin{pspicture}[shift=-0.4](0,0)(2,1.5)
\facegrid{(0,0)}{(2,1)}
\psarc[linewidth=1.5pt,linecolor=blue](1,1){.5}{0}{180}
\psarc[linewidth=1.5pt,linecolor=blue](1,0){.5}{0}{180}
\psarc[linewidth=1.5pt,linecolor=blue](0,1){.5}{-90}{0}
\psarc[linewidth=1.5pt,linecolor=blue](2,1){.5}{180}{270}
\end{pspicture}
\ee
applies to every neighbouring pair of faces in the boundary seam due to the regular shifts by $\lambda$ in the column inhomogeneities. 
This ensures that the requirement, that half-arcs along the lower edge
are projected out, propagates, as a rule, up through the seam and is thus applicable along the upper edge as well.
If the Wenzl-Jones projector~\cite{Jones1983,Wenzl1988,KauffmanLins1994} 
of the appropriate size exists, such a projection rule can be implemented by insertion of the
Wenzl-Jones projector~\cite{PRZ}. But even if the Wenzl-Jones projector does not exist, the propagation of the corresponding rule
follows from the drop-down property~\cite{PRV1210}.

For $w>0$, we now impose that boundary links likewise drop down in the sense that if the rightmost node on the upper edge 
is linked to the boundary, then so is the rightmost node on the lower edge. In the decomposition
\bea
\psset{unit=0.77cm}
\begin{pspicture}[shift=-0.89](0,0)(2,2.5)
\facegrid{(0,0)}{(1,2)}
\pspolygon[fillstyle=solid,fillcolor=lightlightblue](1,1)(2,2)(2,0)(1,1)
\psarc[linewidth=0.025](0,0){0.16}{0}{90}
\psarc[linewidth=0.025](0,1){0.16}{0}{90}
\psline[linecolor=blue,linewidth=1.5pt](1,0.5)(1.5,0.5)
\psline[linecolor=blue,linewidth=1.5pt](1,1.5)(1.5,1.5)
\psarc[linewidth=1.5pt,linecolor=blue](1,2){.5}{90}{180}
\psline[linecolor=blue,linewidth=1.5pt](1,2.5)(2,2.5)
\rput(2,2.5){$\bullet$}
\rput(0.55,0.5){$_{_{u-\xi_1}}$}
\rput(0.55,1.5){$_{_{-\!u\!-\!\xi_0}}$}
\rput(1.65,1){$_{u}$}
\psline[linecolor=blue,linewidth=1.5pt](-0.2,0.5)(0,0.5)
\psline[linecolor=blue,linewidth=1.5pt](-0.2,1.5)(0,1.5)
\psline[linecolor=blue,linewidth=1.5pt](0.5,-0.2)(0.5,0)
\rput(0.51,-0.5){\scriptsize 1}
\rput(-0.5,0.5){\scriptsize 2}
\rput(-0.5,1.5){\scriptsize 3}
\end{pspicture}
&=&\!\!s_0(2u)\Gamma(\xi+\lambda)
\quad
\psset{unit=0.77cm}
\begin{pspicture}[shift=-0.89](0,0)(1,2)
\psline[linecolor=blue,linewidth=1.5pt](0,1.5)(0.7,1.5)
\rput(0.7,1.5){$\bullet$}
\psarc[linewidth=1.5pt,linecolor=blue](0,0.5){.5}{0}{90}
\rput(0.51,0.2){\scriptsize 1}
\rput(-0.2,1){\scriptsize 2}
\rput(-0.2,1.5){\scriptsize 3}
\end{pspicture}
-\,[\Gamma(u)+\beta_2s_0(2u)]s_0(\xi+u)s_2(\xi-u)
\quad
\psset{unit=0.77cm}
\begin{pspicture}[shift=-0.89](0,0)(1,2)
\rput(1,1){$\bullet$}
\psarc[linewidth=1.5pt,linecolor=blue](0,1.5){.5}{-90}{90}
\psarc[linewidth=1.5pt,linecolor=blue](1,0.5){.5}{90}{180}
\rput(0.51,0.2){\scriptsize 1}
\rput(-0.2,1){\scriptsize 2}
\rput(-0.2,2){\scriptsize 3}
\end{pspicture}
\nn
&+&\!\!s_0(2u)s_1(\xi+u)s_2(\xi-u)
\quad
\psset{unit=0.77cm}
\begin{pspicture}[shift=-0.89](0,0)(1,2)
\rput(1,1){$\bullet$}
\rput(1,1.5){$\bullet$}
\rput(1,2){$\bullet$}
\psline[linecolor=blue,linewidth=1.5pt](0,1.5)(1,1.5)
\psline[linecolor=blue,linewidth=1.5pt](0,2)(1,2)
\psarc[linewidth=1.5pt,linecolor=blue](1,0.5){.5}{90}{180}
\rput(0.51,0.2){\scriptsize 1}
\rput(-0.2,1.5){\scriptsize 2}
\rput(-0.2,2){\scriptsize 3}
\end{pspicture}
\eea
we thus require that the first coefficient vanishes for all $u$, that is
\be
 \Gamma(\xi+\lambda)=0
\ee
For $w>0$, this imposes a relation between the constant $\gamma$ and the boundary parameter $\xi$,
\be
 \gamma=\beta_1[s_0(\xi+\lambda)]^2-\beta_2[s_0(\xi+\tfrac{\lambda}{2})]^2
\label{gamma}
\ee
so that
\be
 \Gamma(u)=s_1(\xi-u)\big(\beta_1s_1(\xi+u)-\beta_2 s_0(\xi+u)\big)
\ee
and hence
\be
 \Gamma(0)=s_1(\xi)\big(\beta_1s_1(\xi)-\beta_2s_0(\xi)\big)
\label{G0}
\ee
Combined with the disallowance of half-arcs formed between the $w$ nodes, this is sufficient to ensure that the rule, disallowing 
boundary links emanating from the $w$ nodes on the lower edge, propagates up through the
seam and is thus applicable along the upper edge as well. This is crucial for the construction of the Robin modules
in Section~\ref{Sec:RobinModules}.

The combined projection rule that half-arcs between boundary nodes and boundary links emanating from boundary nodes
are disallowed can be implemented
by the introduction of {\em boundary Wenzl-Jones projectors}~\cite{DubailThesis,MDRT2014}.
We will not discuss this here. Instead, we will follow the approach of~\cite{PRV1210} and incorporate the rule by 
modifying the boundary seam and restricting the vector space of link states in Section~\ref{Sec:RobinModules}.
In preparation for this, we now turn to the description of the relevant link states.

\section{Link states}
\label{Sec:LinkStates}

\subsection{Link states and standard modules}
\label{Sec:LinkStandard}

A boundary link state on $N$ nodes is a planar diagram of non-crossing arc segments.
Such a link state consists of $d\in\{0,\ldots,N\}$ vertical line segments (called defects) attached to individual nodes,
$b\in\{0,\ldots,N\}$ arcs (called boundary links) linking individual nodes to the right boundary, and $\frac{N-d-b}{2}$ 
half-arcs connecting nodes pairwise. An arc segment thus emanates from every node so a link state is 
subject to the parity constraint
\be
 N-d-b\equiv0\mod 2
\label{Ndb02}
\ee
As the defects can be thought of linking nodes to the point above at infinity, the requirement of planarity 
prevents arcs from arching over any of the vertical line segments. 
An example of a boundary link state on $N=10$ nodes with $d=2$ defects and $b=2$ boundary links is given by
\be
\begin{pspicture}[shift=-0.15](-0.0,-0.1)(4.0,0.8)
\psline[linewidth=0.5pt]{-}(-0.2,0)(3.8,0)
\psarc[linewidth=1.5pt,linecolor=darkgreen](0.2,0){0.2}{0}{180}
\psarc[linewidth=1.5pt,linecolor=darkgreen](1.4,0){0.2}{0}{180}
\psarc[linewidth=1.5pt,linecolor=darkgreen](3,0){0.2}{0}{180}
\psarc[linewidth=1.5pt,linecolor=darkgreen](3.8,0){0.2}{90}{180}
\psline[linewidth=1.5pt,linecolor=darkgreen](0.8,0)(0.8,0.6)
\psline[linewidth=1.5pt,linecolor=darkgreen](2.0,0)(2.0,0.6)
\psbezier[linecolor=darkgreen,linewidth=1.5pt](2.4,0)(2.5,0.6)(3,0.6)(3.8,0.6)
\end{pspicture}
\in {\cal V}_{2,2}^{(10)}
\ee
We denote by ${\cal V}_{d,b}^{(N)}$ the linear span of the set of link states on $N$ nodes with $d$ defects and $b$
boundary links, and note that
\be
 \dim {\cal V}_{d,b}^{(N)}=\begin{pmatrix}N \\ \frac{N-d-b}{2}\end{pmatrix}
   -\begin{pmatrix}N \\ \frac{N-d-b-2}{2}\end{pmatrix}
\ee
The link states themselves thus provide a canonical basis for this vector space.
Let ${\cal V}_d^{(N)}$ denote the set of link states for $N$ and $d$ fixed but with $b$ only constrained by (\ref{Ndb02}).
The number of these link states is given by
\be
 \dim {\cal V}_d^{(N)}=\sum_b \dim {\cal V}_{d,b}^{(N)}=\begin{pmatrix}N \\ \lfloor\frac{N-d}{2}\rfloor\end{pmatrix}
\label{dimLdN}
\ee
For given $N$, the total number of link states is
\be
 \dim {\cal V}^{(N)}=\sum_{d,b} \dim {\cal V}_{d,b}^{(N)}=\sum_{d} \dim {\cal V}_{d}^{(N)}=2^N
\label{dimVN}
\ee

Matrix representations of the one-boundary TL algebra
$TL_N(\beta;\beta_1,\beta_2)$ are obtained by letting the algebra generators act on the link states.
A {\em standard module} of $TL_N(\beta;\beta_1,\beta_2)$, in particular, is defined for every $d\in\{0,\ldots,N\}$ and is obtained
by letting the algebra act on ${\cal V}_d^{(N)}$ in a way that preserves the number of defects.
To describe this action, let $c$ be a loop representation of a word in $TL_N(\beta;\beta_1,\beta_2)$ and $v\in {\cal V}_d^{(N)}$\!.
The product $cv$ is then given by concatenating the respective diagrams placing $v$ atop $c$. 
On ${\cal V}_2^{(10)}$\!, this action is illustrated by
\be
\begin{pspicture}[shift=-0.15](-0.0,0.4)(4.0,1.6)
\psline[linewidth=0.5pt](-0.2,-0.2)(3.8,-0.2)
\psarc[linewidth=1.5pt,linecolor=blue](0.2,1){0.2}{180}{360}
\psarc[linewidth=1.5pt,linecolor=blue](1,1){0.2}{180}{360}
\psarc[linewidth=1.5pt,linecolor=blue](3.4,1){0.2}{180}{360}
\psbezier[linecolor=blue,linewidth=1.5pt](2.4,1)(2.5,0.3)(3,0.3)(3.8,0.3)
\psbezier[linecolor=blue,linewidth=1.5pt](2.8,1)(2.84,0.58)(3,0.6)(3.8,0.6)
\psarc[linewidth=1.5pt,linecolor=blue](0.6,-0.2){0.2}{0}{180}
\psarc[linewidth=1.5pt,linecolor=blue](0.6,-0.2){0.6}{0}{180}
\psline[linewidth=1.5pt,linecolor=blue](1.6,-0.2)(1.6,1)
\psarc[linewidth=1.5pt,linecolor=blue](2.2,-0.2){0.2}{0}{180}
\psarc[linewidth=1.5pt,linecolor=blue](3.4,-0.2){0.2}{0}{180}
\psbezier[linecolor=blue,linewidth=1.5pt](2,1)(2.1,0.2)(2.7,0.4)(2.8,-0.2)
\psline[linewidth=0.5pt](-0.2,1)(3.8,1)
\psarc[linewidth=1.5pt,linecolor=darkgreen](0.2,1){0.2}{0}{180}
\psarc[linewidth=1.5pt,linecolor=darkgreen](1.4,1){0.2}{0}{180}
\psarc[linewidth=1.5pt,linecolor=darkgreen](3,1){0.2}{0}{180}
\psarc[linewidth=1.5pt,linecolor=darkgreen](3.8,1){0.2}{90}{180}
\psline[linewidth=1.5pt,linecolor=darkgreen](0.8,1)(0.8,1.6)
\psline[linewidth=1.5pt,linecolor=darkgreen](2.0,1)(2.0,1.6)
\psbezier[linecolor=darkgreen,linewidth=1.5pt](2.4,1)(2.44,1.6)(3,1.6)(3.8,1.6)
\end{pspicture}
=\beta\beta_1\beta_2
\begin{pspicture}[shift=-0.15](-0.35,-0.1)(4.0,0.8)
\psline[linewidth=0.5pt](-0.2,0)(3.8,0)
\psarc[linewidth=1.5pt,linecolor=darkgreen](0.6,0){0.2}{0}{180}
\psarc[linewidth=1.5pt,linecolor=darkgreen](0.6,0){0.6}{0}{180}
\psarc[linewidth=1.5pt,linecolor=darkgreen](2.2,0){0.2}{0}{180}
\psarc[linewidth=1.5pt,linecolor=darkgreen](3.4,0){0.2}{0}{180}
\psline[linewidth=1.5pt,linecolor=darkgreen](1.6,0)(1.6,0.6)
\psline[linewidth=1.5pt,linecolor=darkgreen](2.8,0)(2.8,0.6)
\end{pspicture}
\qquad\quad
\begin{pspicture}[shift=-0.15](-0.0,0.4)(4.0,1.6)
\psline[linewidth=0.5pt](-0.2,-0.2)(3.8,-0.2)
\psarc[linewidth=1.5pt,linecolor=blue](0.2,1){0.2}{180}{360}
\psarc[linewidth=1.5pt,linecolor=blue](1,1){0.2}{180}{360}
\psarc[linewidth=1.5pt,linecolor=blue](3.4,1){0.2}{180}{360}
\psarc[linewidth=1.5pt,linecolor=blue](2.2,1){0.2}{180}{360}
\psbezier[linecolor=blue,linewidth=1.5pt](2.8,1)(2.84,0.58)(3,0.6)(3.8,0.6)
\psbezier[linecolor=blue,linewidth=1.5pt](2.8,-0.2)(2.84,0.22)(3,0.2)(3.8,0.2)
\psarc[linewidth=1.5pt,linecolor=blue](0.6,-0.2){0.2}{0}{180}
\psarc[linewidth=1.5pt,linecolor=blue](0.6,-0.2){0.6}{0}{180}
\psline[linewidth=1.5pt,linecolor=blue](1.6,-0.2)(1.6,1)
\psarc[linewidth=1.5pt,linecolor=blue](2.2,-0.2){0.2}{0}{180}
\psarc[linewidth=1.5pt,linecolor=blue](3.4,-0.2){0.2}{0}{180}
\psline[linewidth=0.5pt](-0.2,1)(3.8,1)
\psarc[linewidth=1.5pt,linecolor=darkgreen](0.2,1){0.2}{0}{180}
\psarc[linewidth=1.5pt,linecolor=darkgreen](1.4,1){0.2}{0}{180}
\psarc[linewidth=1.5pt,linecolor=darkgreen](3,1){0.2}{0}{180}
\psarc[linewidth=1.5pt,linecolor=darkgreen](3.8,1){0.2}{90}{180}
\psline[linewidth=1.5pt,linecolor=darkgreen](0.8,1)(0.8,1.6)
\psline[linewidth=1.5pt,linecolor=darkgreen](2.0,1)(2.0,1.6)
\psbezier[linecolor=darkgreen,linewidth=1.5pt](2.4,1)(2.44,1.6)(3,1.6)(3.8,1.6)
\end{pspicture}
=0
\vspace{0.7cm}
\ee
and is readily seen to give rise to a representation of the algebra.
The ensuing standard module is thus of dimension $\dim {\cal V}_d^{(N)}$\! as given in (\ref{dimLdN}).

\subsection{Robin link states}
\label{Sec:RobinStates}

The link states associated with a Robin boundary condition with boundary seam of width $w$, as described in 
Proposition~\ref{Prop:GenRobin}, constitute a subset of the set of boundary link states ${\cal V}_d^{(N+w)}$\!, and
we denote the linear span of this subset by ${\cal V}_d^{(N,w)}$\!.
To characterise these {\em Robin link states}, we refer to the $N$ leftmost nodes of a link state in ${\cal V}_d^{(N+w)}$
as bulk nodes while the (remaining) $w$ rightmost nodes are called boundary nodes. 
A link state in ${\cal V}_d^{(N,w)}\subset {\cal V}_d^{(N+w)}$ is now defined by requiring that 
\begin{itemize}
\item[(i)] no half-arc is formed between a pair of boundary nodes, 
\item[(ii)] no boundary link emanates from a boundary node. 
\end{itemize}
This implies that every boundary node must be a defect or linked to a bulk node. Examples of vector spaces of Robin link states are
\bea
 {\cal V}_1^{(3,1)}&\!\!=\!\!&\mbox{span}\,\Big\{\ \
\begin{pspicture}[shift=-0.2](0,0)(1.5,1)
\psline[linewidth=0.5pt](-0.2,0)(1.4,0)
\psline[linewidth=0.5pt,linestyle=dashed, dash=1.5pt 1.5pt](1,-0.1)(1,0.8)
\psline[linewidth=1.5pt,linecolor=darkgreen](0,0)(0,0.6)
\psbezier[linecolor=darkgreen,linewidth=1.5pt](0.4,0)(0.45,0.5)(0.9,0.45)(1.4,0.45)
\psarc[linewidth=1.5pt,linecolor=darkgreen](1,0){0.2}{0}{180}
\end{pspicture}
 \Big\}
 \nn
 {\cal V}_1^{(3,2)}&\!\!=\!\!&\mbox{span}\,\Big\{\ \
\begin{pspicture}[shift=-0.2](0,0)(1.9,1)
\psline[linewidth=0.5pt](-0.2,0)(1.8,0)
\psline[linewidth=0.5pt,linestyle=dashed, dash=1.5pt 1.5pt](1,-0.1)(1,0.8)
\psline[linewidth=1.5pt,linecolor=darkgreen](0,0)(0,0.6)
\psbezier[linecolor=darkgreen,linewidth=1.5pt](0.4,0)(0.45,0.5)(1.55,0.5)(1.6,0)
\psarc[linewidth=1.5pt,linecolor=darkgreen](1,0){0.2}{0}{180}
\rput(2,0){,}
\end{pspicture}
\qquad
\begin{pspicture}[shift=-0.2](0,0)(1.9,1)
\psline[linewidth=0.5pt](-0.2,0)(1.8,0)
\psline[linewidth=0.5pt,linestyle=dashed, dash=1.5pt 1.5pt](1,-0.1)(1,0.8)
\psarc[linewidth=1.5pt,linecolor=darkgreen](0.2,0){0.2}{0}{180}
\psarc[linewidth=1.5pt,linecolor=darkgreen](1,0){0.2}{0}{180}
\psline[linewidth=1.5pt,linecolor=darkgreen](1.6,0)(1.6,0.6)
\rput(2,0){,}
\end{pspicture}
\qquad
\begin{pspicture}[shift=-0.2](0,0)(1.9,1)
\psline[linewidth=0.5pt](-0.2,0)(1.8,0)
\psline[linewidth=0.5pt,linestyle=dashed, dash=1.5pt 1.5pt](1,-0.1)(1,0.8)
\psbezier[linecolor=darkgreen,linewidth=1.5pt](0,0)(0.05,0.5)(1.15,0.5)(1.2,0)
\psarc[linewidth=1.5pt,linecolor=darkgreen](0.6,0){0.2}{0}{180}
\psline[linewidth=1.5pt,linecolor=darkgreen](1.6,0)(1.6,0.6)
\end{pspicture}
\Big\} 
\label{LinkStates}
\\
 {\cal V}_0^{(4,2)}&\!\!=\!\!&\mbox{span}\,\Big\{\ \
\begin{pspicture}[shift=-0.2](0,0)(2.3,1)
\psline[linewidth=0.5pt](-0.2,0)(2.2,0)
\psline[linewidth=0.5pt,linestyle=dashed, dash=1.5pt 1.5pt](1.4,-0.1)(1.4,0.8)
\psarc[linewidth=1.5pt,linecolor=darkgreen](0.2,0){0.2}{0}{180}
\psbezier[linecolor=darkgreen,linewidth=1.5pt](0.8,0)(0.85,0.5)(1.95,0.5)(2,0)
\psarc[linewidth=1.5pt,linecolor=darkgreen](1.4,0){0.2}{0}{180}
\rput(2.4,0){,}
\end{pspicture}
\qquad
\begin{pspicture}[shift=-0.2](0,0)(2.3,1)
\psline[linewidth=0.5pt](-0.2,0)(2.2,0)
\psline[linewidth=0.5pt,linestyle=dashed, dash=1.5pt 1.5pt](1.4,-0.1)(1.4,0.8)
\psbezier[linecolor=darkgreen,linewidth=1.5pt](0,0)(0.05,0.7)(1,0.75)(2.2,0.75)
\psbezier[linecolor=darkgreen,linewidth=1.5pt](0.4,0)(0.45,0.5)(0.8,0.55)(2.2,0.55)
\psbezier[linecolor=darkgreen,linewidth=1.5pt](0.8,0)(0.85,0.5)(1.95,0.5)(2,0)
\psarc[linewidth=1.5pt,linecolor=darkgreen](1.4,0){0.2}{0}{180}
\rput(2.4,0){,}
\end{pspicture}
\qquad
\begin{pspicture}[shift=-0.2](0,0)(2.3,1)
\psline[linewidth=0.5pt](-0.2,0)(2.2,0)
\psline[linewidth=0.5pt,linestyle=dashed, dash=1.5pt 1.5pt](1.4,-0.1)(1.4,0.8)
\psbezier[linecolor=darkgreen,linewidth=1.5pt](0,0)(0.1,0.75)(1.9,0.75)(2,0)
\psarc[linewidth=1.5pt,linecolor=darkgreen](0.6,0){0.2}{0}{180}
\psarc[linewidth=1.5pt,linecolor=darkgreen](1.4,0){0.2}{0}{180}
\rput(2.4,0){,}
\end{pspicture}
\qquad
\begin{pspicture}[shift=-0.2](0,0)(2.3,1)
\psline[linewidth=0.5pt](-0.2,0)(2.2,0)
\psline[linewidth=0.5pt,linestyle=dashed, dash=1.5pt 1.5pt](1.4,-0.1)(1.4,0.8)
\psbezier[linecolor=darkgreen,linewidth=1.5pt](0,0)(0.1,0.75)(1.9,0.75)(2,0)
\psbezier[linecolor=darkgreen,linewidth=1.5pt](0.4,0)(0.45,0.5)(1.55,0.5)(1.6,0)
\psarc[linewidth=1.5pt,linecolor=darkgreen](1,0){0.2}{0}{180}
\end{pspicture}
\Big\}
\nonumber
\eea
and the number of these link states is given by
\be
 \dim {\cal V}_d^{(N,w)}=\begin{pmatrix}N \\ \left\lfloor\frac{N-d}{2}\right\rfloor+(-1)^{N-d-w}\left\lceil\frac{w}{2}\right\rceil\end{pmatrix}
\ee
For vanishing seam width $w=0$, this expression correctly reduces to (\ref{dimLdN}).
Let ${\cal V}^{(N,w)}$ denote the space of Robin link states with an arbitrary number of defects.
The number of these link states is given by
\be
 \dim {\cal V}^{(N,w)}=\sum_{d=0}^{N+w}\dim{\cal V}_d^{(N,w)}=2^N
\label{dimVNw}
\ee
and is {\em independent} of $w$.

Requirement (i) above projects onto link states without half-arcs between the boundary nodes. 
As discussed in Section~\ref{Sec:GenRobin}, this is a well-defined prescription due to the drop-down property.
It generalises the role of the Wenzl-Jones projectors used in similar situations,
such as in the construction of $r$-type boundary conditions.
Without requirement (ii), the boundary construction in Section~\ref{Sec:GenRobin} would correspond to the fusion product of a 
Kac boundary condition,
of the form $(r,1)$ and built from a seam of width $w$, and the Robin boundary condition corresponding to a standard module with 
$d$ defects. This is of course also of interest, as indicated in Section~\ref{Sec:Fus}, but does not yield an indecomposable 
Virasoro representation in the continuum scaling limit if $w>0$. 
As discussed in Section~\ref{Sec:GenRobin}, imposing requirement (ii) is well defined.

\section{Transfer matrices}
\label{Sec:Transfer}

\subsection{Double row transfer tangles}
\label{Sec:D}

Focusing on scenarios with Neumann boundary conditions on the left but Robin boundary conditions on the right of the strip lattice,
as described in Section~\ref{Sec:StatLatMod}, we define the double row transfer tangle
\be
 \Db(u):=\quad
\psset{unit=.77cm}
\begin{pspicture}[shift=-0.89](0.4,0)(6.5,2)
\facegrid{(1,0)}{(5,2)}
\pspolygon[fillstyle=solid,fillcolor=lightlightblue](0,0)(1,1)(0,2)(0,0)
\pspolygon[fillstyle=solid,fillcolor=lightlightblue](5,1)(6,2)(6,0)(5,1)
\psarc[linewidth=0.025](1,0){0.16}{0}{90}
\psarc[linewidth=0.025](1,1){0.16}{0}{90}
\psarc[linewidth=0.025](2,0){0.16}{0}{90}
\psarc[linewidth=0.025](2,1){0.16}{0}{90}
\psarc[linewidth=0.025](4,0){0.16}{0}{90}
\psarc[linewidth=0.025](4,1){0.16}{0}{90}
\rput(1.5,0.5){$_{u}$}
\rput(2.5,0.5){$_{u}$}
\rput(4.5,0.5){$_{u}$}
\rput(1.5,1.5){$_{\lambda-u}$}
\rput(2.5,1.5){$_{\lambda-u}$}
\rput(4.5,1.5){$_{\lambda-u}$}
\rput(3.5,0.5){$\ldots$}
\rput(3.5,1.5){$\ldots$}
\rput(5.6,1){$_{u,\xi}$}
\psarc[linewidth=1.5pt,linecolor=blue](1,1){.5}{90}{270}
\psline[linecolor=blue,linewidth=1.5pt]{-}(5,0.5)(5.5,0.5)
\psline[linecolor=blue,linewidth=1.5pt]{-}(5,1.5)(5.5,1.5)
\rput(3,-0.5){$\underbrace{\qquad \qquad \qquad \qquad}_N$}
\end{pspicture}
\vspace{0.5cm}
\label{Duxi}
\ee
It is noted that this is an $(N+w)$-tangle and that the dependence on $N$, $\lambda$, $w$ and $\xi$ has been suppressed.
With multiplication given by vertical concatenation of diagrams, and following~\cite{BPO1996},
it follows from the local inversion relation (\ref{inv}) and the bulk and boundary Yang-Baxter equations (\ref{YBE}) and (\ref{BYBE1}) that
the transfer tangles form a commuting family
\be
 [\Db(u),\Db(v)]=0,\qquad u,v\in\mathbb{C}
\ee
Using (\ref{inv}), (\ref{Nb}) and (\ref{bttxi}), it also follows that they are crossing symmetric
\be
 \Db(\la-u)=\Db(u)
\label{Dcross}
\ee

Acting on link states with an auxiliary half-arc between the nodes in positions $-1$ and $0$, transfer matrix representations follow from
`opening up' the corresponding transfer tangle as
\be
\Db(u)\ =\ 
\psset{unit=.6cm}
\setlength{\unitlength}{.6cm}
\begin{pspicture}[shift=-6](-1,-1)(6,11)
\psline[linewidth=0.5pt,linestyle=dashed, dash=1pt 1pt](-1,10.5)(5,10.5)
\psbezier[linecolor=darkgreen,linewidth=1.5pt](-.5,10.5)(-.5,11)(.5,11)(.5,10.5)
\psline[linecolor=blue,linewidth=1.5pt](1.5,10.5)(1.5,10.9)
\psline[linecolor=blue,linewidth=1.5pt](2.5,10.5)(2.5,10.9)
\psline[linecolor=blue,linewidth=1.5pt](3.5,10.5)(3.5,10.9)
\psline[linecolor=blue,linewidth=1.5pt](4.5,10.5)(4.5,10.9)
\pspolygon[fillstyle=solid,fillcolor=lightlightblue](0,1)(1,0)(5,4)(4,5)(0,1)
\pspolygon[fillstyle=solid,fillcolor=lightlightblue](0,9)(4,5)(5,6)(1,10)(0,9)
\psline(1,2)(2,1)
\psline(2,3)(3,2)
\psline(4,3)(3,4)
\psline(3,6)(4,7)
\psline(2,7)(3,8)
\psline(1,8)(2,9)
\pspolygon[fillstyle=solid,fillcolor=lightlightblue](4,5)(5,6)(5,4)(4,5)
\rput(1,1){$_u$}
\rput(2,2){$_u$}
\rput(2.8,2.8){$.$}\rput(3,3){$.$}\rput(3.2,3.2){$.$}
\rput(2.8,7.2){$.$}\rput(3,7){$.$}\rput(3.2,6.8){$.$}
\rput(4,4){$_u$}
\rput(4,6){$_u$}
\rput(2,8){$_u$}
\rput(1,9){$_u$}
\rput(4.6,5){$_{u,\xi}$}
\psarc[linewidth=0.025](1,0){.16}{45}{135}
\psarc[linewidth=0.025](2,1){.16}{45}{135}
\psarc[linewidth=0.025](4,3){.16}{45}{135}
\psarc[linewidth=0.025](4,5){.16}{45}{135}
\psarc[linewidth=0.025](2,7){.16}{45}{135}
\psarc[linewidth=0.025](1,8){.16}{45}{135}
\psline[linecolor=blue,linewidth=1.5pt](-.5,.5)(-.5,10.5)
\psline[linecolor=blue,linewidth=1.5pt](.5,9.5)(.5,10.5)
\psline[linecolor=blue,linewidth=1.5pt](.5,1.5)(.5,8.5)
\psline[linecolor=blue,linewidth=1.5pt](1.5,2.5)(1.5,7.5)
\psline[linecolor=blue,linewidth=1.5pt](2.5,3.5)(2.5,6.5)
\psline[linecolor=blue,linewidth=1.5pt](3.5,4.5)(3.5,5.5)
\psline[linecolor=blue,linewidth=1.5pt](1.5,-.5)(1.5,.5)
\psline[linecolor=blue,linewidth=1.5pt](2.5,-.5)(2.5,1.5)
\psline[linecolor=blue,linewidth=1.5pt](3.5,-.5)(3.5,2.5)
\psline[linecolor=blue,linewidth=1.5pt](4.5,-.5)(4.5,3.5)
\psline[linecolor=blue,linewidth=1.5pt](1.5,9.5)(1.5,10.5)
\psline[linecolor=blue,linewidth=1.5pt](2.5,8.5)(2.5,10.5)
\psline[linecolor=blue,linewidth=1.5pt](3.5,7.5)(3.5,10.5)
\psline[linecolor=blue,linewidth=1.5pt](4.5,6.5)(4.5,10.5)
\psbezier[linecolor=darkgreen,linewidth=1.5pt](-.5,-.5)(-.5,0)(.5,0)(.5,-.5)
\psbezier[linecolor=blue,linewidth=1.5pt](-.5,.5)(-.5,0)(.5,0)(.5,.5)
\rput(-1.5,-.85){$_{j\,=}$}
\rput(-0.5,-.85){$_{-1}$}
\rput(0.5,-.85){$_0$}
\rput(1.5,-.85){$_1$}
\rput(3,-.85){$\ldots$}
\rput(4.5,-.85){$_N$}
\end{pspicture}
\label{expandedD}
\ee
As an element of $TL_{N+w+2}(\beta;\beta_1,\beta_2)$, this is given by
\be
 \Db(u)=
    e_{-1}X_0(u)X_1(u)\ldots X_{N-1}(u)\,K_N^{(w)}(u,\xi)\,X_{N-1}(u)X_{N-2}(u)\ldots X_0(u)
\ee
where the labelling of the nodes starts at $j=-1$, and where 
\be
\psset{unit=.6cm}
K_N^{(w)}(u,\xi):=\!
\begin{pspicture}[shift=-0.89](0,0)(5,2)
\psline[linecolor=blue,linewidth=1.5pt](0.5,0)(0.5,2)
\psline[linecolor=blue,linewidth=1.5pt](1.5,0)(1.5,2)
\rput(2.5,1){$\ldots$}
\psline[linecolor=blue,linewidth=1.5pt](3.5,0)(3.5,2)
\pspolygon[fillstyle=solid,fillcolor=lightlightblue](4,1)(5,2)(5,0)(4,1)
\psline[linecolor=blue,linewidth=1.5pt](4.5,1.5)(4.5,2)
\psline[linecolor=blue,linewidth=1.5pt](4.5,0)(4.5,0.5)
\rput(4.6,1){$_{_{u,\xi}}$}
\end{pspicture}
\ee
To obtain concrete matrix representations of the transfer tangle, we need to specify the appropriate vector space of link states and
the action thereupon by the transfer tangle.
As discussed in~\cite{PRV1210} and in Section~\ref{Sec:RobinModules} below, this may affect the description of the tangle itself.

Following (\ref{expandedD}), we can also open up the boundary component as
\be
\psset{unit=.77cm}
\begin{pspicture}[shift=-0.89](0,0)(1,2)
\pspolygon[fillstyle=solid,fillcolor=lightlightblue](0,1)(1,2)(1,0)(0,1)
\psline[linecolor=blue,linewidth=1.5pt](0.5,2)(0.5,1.5)
\psline[linecolor=blue,linewidth=1.5pt](0.5,0)(0.5,0.5)
\rput(-0.2,-.35){$_{j\ =}$}
\rput(0.5,-.35){$_{_{N}}$}
\rput(0.6,1){$_{u,\xi}$}
\end{pspicture}
\ =
\psset{unit=.6cm}
\setlength{\unitlength}{.6cm}
\begin{pspicture}[shift=-6](0,-1)(6,11)
\pspolygon[fillstyle=solid,fillcolor=lightlightblue](0,1)(1,0)(5,4)(4,5)(0,1)
\pspolygon[fillstyle=solid,fillcolor=lightlightblue](0,9)(4,5)(5,6)(1,10)(0,9)
\psline(1,2)(2,1)
\psline(2,3)(3,2)
\psline(4,3)(3,4)
\psline(3,6)(4,7)
\psline(2,7)(3,8)
\psline(1,8)(2,9)
\pspolygon[fillstyle=solid,fillcolor=lightlightblue](4,5)(5,6)(5,4)(4,5)
\rput(1,1){$_{u-\xi_w}$}
\rput(2,2){$_{u-\xi_{w\!-\!1}}$}
\rput(2.8,2.8){$.$}\rput(3,3){$.$}\rput(3.2,3.2){$.$}
\rput(2.8,7.2){$.$}\rput(3,7){$.$}\rput(3.2,6.8){$.$}
\rput(4,4){$_{u-\xi_1}$}
\rput(4,6){$_{u+\xi_1}$}
\rput(2,8){$_{u+\xi_{w\!-\!1}}$}
\rput(1,9){$_{u+\xi_w}$}
\rput(4.66,5){$_{u}$}
\psarc[linewidth=0.025](1,0){.16}{45}{135}
\psarc[linewidth=0.025](2,1){.16}{45}{135}
\psarc[linewidth=0.025](4,3){.16}{45}{135}
\psarc[linewidth=0.025](4,5){.16}{45}{135}
\psarc[linewidth=0.025](2,7){.16}{45}{135}
\psarc[linewidth=0.025](1,8){.16}{45}{135}
\psline[linecolor=blue,linewidth=1.5pt](.5,9.5)(.5,10.5)
\psline[linecolor=blue,linewidth=1.5pt](.5,1.5)(.5,8.5)
\psline[linecolor=blue,linewidth=1.5pt](1.5,2.5)(1.5,7.5)
\psline[linecolor=blue,linewidth=1.5pt](2.5,3.5)(2.5,6.5)
\psline[linecolor=blue,linewidth=1.5pt](3.5,4.5)(3.5,5.5)
\psline[linecolor=blue,linewidth=1.5pt](0.5,-.5)(0.5,.5)
\psline[linecolor=blue,linewidth=1.5pt](1.5,-.5)(1.5,.5)
\psline[linecolor=blue,linewidth=1.5pt](2.5,-.5)(2.5,1.5)
\psline[linecolor=blue,linewidth=1.5pt](3.5,-.5)(3.5,2.5)
\psline[linecolor=blue,linewidth=1.5pt](4.5,-.5)(4.5,3.5)
\psline[linecolor=blue,linewidth=1.5pt](1.5,9.5)(1.5,10.5)
\psline[linecolor=blue,linewidth=1.5pt](2.5,8.5)(2.5,10.5)
\psline[linecolor=blue,linewidth=1.5pt](3.5,7.5)(3.5,10.5)
\psline[linecolor=blue,linewidth=1.5pt](4.5,6.5)(4.5,10.5)
\rput(-0.4,-.85){$_{j\ =}$}
\rput(0.5,-.85){$_{_{N}}$}
\rput(1.5,-.85){$_{_{N+1}}$}
\rput(3,-.85){$\ldots$}
\rput(4.5,-.85){$_{_{N+w}}$}
\end{pspicture}
\ee
As an element of $TL_{N+w}(\beta;\beta_1,\beta_2)$ (or $TL_{N+w+2}(\beta;\beta_1,\beta_2)$ if the two auxiliary nodes
are included), this is written as
\bea
 K_N^{(w)}(u,\xi)&\!\!=\!\!&X_N(u-\xi_w)X_{N+1}(u-\xi_{w-1})\ldots X_{N+w-1}(u-\xi_1)\big[\Gamma(u)I+s_0(2u)f_{N+w}\big]\nn
  &&X_{N+w-1}(u+\xi_1)X_{N+w-2}(u+\xi_2)\ldots X_N(u+\xi_w)
\eea
and when viewed as acting on ${\cal V}_d^{(N,w)}$\!, it reduces to
\bea
 K_N^{(w)}(u,\xi)&\!\!\simeq\!\!&\alpha_0^{(w)}I+\alpha_1^{(w)}e_N+\alpha_2^{(w)}e_Ne_{N+1}+\ldots
   +\alpha_w^{(w)}e_Ne_{N+1}\ldots e_{N+w-1}\nn
   &+\!\!&\alpha_{w+1}^{(w)}e_Ne_{N+1}\ldots e_{N+w-1}f_{N+w}
\eea
The fact that the two sides only agree when their actions are restricted to ${\cal V}_d^{(N,w)}$ is reflected in the use of
the similarity sign $\,\simeq\,$ instead of an equality sign.

\subsection{Robin representations}
\label{Sec:RobinModules}

Due to the drop-down properties discussed in Section~\ref{Sec:GenRobin}, the restriction to the Robin link states spanning
${\cal V}_d^{(N,w)}$ (or ${\cal V}_d^{(N+2,w)}$ if we work with (\ref{expandedD})) yields well defined representations 
of the transfer tangle (\ref{Duxi}). 
That is, the drop-down properties ensure that
\be
 \rho_d^{(N,w)}\big(\Db(u)\Db(v)\big)=\rho_d^{(N,w)}\big(\Db(u)\big)\rho_d^{(N,w)}\big(\Db(v)\big)
\ee
where the particular {\em Robin representation} $\rho_d^{(N,w)}(\Db(u))$ is obtained by requiring that the number of defects $d$ is 
preserved in the same way as in the definition of the standard modules in Section~\ref{Sec:LinkStandard}. More general
representations can of course be constructed, but focus here will be on the Robin representations.

In the following, we will thus restrict our considerations to the situation where $\Db(u)$ is meant to act on ${\cal V}_d^{(N,w)}$ 
for some $d$. In a planar decomposition of $K_N^{(w)}(u,\xi)$, we may therefore ignore connectivity diagrams containing half-arcs 
between the $w$ boundary nodes on the lower edge as well as boundary links emanating from these nodes. 
In the spirit of~\cite{PRV1210}, we thus have the decomposition
\begin{align}
\psset{unit=.77cm}
\begin{pspicture}[shift=-0.89](0,0)(1,2)
\pspolygon[fillstyle=solid,fillcolor=lightlightblue](0,1)(1,2)(1,0)(0,1)
\rput(0.6,1){$_{u,\xi}$}
\end{pspicture}
&\;\;\simeq\;
\alpha_0^{(w)}\
\psset{unit=.77cm}
\begin{pspicture}[shift=-0.89](0,0)(6.7,2)
\pspolygon[fillstyle=solid,fillcolor=lightlightblue](0,0)(6.5,0)(6.5,2)(0,2)(0,0)
\psarc[linewidth=1.5pt,linecolor=blue](0,1){.5}{-90}{90}
\multirput(1,0){6}{\psline[linewidth=1.5pt,linecolor=blue](1,0)(1,2)}
\rput(3.5,-0.5){$\underbrace{\qquad \qquad \qquad \qquad\qquad\ }_w$}
\end{pspicture}
+\alpha_1^{(w)}\
\psset{unit=.77cm}
\begin{pspicture}[shift=-0.89](0,0)(6.2,2)
\pspolygon[fillstyle=solid,fillcolor=lightlightblue](0,0)(6,0)(6,2)(0,2)(0,0)
\psarc[linewidth=1.5pt,linecolor=blue](0,0){.5}{0}{90}
\psarc[linewidth=1.5pt,linecolor=blue](0,2){.5}{-90}{0}
\multirput(1,0){5}{\psline[linewidth=1.5pt,linecolor=blue](1.5,0)(1.5,2)}
\rput(3.5,-0.5){$\underbrace{\qquad \qquad \qquad \qquad\ }_{w-1}$}
\end{pspicture}
\nonumber
\\[.8cm]
&+\alpha_2^{(w)}\
\psset{unit=.77cm}
\begin{pspicture}[shift=-0.89](0,0)(6.2,2)
\pspolygon[fillstyle=solid,fillcolor=lightlightblue](0,0)(6,0)(6,2)(0,2)(0,0)
\put(0,0){$\looopb$}
\put(0,1){$\looopb$}
\multirput(1,0){4}{\psline[linewidth=1.5pt,linecolor=blue](2.5,0)(2.5,2)}
\psarc[linewidth=1.5pt,linecolor=blue](1,2){.5}{180}{360}
\psarc[linewidth=1.5pt,linecolor=blue](1,0){.5}{0}{90}
\rput(4,-0.5){$\underbrace{\qquad\qquad\qquad\ }_{w-2}$}
\end{pspicture}
+\cdots+\alpha_k^{(w)}\
\psset{unit=.77cm}
\begin{pspicture}[shift=-0.89](0,0)(6.2,2)
\pspolygon[fillstyle=solid,fillcolor=lightlightblue](0,0)(6,0)(6,2)(0,2)(0,0)
\put(0,0){$\looopb$}
\put(1,0){$\looopb$}
\put(2,0){$\looopb$}
\put(0,1){$\looopb$}
\put(1,1){$\looopb$}
\put(2,1){$\looopb$}
\multirput(1,0){2}{\psline[linewidth=1.5pt,linecolor=blue](4.5,0)(4.5,2)}
\psarc[linewidth=1.5pt,linecolor=blue](3,2){.5}{180}{360}
\psarc[linewidth=1.5pt,linecolor=blue](3,0){.5}{0}{90}
\rput(2,-0.5){$\underbrace{\qquad \qquad \qquad\ }_{k}$}
\rput(5,-0.5){$\underbrace{\qquad\ }_{w-k}$}
\end{pspicture}
+\cdots
\nonumber
\\[.8cm]
&+\alpha_w^{(w)}\
\psset{unit=.77cm}
\begin{pspicture}[shift=-0.89](0,0)(6.2,2)
\pspolygon[fillstyle=solid,fillcolor=lightlightblue](0,0)(6,0)(6,2)(0,2)(0,0)
\put(0,0){$\looopb$}
\put(1,0){$\looopb$}
\put(2,0){$\looopb$}
\put(3,0){$\looopb$}
\put(4,0){$\looopb$}
\put(0,1){$\looopb$}
\put(1,1){$\looopb$}
\put(2,1){$\looopb$}
\put(3,1){$\looopb$}
\put(4,1){$\looopb$}
\psarc[linewidth=1.5pt,linecolor=blue](5,2){.5}{180}{360}
\psarc[linewidth=1.5pt,linecolor=blue](5,0){.5}{0}{90}
\rput(3,-0.5){$\underbrace{\qquad \qquad \qquad \qquad\qquad\ }_w$}
\end{pspicture}
+\alpha_{w+1}^{(w)}\
\psset{unit=.77cm}
\begin{pspicture}[shift=-0.89](0,0)(6.2,2)
\pspolygon[fillstyle=solid,fillcolor=lightlightblue](0,0)(6,0)(6,2)(0,2)(0,0)
\put(0,0){$\looopb$}
\put(1,0){$\looopb$}
\put(2,0){$\looopb$}
\put(3,0){$\looopb$}
\put(4,0){$\looopb$}
\put(5,0){$\looopb$}
\put(0,1){$\looopb$}
\put(1,1){$\looopb$}
\put(2,1){$\looopb$}
\put(3,1){$\looopb$}
\put(4,1){$\looopb$}
\put(5,1){$\looopb$}
\rput(3,-0.5){$\underbrace{\qquad \qquad \qquad \qquad\qquad\ }_w$}
\rput(6,0.5){$\bullet$}
\rput(6,1.5){$\bullet$}
\end{pspicture}
\label{uxidec}
\\[.1cm]
\nonumber
\end{align}
where the decomposition coefficients are functions of $u$ and $\xi$.
\begin{Proposition}
The decomposition coefficients in (\ref{uxidec}) are given by
\bea
 \alpha_0^{(w)}&\!\!\!=\!\!\!&\Gamma(u)\eta^{(w)}(u,\xi)\nn
 \alpha_k^{(w)}&\!\!\!=\!\!\!&\frac{(-1)^{k}s_0(2u)\eta^{(w)}(u,\xi)}{s_0(u+\xi)s_{w+1}(\xi-u)}
   \Big(U_{w-k}\big(\tfrac{\beta}{2}\big)\Gamma(u)-\beta_1s_{w-k+1}(u+\xi)s_{1}(\xi-u)\Big),\qquad k=1,2,\ldots,w \nn
 \alpha_{w+1}^{(w)}&\!\!\!=\!\!\!&\frac{(-1)^ws_0(2u)s_{1}(\xi-u)\eta^{(w)}(u,\xi)}{s_{w+1}(\xi-u)}
\label{al}
\eea
where $U_n(x)$ is the $n$-th Chebyshev polynomial of the second kind and
\be
 \eta^{(w)}(u,\xi):=\prod_{j=1}^w s_{-1}(u+\xi_j)s_{-1}(u-\xi_j)
\label{eta}
\ee
\label{Prop:alpha}
\end{Proposition}
\noindent{\scshape Proof:} 
For $w=0$, the coefficients reduce to
\be
 \alpha_0^{(0)}=\Gamma(u),\qquad \alpha_1^{(0)}=s_0(2u)
\ee
in accordance with the decomposition of the twist boundary condition (\ref{twist}). For $w=1$, we have
\bea
\psset{unit=0.77cm}
\begin{pspicture}[shift=-0.89](0,0)(2,2)
\facegrid{(0,0)}{(1,2)}
\psarc[linewidth=0.025](0,0){0.16}{0}{90}
\psarc[linewidth=0.025](0,1){0.16}{0}{90}
\pspolygon[fillstyle=solid,fillcolor=lightlightblue](1,1)(2,2)(2,0)(1,1)
\psline[linecolor=blue,linewidth=1.5pt](1,0.5)(1.5,0.5)
\psline[linecolor=blue,linewidth=1.5pt](1,1.5)(1.5,1.5)
\rput(0.55,0.5){$_{_{u-\xi_1}}$}
\rput(0.55,1.5){$_{_{-\!u\!-\!\xi_0}}$}
\rput(1.65,1){$_{u}$}
\end{pspicture}
&=&\eta^{(1)}(u,\xi)\,\Gamma(u)
\psset{unit=0.77cm}
\begin{pspicture}[shift=-0.89](-0.2,0)(1.65,2)
\pspolygon[fillstyle=solid,fillcolor=lightlightblue](0,0)(1,0)(1,2)(0,2)(0,0)
\rput(0,0){\looopa}
\rput(0,1){\looopb}
\psarc[linecolor=blue,linewidth=1.5pt](1,1){0.5}{-90}{90}
\end{pspicture}
+s_0(2u)\Big(\Gamma(u)+\beta_1s_0(u+\xi_1)s_0(u-\xi_1)\Big)
\psset{unit=0.77cm}
\begin{pspicture}[shift=-0.89](-0.2,0)(1.15,2)
\pspolygon[fillstyle=solid,fillcolor=lightlightblue](0,0)(1,0)(1,2)(0,2)(0,0)
\psarc[linecolor=blue,linewidth=1.5pt](0,0){0.5}{0}{90}
\psarc[linecolor=blue,linewidth=1.5pt](0,2){0.5}{-90}{0}
\end{pspicture}
\nn
&-&s_0(2u)s_0(u+\xi)s_0(u-\xi_1)
\psset{unit=0.77cm}
\begin{pspicture}[shift=-0.89](-0.2,0)(1.15,2)
\pspolygon[fillstyle=solid,fillcolor=lightlightblue](0,0)(1,0)(1,2)(0,2)(0,0)
\rput(0,0){\looopb}
\rput(0,1){\looopb}
\rput(1,0.5){$\bullet$}
\rput(1,1.5){$\bullet$}
\end{pspicture}
\nn
&+&s_0(2u)\Big(\eta^{(1)}(u,\xi)
\psset{unit=0.77cm}
\begin{pspicture}[shift=-0.89](-0.2,0)(1.15,2)
\pspolygon[fillstyle=solid,fillcolor=lightlightblue](0,0)(1,0)(1,2)(0,2)(0,0)
\rput(0,0){\looopa}
\rput(0,1){\looopb}
\rput(1,0.5){$\bullet$}
\rput(1,1.5){$\bullet$}
\end{pspicture}
-s_1(u+\xi)s_0(u-\xi_2)
\psset{unit=0.77cm}
\begin{pspicture}[shift=-0.89](-0.2,0)(1.15,2)
\pspolygon[fillstyle=solid,fillcolor=lightlightblue](0,0)(1,0)(1,2)(0,2)(0,0)
\rput(0,0){\looopa}
\rput(0,1){\looopa}
\rput(1,0.5){$\bullet$}
\rput(1,1.5){$\bullet$}
\end{pspicture}
\ \Big)
\nonumber
\\[.2cm]
&\simeq&\alpha_0^{(1)}
\psset{unit=0.77cm}
\begin{pspicture}[shift=-0.89](-0.2,0)(1.65,2)
\pspolygon[fillstyle=solid,fillcolor=lightlightblue](0,0)(1,0)(1,2)(0,2)(0,0)
\rput(0,0){\looopa}
\rput(0,1){\looopb}
\psarc[linecolor=blue,linewidth=1.5pt](1,1){0.5}{-90}{90}
\end{pspicture}
+\alpha_1^{(1)}
\psset{unit=0.77cm}
\begin{pspicture}[shift=-0.89](-0.2,0)(1.15,2)
\pspolygon[fillstyle=solid,fillcolor=lightlightblue](0,0)(1,0)(1,2)(0,2)(0,0)
\psarc[linecolor=blue,linewidth=1.5pt](0,0){0.5}{0}{90}
\psarc[linecolor=blue,linewidth=1.5pt](0,2){0.5}{-90}{0}
\end{pspicture}
+\alpha_2^{(1)}
\psset{unit=0.77cm}
\begin{pspicture}[shift=-0.89](-0.2,0)(1.15,2)
\pspolygon[fillstyle=solid,fillcolor=lightlightblue](0,0)(1,0)(1,2)(0,2)(0,0)
\rput(0,0){\looopb}
\rput(0,1){\looopb}
\rput(1,0.5){$\bullet$}
\rput(1,1.5){$\bullet$}
\end{pspicture}
\eea
thus reproducing (\ref{al}). The proof is now completed by induction in $w$. 
Introducing the shorthand notation $v=-u-\xi_{w-1}$, we observe that
\begin{align}
&\psset{unit=0.77cm}
\begin{pspicture}[shift=-0.89](0,0)(1.65,2)
\facegrid{(0,0)}{(1,2)}
\psarc[linewidth=0.025](0,0){0.16}{0}{90}
\psarc[linewidth=0.025](0,1){0.16}{0}{90}
\psarc[linecolor=blue,linewidth=1.5pt](1,1){0.5}{-90}{90}
\rput(0.55,0.5){$_{_{u-\xi_w}}$}
\rput(0.5,1.5){$_{v}$}
\end{pspicture}
=s_{-1}(u+\xi_w)s_{-1}(u-\xi_w)
\psset{unit=0.77cm}
\begin{pspicture}[shift=-0.89](-0.2,0)(1.65,2)
\pspolygon[fillstyle=solid,fillcolor=lightlightblue](0,0)(1,0)(1,2)(0,2)(0,0)
\rput(0,0){\looopa}
\rput(0,1){\looopb}
\psarc[linecolor=blue,linewidth=1.5pt](1,1){0.5}{-90}{90}
\end{pspicture}
+s_0(2u)
\psset{unit=0.77cm}
\begin{pspicture}[shift=-0.89](-0.2,0)(1.15,2)
\pspolygon[fillstyle=solid,fillcolor=lightlightblue](0,0)(1,0)(1,2)(0,2)(0,0)
\psarc[linecolor=blue,linewidth=1.5pt](0,0){0.5}{0}{90}
\psarc[linecolor=blue,linewidth=1.5pt](0,2){0.5}{-90}{0}
\end{pspicture}
\nonumber\\[.2cm]
&\psset{unit=0.77cm}
\begin{pspicture}[shift=-0.89](0,0)(1.65,2)
\facegrid{(0,0)}{(1,2)}
\psarc[linewidth=0.025](0,0){0.16}{0}{90}
\psarc[linewidth=0.025](0,1){0.16}{0}{90}
\psarc[linecolor=blue,linewidth=1.5pt](1,0){0.5}{0}{90}
\psarc[linecolor=blue,linewidth=1.5pt](1,2){0.5}{-90}{0}
\rput(0.55,0.5){$_{_{u-\xi_w}}$}
\rput(0.5,1.5){$_{v}$}
\end{pspicture}
\simeq s_0(u+\xi_w)s_0(u-\xi_w)
\psset{unit=0.77cm}
\begin{pspicture}[shift=-0.89](-0.2,0)(1.65,2)
\pspolygon[fillstyle=solid,fillcolor=lightlightblue](0,0)(1,0)(1,2)(0,2)(0,0)
\rput(0,0){\looopb}
\rput(0,1){\looopa}
\psarc[linecolor=blue,linewidth=1.5pt](1,0){0.5}{0}{90}
\psarc[linecolor=blue,linewidth=1.5pt](1,2){0.5}{-90}{0}
\end{pspicture}
-s_{-1}(u+\xi_w)s_0(u-\xi_w)
\psset{unit=0.77cm}
\begin{pspicture}[shift=-0.89](-0.2,0)(1.65,2)
\pspolygon[fillstyle=solid,fillcolor=lightlightblue](0,0)(1,0)(1,2)(0,2)(0,0)
\rput(0,0){\looopb}
\rput(0,1){\looopb}
\psarc[linecolor=blue,linewidth=1.5pt](1,0){0.5}{0}{90}
\psarc[linecolor=blue,linewidth=1.5pt](1,2){0.5}{-90}{0}
\end{pspicture}
\nonumber\\[.2cm]
&\psset{unit=0.77cm}
\begin{pspicture}[shift=-0.89](0,0)(2.5,2)
\facegrid{(0,0)}{(1,2)}
\psarc[linewidth=0.025](0,0){0.16}{0}{90}
\psarc[linewidth=0.025](0,1){0.16}{0}{90}
\psarc[linecolor=blue,linewidth=1.5pt](1,0){0.5}{0}{90}
\psarc[linecolor=blue,linewidth=1.5pt](1,1){0.5}{0}{90}
\psarc[linecolor=blue,linewidth=1.5pt](2,1){0.5}{180}{270}
\psarc[linecolor=blue,linewidth=1.5pt](2,0){0.5}{0}{90}
\rput(0.55,0.5){$_{_{u-\xi_w}}$}
\rput(0.5,1.5){$_{v}$}
\end{pspicture}
\simeq -s_{-1}(u+\xi_w)s_0(u-\xi_w)
\psset{unit=0.77cm}
\begin{pspicture}[shift=-0.89](-0.2,0)(1.65,2)
\pspolygon[fillstyle=solid,fillcolor=lightlightblue](0,0)(1,0)(1,2)(0,2)(0,0)
\rput(0,0){\looopb}
\rput(0,1){\looopb}
\psarc[linecolor=blue,linewidth=1.5pt](1,0){0.5}{0}{90}
\psarc[linecolor=blue,linewidth=1.5pt](1,1){0.5}{0}{90}
\psarc[linecolor=blue,linewidth=1.5pt](2,1){0.5}{180}{270}
\psarc[linecolor=blue,linewidth=1.5pt](2,0){0.5}{0}{90}
\end{pspicture}
\end{align}
and
\be
\psset{unit=0.77cm}
\begin{pspicture}[shift=-0.89](0,0)(2.5,2)
\facegrid{(0,0)}{(1,2)}
\psarc[linewidth=0.025](0,0){0.16}{0}{90}
\psarc[linewidth=0.025](0,1){0.16}{0}{90}
\psarc[linecolor=blue,linewidth=1.5pt](1,0){0.5}{0}{90}
\psarc[linecolor=blue,linewidth=1.5pt](1,1){0.5}{0}{90}
\psarc[linecolor=blue,linewidth=1.5pt](2,1){0.5}{180}{270}
\psarc[linecolor=blue,linewidth=1.5pt](2,2){0.5}{180}{270}
\rput(0.55,0.5){$_{_{u-\xi_2}}$}
\rput(0.55,1.5){$_{_{-\!u\!-\!\xi_{1}}}$}
\rput(2,0.5){$\bullet$}
\rput(2,1.5){$\bullet$}
\end{pspicture}
\simeq -s_{0}(u+\xi_1)s_0(u-\xi_2)
\psset{unit=0.77cm}
\begin{pspicture}[shift=-0.89](-0.2,0)(1.65,2)
\pspolygon[fillstyle=solid,fillcolor=lightlightblue](0,0)(1,0)(1,2)(0,2)(0,0)
\rput(0,0){\looopb}
\rput(0,1){\looopb}
\psarc[linecolor=blue,linewidth=1.5pt](1,0){0.5}{0}{90}
\psarc[linecolor=blue,linewidth=1.5pt](1,1){0.5}{0}{90}
\psarc[linecolor=blue,linewidth=1.5pt](2,1){0.5}{180}{270}
\psarc[linecolor=blue,linewidth=1.5pt](2,2){0.5}{180}{270}
\rput(2,0.5){$\bullet$}
\rput(2,1.5){$\bullet$}
\end{pspicture}
\label{rel2}
\ee
where the relation (\ref{rel2}) is relevant for $w=2$ only. For general $w\geq2$, we then deduce that
\bea
 \alpha_0^{(w)}&\!\!\!=\!\!\!&s_0(u+\xi_{w-1})s_0(u-\xi_{w+1})\alpha_0^{(w-1)}\nn
 \alpha_1^{(w)}&\!\!\!=\!\!\!&s_0(2u)\alpha_0^{(w-1)}+s_0(u+\xi_w)s_0(u-\xi_w)\alpha_1^{(w-1)}\nn
 \alpha_k^{(w)}&\!\!\!=\!\!\!&-s_0(u+\xi_{w-1})s_0(u-\xi_w)\alpha_{k-1}^{(w-1)},\qquad\qquad k=2,\ldots,w+1
\label{alal}
\eea
Using that
\be
 s_0(u+\xi)+U_{w-2}\big(\tfrac{\beta}{2}\big)s_0(u+\xi_w)=U_{w-1}\big(\tfrac{\beta}{2}\big)s_0(u+\xi_{w-1})
\ee 
the recursion relations (\ref{alal}) are seen to be satisfied by (\ref{al}). This completes the induction step and hence the proof.
\hfill $\square$

\subsection{Renormalised transfer tangles for $\beta\neq0$}
\label{Sec:Ren}

Here we assume $\beta\neq0$. The case $\beta=0$ corresponds to critical dense polymers ${\cal LM}(1,2)$ and
is treated separately in Section~\ref{Sec:CritDensePol}.

It is convenient to introduce the renormalised transfer tangle
\be
 \db(u):=\frac{1}{\eta(u)}\Db(u),\qquad 
  \eta(u):=\frac{\beta\,\Gamma(0)s_w(u+\xi)s_{0}(\xi)\eta^{(w)}(u,\xi)}{s_0(u+\xi)s_{w}(\xi)}
\label{eta0}
\ee
noting that the normalisation function reduces to $\eta(u)=\beta\,\Gamma(0)$ for $w=0$.
Renormalising the decomposition coefficients (\ref{al}) accordingly
\be
 \hat\alpha_k^{(w)}:=\frac{\alpha_k^{(w)}}{\eta(u)}
\ee
yields
\begin{align}
 \hat\alpha_0^{(w)}&=\frac{\Gamma(u)s_0(u+\xi)s_w(\xi)}{\beta\,\Gamma(0)s_w(u+\xi)s_0(\xi)}
\nonumber\\[.2cm]
 \hat\alpha_k^{(w)}&=\frac{(-1)^ks_0(2u)s_w(\xi)\big[U_{w-k}\big(\tfrac{\beta}{2}\big)\Gamma(u)
   -\beta_1s_{w-k+1}(u+\xi)s_1(\xi-u)\big]}{\beta\,\Gamma(0)s_{w+1}(\xi-u)s_w(u+\xi)s_0(\xi)},\qquad k=1,\ldots,w
\nonumber\\[.2cm]
 \hat\alpha_{w+1}^{(w)}&=\frac{(-1)^w s_0(2u)s_1(\xi-u)s_0(u+\xi)s_w(\xi)}{\beta\,\Gamma(0) s_{w+1}(\xi-u)s_w(u+\xi)s_0(\xi)}
\end{align}
This renormalisation ensures that
\be
 \lim_{u\to0}\db(u)\simeq I
\label{dI}
\ee
since
\be
 \lim_{u\to0}\db(u)\simeq\frac{1}{\beta}\lim_{u\to0}\,
\psset{unit=.6cm}
\begin{pspicture}[shift=-0.89](0.5,0)(9.5,2)
\facegrid{(1,0)}{(9,2)}
\psarc[linewidth=0.025](1,0){0.16}{0}{90}
\psarc[linewidth=0.025](1,1){0.16}{0}{90}
\psarc[linewidth=0.025](2,0){0.16}{0}{90}
\psarc[linewidth=0.025](2,1){0.16}{0}{90}
\psarc[linewidth=0.025](3,0){0.16}{0}{90}
\psarc[linewidth=0.025](3,1){0.16}{0}{90}
\psarc[linewidth=0.025](4,0){0.16}{0}{90}
\psarc[linewidth=0.025](4,1){0.16}{0}{90}
\rput(1.5,0.5){$_{_u}$}
\rput(2.5,0.5){$_{_u}$}
\rput(3.5,0.5){$\ldots$}
\rput(4.5,0.5){$_{_u}$}
\rput(1.5,1.5){$_{_{\lambda-u}}$}
\rput(2.5,1.5){$_{_{\lambda-u}}$}
\rput(3.5,1.5){$\ldots$}
\rput(4.5,1.5){$_{_{\lambda-u}}$}
\rput(5,0){$\loopa$}\rput(5,1){$\loopb$}
\rput(6,0){$\loopa$}\rput(6,1){$\loopb$}
\rput(8,0){$\loopa$}\rput(8,1){$\loopb$}
\rput(7.5,0.5){$\ldots$}
\rput(7.5,1.5){$\ldots$}
\psarc[linewidth=1.5pt,linecolor=blue](1,1){.5}{90}{270}
\psarc[linewidth=1.5pt,linecolor=blue](9,1){.5}{-90}{90}
\rput(3,-0.6){$\underbrace{\qquad \qquad \qquad\,}_N$}
\rput(7,-0.6){$\underbrace{\qquad \qquad \qquad\,}_w$}
\end{pspicture}
\, =\frac{1}{\beta}\;
\psset{unit=.6cm}
\begin{pspicture}[shift=-0.89](0.5,0)(9.5,2)
\facegrid{(1,0)}{(9,2)}
\rput(3.5,0.5){$\ldots$}
\rput(3.5,1.5){$\ldots$}
\rput(1,0){$\loopa$}\rput(1,1){$\loopb$}
\rput(2,0){$\loopa$}\rput(2,1){$\loopb$}
\rput(4,0){$\loopa$}\rput(4,1){$\loopb$}
\rput(5,0){$\loopa$}\rput(5,1){$\loopb$}
\rput(6,0){$\loopa$}\rput(6,1){$\loopb$}
\rput(8,0){$\loopa$}\rput(8,1){$\loopb$}
\rput(7.5,0.5){$\ldots$}
\rput(7.5,1.5){$\ldots$}
\psarc[linewidth=1.5pt,linecolor=blue](1,1){.5}{90}{270}
\psarc[linewidth=1.5pt,linecolor=blue](9,1){.5}{-90}{90}
\rput(3,-0.6){$\underbrace{\qquad \qquad \qquad\,}_N$}
\rput(7,-0.6){$\underbrace{\qquad \qquad \qquad\,}_w$}
\end{pspicture}
\\[.4cm]
\ee
where the first relation is a consequence of
\be
 \lim_{u\to0}\hat\alpha_0^{(w)}=\frac{1}{\beta},\qquad \lim_{u\to0}\hat\alpha_k^{(w)}=0,\qquad k=1,\ldots,w+1
\ee
Due to
\be
 \eta(\lambda-u)=\eta(u)
\ee
the renormalisation also preserves the crossing symmetry 
\be
 \db(\lambda-u)=\db(u)
\label{dcross}
\ee
It is noted that the normalisation (\ref{eta0}) is not uniquely determined by requiring the renormalised transfer tangle
to have the properties (\ref{dI}) and (\ref{dcross}).

\subsection{Hamiltonians for $\beta\neq0$}

The Hamiltonian is obtained by expanding the renormalised double row transfer tangle as
\be
 \db(u)=I-\frac{2u}{\sin\lambda}(\Hb+hI)+{\cal O}(u^2)
\label{dH}
\ee
where $h$ measures a convenient shift in the groundstate energy.
Recalling the Robin representation $\rho_d^{(N,w)}(\Db(u))$ discussed in Section~\ref{Sec:RobinModules},
the matrix representation $\rho_d^{(N,w)}(\Hb)$ of $\Hb$ is introduced as
\be
 \rho_d^{(N,w)}(\Hb)=-\frac{\sin\lambda}{2}\frac{\partial}{\partial u}\rho_d^{(N,w)}\big(\db(u)\big)\Big|_{u=0}-hI
\label{rhoH}
\ee
where $I$ is the identity matrix of the appropriate size.

We now assume $\beta\neq0$. The case $\beta=0$ corresponds to critical dense polymers ${\cal LM}(1,2)$ and
is treated separately in Section~\ref{Sec:CritDensePol}. For $\beta\neq0$, we set
\be
 h:=\frac{N\beta}{2}-\frac{1}{\beta}+\frac{\beta_2}{2\,\Gamma(0)}-\frac{U_{w-1}\big(\frac{\beta}{2}\big)}{2s_0(\xi)s_w(\xi)}
\label{h}
\ee
and note that the last contribution vanishes for $w=0$ since $U_{-1}(x)\equiv0$.
\begin{Proposition}
As an element of $TL_{N+w}(\beta;\beta_1,\beta_2)$ 
designed to act on ${\cal V}_d^{(N,w)}$ for any $d$, the Hamiltonian for $\beta\neq0$ is given by
\bea
 &&\Hb\simeq -\sum_{j=1}^{N-1}e_j-\sum_{k=1}^{w}\frac{(-1)^k}{s_0(\xi)s_{w+1}(\xi)}
  \Big(U_{w-k}\big(\tfrac{\beta}{2}\big)-\frac{\beta_1 s_1(\xi)s_{w-k+1}(\xi)}{\Gamma(0)}\Big)
  e_Ne_{N+1}\ldots e_{N+k-1}\nn
 &&\qquad\qquad\qquad\qquad -(-1)^w\frac{s_1(\xi)}{s_{w+1}(\xi)\Gamma(0)}e_Ne_{N+1}\ldots e_{N+w-1}f_{N+w}
\label{Hbeta}
\eea
\label{Prop:Hbeta}
\end{Proposition}
\noindent{\scshape Proof:}
To keep the presentation simple, we first focus on the situation where $w=0$. In this case, we have
\bea
 \db(u)&\!\!\!=\!\!\!&\frac{1}{\eta(u)}
\psset{unit=.6cm}
\left(
\Gamma(u)\;
\begin{pspicture}[shift=-0.89](0.5,0)(7.5,2)
\facegrid{(1,0)}{(7,2)}
\psarc[linewidth=0.025](1,0){0.16}{0}{90}
\psarc[linewidth=0.025](1,1){0.16}{0}{90}
\psarc[linewidth=0.025](2,0){0.16}{0}{90}
\psarc[linewidth=0.025](2,1){0.16}{0}{90}
\psarc[linewidth=0.025](3,0){0.16}{0}{90}
\psarc[linewidth=0.025](3,1){0.16}{0}{90}
\psarc[linewidth=0.025](4,0){0.16}{0}{90}
\psarc[linewidth=0.025](4,1){0.16}{0}{90}
\psarc[linewidth=0.025](6,0){0.16}{0}{90}
\psarc[linewidth=0.025](6,1){0.16}{0}{90}
\rput(1.5,0.5){$_{_u}$}
\rput(2.5,0.5){$_{_u}$}
\rput(3.5,0.5){$_{_u}$}
\rput(4.5,0.5){$_{_u}$}
\rput(6.5,0.5){$_{_u}$}
\rput(1.5,1.5){$_{_{\lambda-u}}$}
\rput(2.5,1.5){$_{_{\lambda-u}}$}
\rput(3.5,1.5){$_{_{\lambda-u}}$}
\rput(4.5,1.5){$_{_{\lambda-u}}$}
\rput(6.5,1.5){$_{_{\lambda-u}}$}
\rput(5.5,0.5){$\ldots$}
\rput(5.5,1.5){$\ldots$}
\psarc[linewidth=1.5pt,linecolor=blue](1,1){.5}{90}{270}
\psarc[linewidth=1.5pt,linecolor=blue](7,1){.5}{-90}{90}
\end{pspicture}
\;+s_0(2u)\;
\begin{pspicture}[shift=-0.89](0.5,0)(7.5,2)
\facegrid{(1,0)}{(7,2)}
\psarc[linewidth=0.025](1,0){0.16}{0}{90}
\psarc[linewidth=0.025](1,1){0.16}{0}{90}
\psarc[linewidth=0.025](2,0){0.16}{0}{90}
\psarc[linewidth=0.025](2,1){0.16}{0}{90}
\psarc[linewidth=0.025](3,0){0.16}{0}{90}
\psarc[linewidth=0.025](3,1){0.16}{0}{90}
\psarc[linewidth=0.025](4,0){0.16}{0}{90}
\psarc[linewidth=0.025](4,1){0.16}{0}{90}
\psarc[linewidth=0.025](6,0){0.16}{0}{90}
\psarc[linewidth=0.025](6,1){0.16}{0}{90}
\rput(1.5,0.5){$_{_u}$}
\rput(2.5,0.5){$_{_u}$}
\rput(3.5,0.5){$_{_u}$}
\rput(4.5,0.5){$_{_u}$}
\rput(6.5,0.5){$_{_u}$}
\rput(1.5,1.5){$_{_{\lambda-u}}$}
\rput(2.5,1.5){$_{_{\lambda-u}}$}
\rput(3.5,1.5){$_{_{\lambda-u}}$}
\rput(4.5,1.5){$_{_{\lambda-u}}$}
\rput(6.5,1.5){$_{_{\lambda-u}}$}
\rput(5.5,0.5){$\ldots$}
\rput(5.5,1.5){$\ldots$}
\psarc[linewidth=1.5pt,linecolor=blue](1,1){.5}{90}{270}
\psline[linecolor=blue,linewidth=1.5pt]{-}(7,0.5)(7.5,0.5)
\psline[linecolor=blue,linewidth=1.5pt]{-}(7,1.5)(7.5,1.5)
\rput(7.5,0.5){$\bullet$}
\rput(7.5,1.5){$\bullet$}
\end{pspicture}
\,\right)\nonumber\\[.15cm]
&\!\!\!=\!\!\!&\frac{[s_1(-u)]^{2N}}{\eta(u)}
\psset{unit=.6cm}
\left(
\Gamma(u)\;
\begin{pspicture}[shift=-0.89](0.5,0)(7.5,2)
\facegrid{(1,0)}{(7,2)}
\rput(1,0){$\loopa$}\rput(1,1){$\loopb$}
\rput(2,0){$\loopa$}\rput(2,1){$\loopb$}
\rput(3,0){$\loopa$}\rput(3,1){$\loopb$}
\rput(4,0){$\loopa$}\rput(4,1){$\loopb$}
\rput(5.5,0.5){$\ldots$}
\rput(5.5,1.5){$\ldots$}
\rput(6,0){$\loopa$}\rput(6,1){$\loopb$}
\psarc[linewidth=1.5pt,linecolor=blue](1,1){.5}{90}{270}
\psarc[linewidth=1.5pt,linecolor=blue](7,1){.5}{-90}{90}
\end{pspicture}
\;+s_0(2u)\;
\begin{pspicture}[shift=-0.89](0.5,0)(7.5,2)
\facegrid{(1,0)}{(7,2)}
\rput(1,0){$\loopa$}\rput(1,1){$\loopb$}
\rput(2,0){$\loopa$}\rput(2,1){$\loopb$}
\rput(3,0){$\loopa$}\rput(3,1){$\loopb$}
\rput(4,0){$\loopa$}\rput(4,1){$\loopb$}
\rput(5.5,0.5){$\ldots$}
\rput(5.5,1.5){$\ldots$}
\rput(6,0){$\loopa$}\rput(6,1){$\loopb$}
\psarc[linewidth=1.5pt,linecolor=blue](1,1){.5}{90}{270}
\psline[linecolor=blue,linewidth=1.5pt]{-}(7,0.5)(7.5,0.5)
\psline[linecolor=blue,linewidth=1.5pt]{-}(7,1.5)(7.5,1.5)
\rput(7.5,0.5){$\bullet$}
\rput(7.5,1.5){$\bullet$}
\end{pspicture}
\,\right)\nonumber\\[.15cm]
&\!\!\!+\!\!\!&\frac{s_0(u)[s_1(-u)]^{2N-1}\Gamma(u)}{\eta(u)}
\psset{unit=.6cm}
\left(\,
\begin{pspicture}[shift=-0.89](0.5,0)(7.5,2)
\facegrid{(1,0)}{(7,2)}
\rput(1,0){$\loopb$}\rput(1,1){$\loopb$}
\rput(2,0){$\loopa$}\rput(2,1){$\loopb$}
\rput(3,0){$\loopa$}\rput(3,1){$\loopb$}
\rput(4,0){$\loopa$}\rput(4,1){$\loopb$}
\rput(5.5,0.5){$\ldots$}
\rput(5.5,1.5){$\ldots$}
\rput(6,0){$\loopa$}\rput(6,1){$\loopb$}
\psarc[linewidth=1.5pt,linecolor=blue](1,1){.5}{90}{270}
\psarc[linewidth=1.5pt,linecolor=blue](7,1){.5}{-90}{90}
\end{pspicture}
\;+\;
\begin{pspicture}[shift=-0.89](0.5,0)(7.5,2)
\facegrid{(1,0)}{(7,2)}
\rput(1,0){$\loopa$}\rput(1,1){$\loopa$}
\rput(2,0){$\loopa$}\rput(2,1){$\loopb$}
\rput(3,0){$\loopa$}\rput(3,1){$\loopb$}
\rput(4,0){$\loopa$}\rput(4,1){$\loopb$}
\rput(5.5,0.5){$\ldots$}
\rput(5.5,1.5){$\ldots$}
\rput(6,0){$\loopa$}\rput(6,1){$\loopb$}
\psarc[linewidth=1.5pt,linecolor=blue](1,1){.5}{90}{270}
\psarc[linewidth=1.5pt,linecolor=blue](7,1){.5}{-90}{90}
\end{pspicture}
\,\right.\nonumber\\[.15cm]
&&
\hspace{3.5cm}+\
\psset{unit=.6cm}
\begin{pspicture}[shift=-0.89](0.5,0)(7.5,2)
\facegrid{(1,0)}{(7,2)}
\rput(1,0){$\loopa$}\rput(1,1){$\loopb$}
\rput(2,0){$\loopb$}\rput(2,1){$\loopb$}
\rput(3,0){$\loopa$}\rput(3,1){$\loopb$}
\rput(4,0){$\loopa$}\rput(4,1){$\loopb$}
\rput(5.5,0.5){$\ldots$}
\rput(5.5,1.5){$\ldots$}
\rput(6,0){$\loopa$}\rput(6,1){$\loopb$}
\psarc[linewidth=1.5pt,linecolor=blue](1,1){.5}{90}{270}
\psarc[linewidth=1.5pt,linecolor=blue](7,1){.5}{-90}{90}
\end{pspicture}
\;+\;
\begin{pspicture}[shift=-0.89](0.5,0)(7.5,2)
\facegrid{(1,0)}{(7,2)}
\rput(1,0){$\loopa$}\rput(1,1){$\loopb$}
\rput(2,0){$\loopa$}\rput(2,1){$\loopa$}
\rput(3,0){$\loopa$}\rput(3,1){$\loopb$}
\rput(4,0){$\loopa$}\rput(4,1){$\loopb$}
\rput(5.5,0.5){$\ldots$}
\rput(5.5,1.5){$\ldots$}
\rput(6,0){$\loopa$}\rput(6,1){$\loopb$}
\psarc[linewidth=1.5pt,linecolor=blue](1,1){.5}{90}{270}
\psarc[linewidth=1.5pt,linecolor=blue](7,1){.5}{-90}{90}
\end{pspicture}
\\[.15cm]
&&
\hspace{3.5cm}+\
\psset{unit=.6cm}
\begin{pspicture}[shift=-0.89](0.5,0)(7.5,2)
\facegrid{(1,0)}{(7,2)}
\rput(1,0){$\loopa$}\rput(1,1){$\loopb$}
\rput(2,0){$\loopa$}\rput(2,1){$\loopb$}
\rput(3,0){$\loopb$}\rput(3,1){$\loopb$}
\rput(4,0){$\loopa$}\rput(4,1){$\loopb$}
\rput(5.5,0.5){$\ldots$}
\rput(5.5,1.5){$\ldots$}
\rput(6,0){$\loopa$}\rput(6,1){$\loopb$}
\psarc[linewidth=1.5pt,linecolor=blue](1,1){.5}{90}{270}
\psarc[linewidth=1.5pt,linecolor=blue](7,1){.5}{-90}{90}
\end{pspicture}
\;+\;
\begin{pspicture}[shift=-0.89](0.5,0)(7.5,2)
\facegrid{(1,0)}{(7,2)}
\rput(1,0){$\loopa$}\rput(1,1){$\loopb$}
\rput(2,0){$\loopa$}\rput(2,1){$\loopb$}
\rput(3,0){$\loopa$}\rput(3,1){$\loopa$}
\rput(4,0){$\loopa$}\rput(4,1){$\loopb$}
\rput(5.5,0.5){$\ldots$}
\rput(5.5,1.5){$\ldots$}
\rput(6,0){$\loopa$}\rput(6,1){$\loopb$}
\psarc[linewidth=1.5pt,linecolor=blue](1,1){.5}{90}{270}
\psarc[linewidth=1.5pt,linecolor=blue](7,1){.5}{-90}{90}
\end{pspicture}
\nonumber\\[.15cm]
&&
\hspace{3.5cm}+\
\psset{unit=.6cm}
\begin{pspicture}[shift=-0.89](0.5,0)(7.5,2)
\facegrid{(1,0)}{(7,2)}
\rput(1,0){$\loopa$}\rput(1,1){$\loopb$}
\rput(2,0){$\loopa$}\rput(2,1){$\loopb$}
\rput(3,0){$\loopa$}\rput(3,1){$\loopb$}
\rput(4,0){$\loopb$}\rput(4,1){$\loopb$}
\rput(5.5,0.5){$\ldots$}
\rput(5.5,1.5){$\ldots$}
\rput(6,0){$\loopa$}\rput(6,1){$\loopb$}
\psarc[linewidth=1.5pt,linecolor=blue](1,1){.5}{90}{270}
\psarc[linewidth=1.5pt,linecolor=blue](7,1){.5}{-90}{90}
\end{pspicture}
\;+\;
\begin{pspicture}[shift=-0.89](0.5,0)(7.5,2)
\facegrid{(1,0)}{(7,2)}
\rput(1,0){$\loopa$}\rput(1,1){$\loopb$}
\rput(2,0){$\loopa$}\rput(2,1){$\loopb$}
\rput(3,0){$\loopa$}\rput(3,1){$\loopb$}
\rput(4,0){$\loopa$}\rput(4,1){$\loopa$}
\rput(5.5,0.5){$\ldots$}
\rput(5.5,1.5){$\ldots$}
\rput(6,0){$\loopa$}\rput(6,1){$\loopb$}
\psarc[linewidth=1.5pt,linecolor=blue](1,1){.5}{90}{270}
\psarc[linewidth=1.5pt,linecolor=blue](7,1){.5}{-90}{90}
\end{pspicture}
\nonumber\\[.15cm]
&&
\left.
\hspace{2.47cm}+\ldots+\
\psset{unit=.6cm}
\begin{pspicture}[shift=-0.89](0.5,0)(7.5,2)
\facegrid{(1,0)}{(7,2)}
\rput(1,0){$\loopa$}\rput(1,1){$\loopb$}
\rput(2,0){$\loopa$}\rput(2,1){$\loopb$}
\rput(3,0){$\loopa$}\rput(3,1){$\loopb$}
\rput(4,0){$\loopa$}\rput(4,1){$\loopb$}
\rput(5.5,0.5){$\ldots$}
\rput(5.5,1.5){$\ldots$}
\rput(6,0){$\loopb$}\rput(6,1){$\loopb$}
\psarc[linewidth=1.5pt,linecolor=blue](1,1){.5}{90}{270}
\psarc[linewidth=1.5pt,linecolor=blue](7,1){.5}{-90}{90}
\end{pspicture}
\;+\;
\begin{pspicture}[shift=-0.89](0.5,0)(7.5,2)
\facegrid{(1,0)}{(7,2)}
\rput(1,0){$\loopa$}\rput(1,1){$\loopb$}
\rput(2,0){$\loopa$}\rput(2,1){$\loopb$}
\rput(3,0){$\loopa$}\rput(3,1){$\loopb$}
\rput(4,0){$\loopa$}\rput(4,1){$\loopb$}
\rput(5.5,0.5){$\ldots$}
\rput(5.5,1.5){$\ldots$}
\rput(6,0){$\loopa$}\rput(6,1){$\loopa$}
\psarc[linewidth=1.5pt,linecolor=blue](1,1){.5}{90}{270}
\psarc[linewidth=1.5pt,linecolor=blue](7,1){.5}{-90}{90}
\end{pspicture}
\,\right)
+\mathcal O(u^2)
\nonumber
\eea
and using the explicit power series expansions
\be
 [s_1(-u)]^n=1-n(\cot\lambda)u+\mathcal O(u^2),\qquad
 s_0(nu)=\frac{n}{\sin\lambda}\,u+\mathcal O(u^2),\qquad
 \Gamma(u)=\Gamma(0)-\frac{\beta_2}{\sin\lambda}\,u
  +\mathcal O(u^2)
\ee
we identify the Hamiltonian (\ref{Hbeta})
\be
 \Hb=-\sum_{j=1}^{N-1}e_j-\frac{1}{\Gamma(0)}\,f_N,\qquad w=0
\ee
The generalisation to $w>0$ is straightforward and follows from
\begin{align}
 \hat\alpha_0^{(w)}&=\frac{1}{\beta}+\frac{u}{\sin2\lambda}\Big(\frac{U_{w-1}\big(\frac{\beta}{2}\big)}{s_0(\xi)s_w(\xi)}
   -\frac{\beta_2}{\Gamma(0)}\Big) +{\cal O}(u^2)
 \nonumber\\[.2cm]
 \hat\alpha_k^{(w)}&=\frac{2u}{\sin2\lambda}\frac{(-1)^k}{s_0(\xi)s_{w+1}(\xi)}\Big(
   U_{w-k}\big(\tfrac{\beta}{2}\big)-\frac{\beta_1s_1(\xi)s_{w-k+1}(\xi)}{\Gamma(0)}\Big)+{\cal O}(u^2),
    \qquad k=1,\ldots,w
 \nonumber\\[.2cm]
 \hat\alpha_{w+1}^{(w)}&=\frac{2u}{\sin2\lambda}\frac{(-1)^w s_1(\xi)}{\Gamma(0)s_{w+1}(\xi)}+{\cal O}(u^2)\end{align}
\hfill $\square$

\section{Critical dense polymers}
\label{Sec:CritDensePol}

Critical dense polymers is described by the logarithmic minimal model ${\cal LM}(1,2)$.
Since $\lambda=\frac{\pi}{2}$ in this case, the bulk loop fugacity is zero,
\be
 \beta=0
\ee
thus disallowing closed loops in the bulk. In Section~\ref{Sec:H0}, we will keep the boundary loop fugacities $\beta_1$ and $\beta_2$
arbitrary, but set them equal to $1$ in Section~\ref{Sec:BdyLoops} and subsequent sections. For those special values, 
one can simply ignore all boundary loops as they merely contribute factors of $1$.

\subsection{Hamiltonian limit}
\label{Sec:H0}

For $\beta=0$, we introduce the renormalised transfer tangle
\be
 \db(u):=\frac{1}{\eta(u)}\Db(u),\qquad 
  \eta(u):=\frac{\Gamma(0)s_0(2u)s_w(u+\xi)s_{0}(\xi)\eta^{(w)}(u,\xi)}{s_0(u+\xi)s_{w}(\xi)}
\label{db0}
\ee
noting that the normalisation function reduces to $\eta(u)=\Gamma(0)s_0(2u)$ for $w=0$.
As for $\beta\neq0$ in Section~\ref{Sec:Ren}, this normalisation ensures that
\be
 \lim_{u\to0}\db(u)\simeq I,\qquad \db(\lambda-u)=\db(u)
\label{limdI}
\ee
The Hamiltonian is obtained as in (\ref{dH}), but with $h$ in (\ref{h}) replaced by
\be
 h:=\frac{\beta_2}{2\,\Gamma(0)}+\frac{(-1)^w-1}{2s_0(2\xi)}
\label{h0}
\ee
\begin{Proposition}
As an element of $TL_{N+w}(0;\beta_1,\beta_2)$ 
designed to act on ${\cal V}_d^{(N,w)}$ for any $d$, the Hamiltonian is given by
\bea
 \Hb&\!\!\!\simeq\!\!\!&-\sum_{j=1}^{N-1}e_j+\sum_{k=1}^{w}\frac{(-1)^ks_1(\xi)}{\Gamma(0)s_{w+1}(\xi)}
  \left(\beta_1\cos\frac{(w-k+1)\pi}{2}+\beta_2\sin\frac{(w-k+1)\pi}{2}\right)
  e_Ne_{N+1}\ldots e_{N+k-1}\nn
  &&\qquad\qquad -\frac{(-1)^ws_1(\xi)}{\Gamma(0)s_{w+1}(\xi)}\,e_Ne_{N+1}\ldots e_{N+w-1}f_{N+w}
\label{H0}
\eea
\end{Proposition}
\noindent{\scshape Proof:} 
Because $\beta=0$, we need to expand $\Db(u)$ to second order in $u$. Using techniques similar to the ones employed in the
proof of Proposition~\ref{Prop:Hbeta}, we first expand the bulk part of the transfer tangle
\begin{align}
\hspace{1cm}&
\psset{unit=.77cm}
\begin{pspicture}[shift=-0.89](0,0)(6.5,2)
\psarc[linewidth=1.5pt,linecolor=blue](0,1){.5}{90}{270}
\facegrid{(0,0)}{(6,2)}
\psarc[linewidth=0.025](0,0){0.16}{0}{90}
\psarc[linewidth=0.025](0,1){0.16}{0}{90}
\psarc[linewidth=0.025](1,0){0.16}{0}{90}
\psarc[linewidth=0.025](1,1){0.16}{0}{90}
\psarc[linewidth=0.025](3,0){0.16}{0}{90}
\psarc[linewidth=0.025](3,1){0.16}{0}{90}
\psarc[linewidth=0.025](4,0){0.16}{0}{90}
\psarc[linewidth=0.025](4,1){0.16}{0}{90}
\psarc[linewidth=0.025](5,0){0.16}{0}{90}
\psarc[linewidth=0.025](5,1){0.16}{0}{90}
\rput(0.5,0.5){$_{u}$}
\rput(1.5,0.5){$_{u}$}
\rput(2.5,0.5){$\ldots$}
\rput(3.5,0.5){$_{u}$}
\rput(4.5,0.5){$_{u}$}
\rput(5.5,0.5){$_{u}$}
\rput(0.5,1.5){$_{\lambda-u}$}
\rput(1.5,1.5){$_{\lambda-u}$}
\rput(2.5,1.5){$\ldots$}
\rput(3.5,1.5){$_{\lambda-u}$}
\rput(4.5,1.5){$_{\lambda-u}$}
\rput(5.5,1.5){$_{\lambda-u}$}
\psline[linewidth=1.5pt,linecolor=blue](6,0.5)(6.3,0.5)
\psline[linewidth=1.5pt,linecolor=blue](6,1.5)(6.3,1.5)
\end{pspicture}
=2u\ \
\psset{unit=.77cm}
\begin{pspicture}[shift=-0.89](0,0)(6.3,2)
\pspolygon[fillstyle=solid,fillcolor=lightlightblue](0,0)(6,0)(6,2)(0,2)(0,0)
\psline[linewidth=1.5pt,linecolor=blue](0.5,0)(0.5,2)
\psline[linewidth=1.5pt,linecolor=blue](1.5,0)(1.5,2)
\rput(2.5,0.5){$\ldots$}
\rput(2.5,1.5){$\ldots$}
\psline[linewidth=1.5pt,linecolor=blue](3.5,0)(3.5,2)
\psline[linewidth=1.5pt,linecolor=blue](4.5,0)(4.5,2)
\psarc[linewidth=1.5pt,linecolor=blue](6,0){.5}{90}{180}
\psarc[linewidth=1.5pt,linecolor=blue](6,2){.5}{180}{270}
\psline[linewidth=1.5pt,linecolor=blue](6,0.5)(6.3,0.5)
\psline[linewidth=1.5pt,linecolor=blue](6,1.5)(6.3,1.5)
\end{pspicture}
\nonumber\\[.4cm]
+&\,4u^2\Bigg(\ \
\psset{unit=.77cm}
\begin{pspicture}[shift=-0.89](0,0)(6.3,2)
\pspolygon[fillstyle=solid,fillcolor=lightlightblue](0,0)(6,0)(6,2)(0,2)(0,0)
\psarc[linewidth=1.5pt,linecolor=blue](1,0){.5}{0}{180}
\psarc[linewidth=1.5pt,linecolor=blue](1,2){.5}{180}{0}
\psline[linewidth=1.5pt,linecolor=blue](2.5,0)(2.5,2)
\rput(3.5,0.5){$\ldots$}
\rput(3.5,1.5){$\ldots$}
\psline[linewidth=1.5pt,linecolor=blue](4.5,0)(4.5,2)
\psarc[linewidth=1.5pt,linecolor=blue](6,0){.5}{90}{180}
\psarc[linewidth=1.5pt,linecolor=blue](6,2){.5}{180}{270}
\psline[linewidth=1.5pt,linecolor=blue](6,0.5)(6.3,0.5)
\psline[linewidth=1.5pt,linecolor=blue](6,1.5)(6.3,1.5)
\end{pspicture}
+\ldots+\ \,
\psset{unit=.77cm}
\begin{pspicture}[shift=-0.89](0,0)(6.3,2)
\pspolygon[fillstyle=solid,fillcolor=lightlightblue](0,0)(6,0)(6,2)(0,2)(0,0)
\psline[linewidth=1.5pt,linecolor=blue](0.5,0)(0.5,2)
\rput(1.5,0.5){$\ldots$}
\rput(1.5,1.5){$\ldots$}
\psline[linewidth=1.5pt,linecolor=blue](2.5,0)(2.5,2)
\psarc[linewidth=1.5pt,linecolor=blue](4,0){.5}{0}{180}
\psarc[linewidth=1.5pt,linecolor=blue](4,2){.5}{180}{0}
\psarc[linewidth=1.5pt,linecolor=blue](6,0){.5}{90}{180}
\psarc[linewidth=1.5pt,linecolor=blue](6,2){.5}{180}{270}
\psline[linewidth=1.5pt,linecolor=blue](6,0.5)(6.3,0.5)
\psline[linewidth=1.5pt,linecolor=blue](6,1.5)(6.3,1.5)
\end{pspicture}
\nonumber\\[.4cm]
&\quad+\frac{1}{2}\ \
\psset{unit=.77cm}
\begin{pspicture}[shift=-0.89](0,0)(6.3,2)
\pspolygon[fillstyle=solid,fillcolor=lightlightblue](0,0)(6,0)(6,2)(0,2)(0,0)
\psline[linewidth=1.5pt,linecolor=blue](0.5,0)(0.5,2)
\psline[linewidth=1.5pt,linecolor=blue](1.5,0)(1.5,2)
\rput(2.5,0.5){$\ldots$}
\rput(2.5,1.5){$\ldots$}
\psline[linewidth=1.5pt,linecolor=blue](3.5,0)(3.5,2)
\psarc[linewidth=1.5pt,linecolor=blue](5,0){.5}{90}{180}
\psarc[linewidth=1.5pt,linecolor=blue](5,2){.5}{180}{270}
\put(5,0){$\looopb$}
\put(5,1){$\looopb$}
\psline[linewidth=1.5pt,linecolor=blue](6,0.5)(6.3,0.5)
\psline[linewidth=1.5pt,linecolor=blue](6,1.5)(6.3,1.5)
\end{pspicture}
+\frac{1}{2}\ \
\psset{unit=.77cm}
\begin{pspicture}[shift=-0.89](0,0)(6.3,2)
\pspolygon[fillstyle=solid,fillcolor=lightlightblue](0,0)(6,0)(6,2)(0,2)(0,0)
\psline[linewidth=1.5pt,linecolor=blue](0.5,0)(0.5,2)
\psline[linewidth=1.5pt,linecolor=blue](1.5,0)(1.5,2)
\rput(2.5,0.5){$\ldots$}
\rput(2.5,1.5){$\ldots$}
\psline[linewidth=1.5pt,linecolor=blue](3.5,0)(3.5,2)
\psarc[linewidth=1.5pt,linecolor=blue](5,0){.5}{90}{180}
\psarc[linewidth=1.5pt,linecolor=blue](5,2){.5}{180}{270}
\put(5,0){$\looopa$}
\put(5,1){$\looopa$}
\psline[linewidth=1.5pt,linecolor=blue](6,0.5)(6.3,0.5)
\psline[linewidth=1.5pt,linecolor=blue](6,1.5)(6.3,1.5)
\end{pspicture}
\; \Bigg)+\mathcal O(u^3)
\label{diag1}
\end{align}
Next, we glue these diagrams together with the diagrams in the decomposition (\ref{uxidec}) to form $\Db(u)$ in (\ref{Duxi}).
This is illustrated here by gluing together the last diagram in (\ref{diag1}) with the diagram whose coefficient in (\ref{uxidec}) is
$\alpha_1^{(w)}$,
\be
\psset{unit=.77cm}
\begin{pspicture}[shift=-0.89](0,0)(12,2)
\pspolygon[fillstyle=solid,fillcolor=lightlightblue](0,0)(6,0)(6,2)(0,2)(0,0)
\psline[linewidth=1.5pt,linecolor=blue](0.5,0)(0.5,2)
\psline[linewidth=1.5pt,linecolor=blue](1.5,0)(1.5,2)
\rput(2.5,0.5){$\ldots$}
\rput(2.5,1.5){$\ldots$}
\psline[linewidth=1.5pt,linecolor=blue](3.5,0)(3.5,2)
\psarc[linewidth=1.5pt,linecolor=blue](5,0){.5}{90}{180}
\psarc[linewidth=1.5pt,linecolor=blue](5,2){.5}{180}{270}
\put(5,0){$\looopa$}
\put(5,1){$\looopa$}
\psline[linewidth=1.5pt,linecolor=blue,linestyle=dashed](6,0.5)(7,0.5)
\psline[linewidth=1.5pt,linecolor=blue,linestyle=dashed](6,1.5)(7,1.5)
\pspolygon[fillstyle=solid,fillcolor=lightlightblue](7,0)(12,0)(12,2)(7,2)(7,0)
\psarc[linewidth=1.5pt,linecolor=blue](7,0){.5}{0}{90}
\psarc[linewidth=1.5pt,linecolor=blue](7,2){.5}{-90}{0}
\psline[linewidth=1.5pt,linecolor=blue](8.5,0)(8.5,2)
\psline[linewidth=1.5pt,linecolor=blue](9.5,0)(9.5,2)
\rput(10.5,0.5){$\ldots$}
\rput(10.5,1.5){$\ldots$}
\psline[linewidth=1.5pt,linecolor=blue](11.5,0)(11.5,2)
\rput(3,-0.6){$\underbrace{\qquad \qquad \qquad\qquad\qquad\,}_N$}
\rput(9.5,-0.6){$\underbrace{\qquad \qquad \qquad\qquad\,}_w$}
\end{pspicture}
\ =e_Ne_{N-1}
\\[.6cm]
\ee
As it turns out, this composite diagram does not contribute to the Hamiltonian.
To see this and determine the ones that do contribute, 
we must combine the power series expansion of $\Db(u)$ with that of the normalisation function $\eta(u)$.
Thus, writing the expansion of the $u$-dependent coefficients $\alpha_k^{(w)}$ as
\be
 \alpha_k^{(w)}=\alpha_{k,0}^{(w)}+\alpha_{k,1}^{(w)}u+\mathcal O(u^2),\qquad \alpha_{k,0}^{(w)},\alpha_{k,1}^{(w)}\in\oR,
  \qquad k=0,\ldots,w+1
\ee
we observe that
\be
 \alpha_{k,0}^{(w)}=0,\qquad k=1,\ldots,w+1
\ee
while writing
\be
 \frac{1}{\eta(u)}=\frac{\eta_{-1}}{u}+\eta_0+\mathcal O(u),\qquad \eta_{-1},\eta_0\in\oR
\ee 
we note that 
\be
 \eta_{-1}=\frac{1}{2\alpha_{0,0}^{(w)}}
\ee
This last relation must be satisfied to ensure the limit in (\ref{limdI}) and is readily verified.
Combining these results yields 
\be
 h=-\frac{\alpha_{0,1}^{(w)}}{2\alpha_{0,0}^{(w)}}-\eta_0\,\alpha_{0,0}^{(w)},\qquad
 \Hb\simeq-\sum_{j=1}^{N-1}e_j-\sum_{k=1}^w\frac{\alpha_{k,1}^{(w)}}{2\al_{0,0}^{(w)}}\,e_N\ldots e_{N+k-1}
  -\frac{\alpha_{w+1,1}^{(w)}}{2\alpha_{0,0}^{(w)}}\,e_N\ldots e_{N+w-1}f_{N+w}
\ee
where we have identified $h$ as the coefficient to $-2uI$ in the expansion of $\db(u)$. It is noted that this expression for
$\Hb$ is independent of $\eta_0$.
With the concrete expressions for $\alpha_k^{(w)}$, $k=0,\ldots,w+1$, in (\ref{al}) and $\eta(u)$ in (\ref{db0}), 
we finally obtain (\ref{h0}) and (\ref{H0}).
\hfill $\square$
\medskip

\noindent 
For $w=0$, (\ref{H0}) becomes the equality
\be
 \Hb=-\sum_{j=1}^{N-1}e_j-\frac{1}{\Gamma(0)}f_N
\label{H00}
\ee

\subsection{Special parameter values}
\label{Sec:BdyLoops}

From here onwards, we fix the boundary loop fugacities to the values
\be
 \beta_1=\beta_2=1
\label{b1b2}
\ee
allowing us to ignore all boundary loops.
For $w>0$, we furthermore restrict our considerations to the special value
\be
 \xi=-\frac{\lambda}{2}=-\frac{\pi}{4}
\label{xi}
\ee
in which case (\ref{gamma}) implies that $\gamma=\frac{1}{2}$. Extending this to $w=0$, we thus have
\be
 \Gamma(u)=\cos u\,(\cos u-\sin u),\qquad \Gamma(0)=1
\ee
for all seam widths $w\in\oN_0$. Since
\be
 \eta^{(w)}(u,-\tfrac{\pi}{4})=\big[\tfrac{1}{2}\cos2u\big]^w
\label{minuspi4}
\ee
the renormalised transfer tangle is given by
\be
 \db(u)=\frac{2^w\big(\!\cos u-\sin u\big)}{\sin2u\,[\cos2u]^w\big(\!\cos u-(-1)^w\sin u\big)}\,\Db(u),\qquad
  w\in\oN_0
\ee
The corresponding shift in groundstate energy is
\be
 h=1-\tfrac{1}{2}(-1)^w,\qquad w\in\oN_0
\ee
\begin{Corollary}
As an element of $TL_{N+w}(0;1,1)$ 
designed to act on ${\cal V}_d^{(N,w)}$ for any $d$, the Hamiltonian is given by
\be
 \Hb\simeq-\sum_{j=1}^{N}e_j+\sum_{k=1}^{w}(-1)^{\floor{\frac{k+1}{2}}+(k-1)w}
  e_Ne_{N+1}\ldots e_{N+k}
\label{H0spec}
\ee
where $e_{N+w}\equiv f_{N+w}$. For $w=0$, this becomes the equality
\be
 \Hb=-\sum_{j=1}^{N-1}e_j-f_N
\label{H0spec0}
\ee
\end{Corollary}

The matrix representation $\rho_d^{(N,w)}(\Hb)$ (\ref{rhoH}) of $\Hb$ given in (\ref{H0spec}) can be computed by acting with $\Hb$ 
on $\Vc_d^{(N,w)}$\!. With bases of $\Vc_1^{(3,1)}$\!, $\Vc_1^{(3,2)}$ and $\Vc_0^{(4,2)}$ as indicated in (\ref{LinkStates}),
we thus find
\be
 \rho_1^{(3,1)}(\Hb)=-(0)-(0)-(0)+(1)=(1)
\ee
\be
 \rho_1^{(3,2)}(\Hb)=-\begin{pmatrix} 0&0&0\\1&0&1\\0&0&0\end{pmatrix}-\begin{pmatrix} 0&0&0\\0&0&0\\0&1&0\end{pmatrix}
  -\begin{pmatrix} 0&0&0\\0&0&1\\0&0&0\end{pmatrix}-\begin{pmatrix} 1&0&1\\0&1&0\\0&0&0\end{pmatrix}
   +\begin{pmatrix} 0&0&0\\0&0&0\\0&0&0\end{pmatrix}
  = -\begin{pmatrix} 1&0&1\\1&1&2\\0&1&0\end{pmatrix}
\ee
and
\begin{align}
 \rho_0^{(4,2)}(\Hb)
 =&-\begin{pmatrix} 0&1&1&0\\0&0&0&0\\0&0&0&0\\0&0&0&0\end{pmatrix}
  -\begin{pmatrix} 0&0&0&0\\0&0&0&0\\1&0&0&1\\0&0&0&0\end{pmatrix}
  -\begin{pmatrix} 0&0&0&0\\0&0&0&0\\0&0&0&0\\0&0&1&0\end{pmatrix}
  -\begin{pmatrix} 0&0&0&0\\0&0&0&0\\0&0&0&1\\0&0&0&0\end{pmatrix}\nn
 &-\begin{pmatrix} 1&0&0&1\\0&1&0&0\\0&0&1&0\\0&0&0&0\end{pmatrix}
  +\begin{pmatrix} 0&0&0&0\\0&0&0&1\\0&0&0&0\\0&0&0&0\end{pmatrix}
  =-\begin{pmatrix} 1&1&1&1\\0&1&0&-1\\1&0&1&2\\0&0&1&0\end{pmatrix}
\end{align}

\subsection{Inversion identity}
\label{Sec:InvIdentity}

With the parameters $\beta_1$, $\beta_2$ and $\xi$ fixed as in (\ref{b1b2}) and (\ref{xi}), the renormalised
transfer tangle $\db(u)$ satisfies the inversion identity given in Proposition~\ref{Prop:InvIdSpec} below.
This is generalised to all $\beta_1$, $\beta_2$ and $\xi$ in Appendix~\ref{App:GenInv} 
where we also present a proof of the inversion identity in the general case.
\begin{Proposition}
As elements of $TL_{N+w}(0;1,1)$, the renormalised transfer tangles with $\xi=-\frac{\pi}{4}$ satisfy the inversion identity
\be
 \db(u)\db(u+\tfrac{\pi}{2})=\frac{\cos^{4N+2}\!u-\sin^{4N+2}\!u}{\cos^2u-\sin^2u}\,I
\label{ddI}
\ee
\label{Prop:InvIdSpec}
\end{Proposition}
It is noted that the exact same inversion identity holds for the corresponding algebra elements designed to act on
${\cal V}_d^{(N,w)}$\!, as described in Section~\ref{Sec:RobinModules}.
This is thanks to the drop-down properties of Section~\ref{Sec:GenRobin}.
It is stressed that the inversion identity is independent of the width $w$ of the boundary seam.

\subsection{Exact solution for eigenvalues}

It follows from Proposition~\ref{Prop:InvIdSpec} and the properties (\ref{limdI}) that the eigenvalues $\La_n(u)$ of 
$\rho_d^{(N,w)}(\db(u))$ are of the form
\be
 \La_n(u)=\frac{1}{2^N}\prod_{j=1}^N\big(\epsilon^{(n)}_j\sin2u+\cosec t_j\big)
  =\prod_{j=1}^N\big(1+\epsilon^{(n)}_j\sin t_j\sin2u\big),
\qquad n=0,1,2,\ldots\label{eigfactor}
\ee
where 
\be
 \epsilon^{(n)}_j=\pm 1,\qquad t_j=\frac{(j-\frac{1}{2})\pi}{2N+1},\qquad  j=1,\ldots,N
\label{tj}
\ee
Selection rules specifying the signs $\epsilon^{(n)}_j$ are found empirically and discussed in the following. 

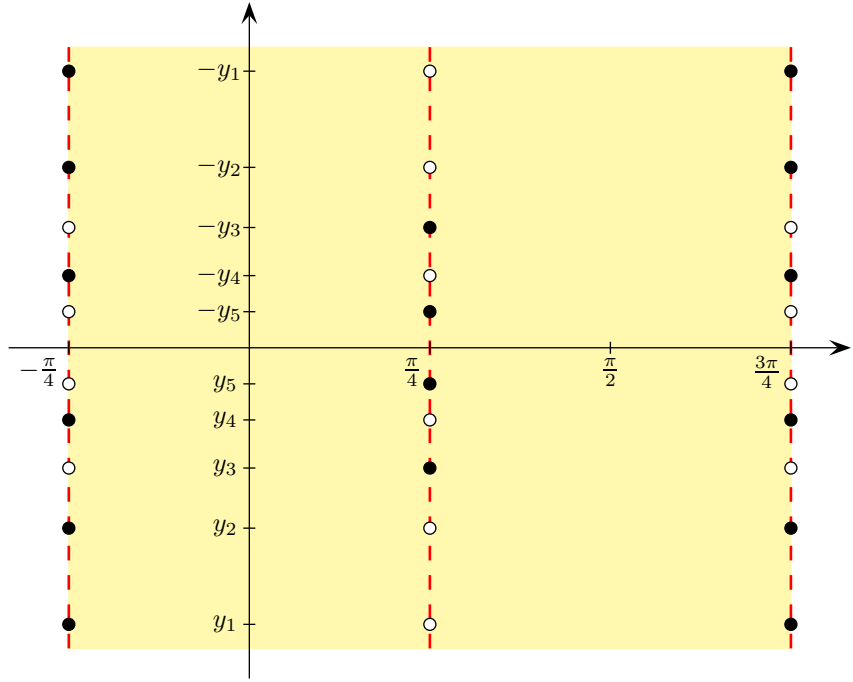
\begin{figure}[htb]
\psset{unit=.8cm}
\setlength{\unitlength}{.8cm}
\begin{center}
\begin{pspicture}(-.25,1)(14,12)
\psframe[linecolor=yellow!40!white,linewidth=0pt,fillstyle=solid,
   fillcolor=yellow!40!white](1,1)(13,11)
\psline[linecolor=black,linewidth=.5pt,arrowsize=6pt]{->}(4,0.5)(4,11.75)
\psline[linecolor=black,linewidth=.5pt,arrowsize=6pt]{->}(0,6)(14,6)
\psline[linecolor=red,linewidth=1pt,linestyle=dashed,dash=.25 .25](1,1)(1,11)
\psline[linecolor=red,linewidth=1pt,linestyle=dashed,dash=.25 .25](7,1)(7,11)
\psline[linecolor=red,linewidth=1pt,linestyle=dashed,dash=.25 .25](13,1)(13,11)
\psline[linecolor=black,linewidth=.5pt](1,5.9)(1,6.1)
\psline[linecolor=black,linewidth=.5pt](7,5.9)(7,6.1)
\psline[linecolor=black,linewidth=.5pt](10,5.9)(10,6.1)
\psline[linecolor=black,linewidth=.5pt](13,5.9)(13,6.1)
\rput(.5,5.6){\small $-\frac{\pi}{4}$}
\rput(6.7,5.6){\small $\frac{\pi}{4}$}
\rput(10,5.6){\small $\frac{\pi}{2}$}
\rput(12.6,5.6){\small $\frac{3\pi}{4}$}
\psline[linecolor=black,linewidth=.5pt](3.9,6.6)(4.1,6.6)
\psline[linecolor=black,linewidth=.5pt](3.9,7.2)(4.1,7.2)
\psline[linecolor=black,linewidth=.5pt](3.9,8.0)(4.1,8.0)
\psline[linecolor=black,linewidth=.5pt](3.9,9.0)(4.1,9.0)
\psline[linecolor=black,linewidth=.5pt](3.9,10.6)(4.1,10.6)
\psline[linecolor=black,linewidth=.5pt](3.9,5.4)(4.1,5.4)
\psline[linecolor=black,linewidth=.5pt](3.9,4.8)(4.1,4.8)
\psline[linecolor=black,linewidth=.5pt](3.9,4.0)(4.1,4.0)
\psline[linecolor=black,linewidth=.5pt](3.9,3.0)(4.1,3.0)
\psline[linecolor=black,linewidth=.5pt](3.9,1.4)(4.1,1.4)
\rput(3.5,6.6){\small $-y_5$}
\rput(3.5,7.2){\small $-y_4$}
\rput(3.5,8.0){\small $-y_3$}
\rput(3.5,9.0){\small $-y_2$}
\rput(3.5,10.6){\small $-y_1$}
\rput(3.6,5.4){\small $y_5$}
\rput(3.6,4.8){\small $y_4$}
\rput(3.6,4.0){\small $y_3$}
\rput(3.6,3.0){\small $y_2$}
\rput(3.6,1.4){\small $y_1$}
\psarc[linecolor=black,linewidth=.5pt,fillstyle=solid,fillcolor=white](1,6.6){.1}{0}{360}
\psarc[linecolor=black,linewidth=.5pt,fillstyle=solid,fillcolor=black](1,7.2){.1}{0}{360}
\psarc[linecolor=black,linewidth=.5pt,fillstyle=solid,fillcolor=white](1,8.0){.1}{0}{360}
\psarc[linecolor=black,linewidth=.5pt,fillstyle=solid,fillcolor=black](1,9.0){.1}{0}{360}
\psarc[linecolor=black,linewidth=.5pt,fillstyle=solid,fillcolor=black](1,10.6){.1}{0}{360}
\psarc[linecolor=black,linewidth=.5pt,fillstyle=solid,fillcolor=black](7,6.6){.1}{0}{360}
\psarc[linecolor=black,linewidth=.5pt,fillstyle=solid,fillcolor=white](7,7.2){.1}{0}{360}
\psarc[linecolor=black,linewidth=.5pt,fillstyle=solid,fillcolor=black](7,8.0){.1}{0}{360}
\psarc[linecolor=black,linewidth=.5pt,fillstyle=solid,fillcolor=white](7,9.0){.1}{0}{360}
\psarc[linecolor=black,linewidth=.5pt,fillstyle=solid,fillcolor=white](7,10.6){.1}{0}{360}
\psarc[linecolor=black,linewidth=.5pt,fillstyle=solid,fillcolor=white](13,6.6){.1}{0}{360}
\psarc[linecolor=black,linewidth=.5pt,fillstyle=solid,fillcolor=black](13,7.2){.1}{0}{360}
\psarc[linecolor=black,linewidth=.5pt,fillstyle=solid,fillcolor=white](13,8.0){.1}{0}{360}
\psarc[linecolor=black,linewidth=.5pt,fillstyle=solid,fillcolor=black](13,9.0){.1}{0}{360}
\psarc[linecolor=black,linewidth=.5pt,fillstyle=solid,fillcolor=black](13,10.6){.1}{0}{360}
\psarc[linecolor=black,linewidth=.5pt,fillstyle=solid,fillcolor=white](1,5.4){.1}{0}{360}
\psarc[linecolor=black,linewidth=.5pt,fillstyle=solid,fillcolor=black](1,4.8){.1}{0}{360}
\psarc[linecolor=black,linewidth=.5pt,fillstyle=solid,fillcolor=white](1,4.0){.1}{0}{360}
\psarc[linecolor=black,linewidth=.5pt,fillstyle=solid,fillcolor=black](1,3.0){.1}{0}{360}
\psarc[linecolor=black,linewidth=.5pt,fillstyle=solid,fillcolor=black](1,1.4){.1}{0}{360}
\psarc[linecolor=black,linewidth=.5pt,fillstyle=solid,fillcolor=black](7,5.4){.1}{0}{360}
\psarc[linecolor=black,linewidth=.5pt,fillstyle=solid,fillcolor=white](7,4.8){.1}{0}{360}
\psarc[linecolor=black,linewidth=.5pt,fillstyle=solid,fillcolor=black](7,4.0){.1}{0}{360}
\psarc[linecolor=black,linewidth=.5pt,fillstyle=solid,fillcolor=white](7,3.0){.1}{0}{360}
\psarc[linecolor=black,linewidth=.5pt,fillstyle=solid,fillcolor=white](7,1.4){.1}{0}{360}
\psarc[linecolor=black,linewidth=.5pt,fillstyle=solid,fillcolor=white](13,5.4){.1}{0}{360}
\psarc[linecolor=black,linewidth=.5pt,fillstyle=solid,fillcolor=black](13,4.8){.1}{0}{360}
\psarc[linecolor=black,linewidth=.5pt,fillstyle=solid,fillcolor=white](13,4.0){.1}{0}{360}
\psarc[linecolor=black,linewidth=.5pt,fillstyle=solid,fillcolor=black](13,3.0){.1}{0}{360}
\psarc[linecolor=black,linewidth=.5pt,fillstyle=solid,fillcolor=black](13,1.4){.1}{0}{360}
\end{pspicture}
\end{center}
\caption{A typical pattern of zeros in the complex $u$-plane for $N=5$. 
The pattern of zeros is symmetric under complex conjugation of $u$.
The ordinates of the locations of the zeros $u_j$ in the lower-half plane are
$y_j=\half \log \tan{\frac{(2j-1)\pi}{4N+2}}, j=1,2,\ldots,N$. At each position $j$, there is either 
a 1-string with $\Re u_j=\pi/4$ or a 2-string with $\Re u_j=-\pi/4, 3\pi/4$. Each such pattern is 
encoded by the subset ${\cal E}_n$ of $j$ indices for which $\epsilon_j=-1$. For each $j\in{\cal E}_n$, the eigenvalue 
$\La_n(u)$ has a 1-string in the fundamental strip with ordinate $y=y_j$ with excitation energy $E_j=\frac{1}{2}(j-\frac{1}{2})$. 
For the eigenvalue shown, ${\cal E}_n=\{3,5\}$ and $E({\cal E}_n)=E_3+E_5=\frac{5}{4}+\frac{9}{4}=\frac{7}{2}$.
\label{uplanezeros}}
\end{figure}

From the crossing symmetry and periodicity, the zeros $u_j$ of an eigenvalue $\La_n(u)$ occur in complex conjugate pairs 
in the complex $u$-plane and appear with a periodicity $\pi$ in the real part of $u$. 
For these reasons, we restrict our attention to the fundamental strip in the lower-half plane
\be
 -\frac{\pi}{4}<\Re u\le \frac{3\pi}{4},\qquad \Im u\le 0
\ee
From (\ref{eigfactor}), the zeros $u_j$ inside or on the boundary of the fundamental strip are located at
\be
 u_j=(2+\epsilon_j)\frac{\pi}{4}+\frac{i}{2}\ln\tan\frac{t_j}{2},\qquad \epsilon_j=\pm 1
\label{uzeros}
\ee
Thus, the ordinates of the locations of these zeros are
\be
 y_j\,=\,\half \ln \tan\half t_j, \qquad j=1,2,\ldots,N
\ee
If $\epsilon_j=-1$, there is a single zero or ``1-string" inside the fundamental strip at $u=\frac{\pi}{4}+iy_j$. If $\epsilon_j=+1$, 
there are zeros at $u=-\frac{\pi}{4}+i y_j$ and $u=\frac{3\pi}{4}+i y_j$ on the boundary of the strip, and 
we refer to them as forming a ``2-string". A typical pattern of zeros for $N=5$ is shown in Figure~\ref{uplanezeros}.

\subsection{Finite-size corrections}

The eigenvalues $\La_n(u)$ of the transfer matrix $\rho_d^{(N,w)}(\db(u))$ are of the form
\be
 \La_n(u)=\prod_{j=1}^N(1+\epsilon^{(n)}_j\sin t_j \sin2u),\qquad n=0,1,2,\ldots
\label{tm-eig}
\ee
where $t_j$ is defined in (\ref{tj}). Let ${\cal E}_n$ be the subset of $j$ indices for which $\eps_j=-1$. 
A particular eigenvalue $\La_n(u)$ is determined by the pattern ${\cal E}_n$ of values for $\epsilon^{(n)}_j$. 
In principle, there are $2^N$ different patterns giving $2^N$ possible
eigenvalues but only a subset of these occur as eigenvalues of the transfer matrix for a given Robin boundary condition. 
The set of allowed patterns ${\cal E}_n$ that actually occur thus depends on $d$ and $w$ and its specification is encoded in 
selection rules.

Conformal invariance dictates that, for large $N$, the transfer matrix eigenenergies take the form
\be
 E_n(u)=-\ln \La_n(u)= 2Nf_{bulk}(u)+f_{bdy}(u)+\frac{2\pi\sin2u}{N}\big(\!-\frac{c}{24}+\Delta+k\big)
 +\mathcal{O}\big(\frac{1}{N^2}\big),\quad n=0,1,2,\ldots
\label{N-dec}
\ee
where $f_{bulk}(u)$ is the bulk free energy per face, $f_{bdy}(u)$ the boundary free energy, $c$ is the central charge and 
$\Delta$ is a conformal dimension determined by the boundary condition.  
The non-negative integer $k$ is associated with descendants in the tower of eigenenergies and depends on the transfer matrix 
eigenvalue label $n$ with $k=0$ for $n=0$.

Let us introduce the function
\be
 F(t)=\ln(1+\sin t \sin 2u)
\label{F}
\ee
where for simplicity the $u$ dependence has been suppressed.
As a partial evaluation of the energies (\ref{N-dec}), we find
\be
 \ln\prod_{j=1}^N\big(1+\eps^{(n)}_j\sin t_j\,\sin 2u\big)
  =\tfrac{1}{2}\sum_{j=1}^{2N+1} F(t_j)-\tfrac{1}{2}F(\tfrac{\pi}{2})-\frac{2\pi\sin 2u}{N}
   \sum_{j\in{\cal E}_n}E_j+{\cal O}\big(\frac{1}{N^2}\big)
\label{sumE}
\ee
where $F(\frac{\pi}{2})=\ln(1+\sin 2u)$ and the conformal energies of elementary finite excitations are
\be
 E_j=\half(j-\half)
\label{xenergies}
\ee
This formula is valid for finite excitations for which the subset ${\cal E}_n$ remains finite as $N\rightarrow\infty$.

The finite-size corrections in (\ref{N-dec}) are determined by applying an appropriate Euler-Maclaurin formula. 
The midpoint Euler-Maclaurin formula is
\be
 \sum_{j=1}^{m}F\big(a+(j-\half)h\big)=\frac{1}{h}\int_a^b F(t)\,dt-\frac{h}{24} [F'(b)-F'(a)]+{\cal O}\big(h^2\big)
\label{EM}
\ee
where $a=0$, $b=\pi$, $m=2N+1$ and $h=(b-a)/m=\pi/(2N+1)$. This formula is valid since $F(t)$ and its first two derivatives are 
continuous on the closed interval $[a,b]=[0,\pi/2]$. 
Applying this formula, we find
\be
 E_{n}(u)=2Nf_{bulk}(u)+f_{bdy}(u)+\frac{2\pi\sin 2u}{N}
  \disp\,\big(\!-\frac{1}{96}+\sum_{j\in{\cal E}_{n}}E_j\big)+{\cal O}\big(\frac{1}{N^2}\big)
\label{logLafin}
\ee
In these expressions, the bulk free energy per face is given by
\be
  f_{bulk}(u)=-\frac{1}{\pi}\int_{0}^{\pi/2}\ln(1+\sin t\,\sin 2u)dt
\label{fb}
\ee
whereas the boundary free energy is given by
\be
 f_{bdy}(u)\,=\,f_{bulk}(u)+\tfrac{1}{2}F(\tfrac{\pi}{2})
\label{fs}
\ee
We conclude that
\be
-\frac{c}{24}+\Delta=-\frac{1}{96}+\sum_{j\in{\cal E}_{n}}E_j,\qquad \Delta=-\tfrac{3}{32}+\sum_{j\in{\cal E}_{n}}\tfrac{1}{2}(j-\tfrac{1}{2})
\ee
where
\be
 c=-2
\ee
From Section~\ref{PhysComb}, the minimum energy for a twist boundary condition ($w=0$) with a fixed number 
of defects $d$ is $\frac{1}{8}d(d+1)$. It follows that for these groundstate patterns ${\cal E}_n$, as in Figure~\ref{EsigmaZ4},
\be
 \Delta_{0,d+\frac{1}{2}}=
-\tfrac{3}{32}+\tfrac{1}{8}d(d+1)=-\tfrac{3}{32},\tfrac{5}{32},\tfrac{21}{32},\tfrac{45}{32},\tfrac{77}{32},\ldots,\quad d=0,1,2,3,4,\ldots
\ee

The link between the lattice model and the CFT characters is governed by the modular nome 
\be
 q=e^{-2\pi\tau},\qquad\tau=\delta\sin \vartheta
\ee
where $\delta=M/N$ is the aspect ratio and $\vartheta=2u$ is the 
anisotropy angle~\cite{KimP} related to the geometry of the lattice.
For a given boundary condition, the excitations form a conformal tower, indexed by $k$ in (\ref{N-dec}), 
with integer spacings above the lowest energy state given by the corresponding conformal weight $\Delta$. 
The lowest groundstate energy, with $\Delta=-\frac{3}{32}$, occurs when $\epsilon_j=+1$ for all $j$. Consider an elementary 
finite excitation where only one $\epsilon_j=-1$. 
This corresponds to a single 1-string in the fundamental strip. 
Taking the ratio of (\ref{tm-eig}) with precisely one $\epsilon_j=-1$ to (\ref{tm-eig}) with all 
$\epsilon_j=+1$, and taking the limit $M,N\to\infty$ with a fixed aspect ratio $\delta=M/N$ gives
\be
 \lim_{M,N\to\infty}\bigg(\frac{1-\sin\frac{(2j-1)\pi}{4N+2}\,\sin 2u}{{1+\sin\frac{(2j-1)\pi}{4N+2}\,\sin 2u}}\bigg)^{M}\!\!
 =\exp[-(j-\tfrac{1}{2}) \pi\,\delta\,\sin 2u]\;=\;q^{E_j}
 \ee
where the conformal energies of elementary finite excitations are given by (\ref{xenergies}).

It follows that the conformal partition functions take the form
\be
 Z(q)=q^{-\frac{c}{24}+\Delta}\sum_\text{allowed ${\cal E}_n$} q^{E({\cal E}_n)}
\ee
where the total conformal energy of an allowed pattern ${\cal E}_n$ with $n=0,1,2,\ldots$ is
\be
 E({\cal E}_n)=\sum_{j\in {\cal E}_n} E_j
\ee

\subsection{Physical combinatorics and finitized characters}
\label{PhysComb}

\def\sm#1{\mbox{\small $#1$}}
\nc{\qbar}{\bar{q}}
\def\qbin#1#2#3{{\genfrac{[}{]}{0pt}{}{#1}{#2}}_{#3}}
\def\sqbin#1#2#3{\mbox{$\genfrac{[}{]}{0pt}{}{#1}{#2}$}_{#3}}

\begin{figure}[htbp]
\psset{unit=.6cm}
\setlength{\unitlength}{.6cm}
\begin{center}
\begin{pspicture}[shift=-3](-.25,-.25)(.5,6.0)
\rput(0,.5){\scriptsize $j=1$}
\rput(0,1.5){\scriptsize $j=2$}
\rput(0,2.5){\scriptsize $j=3$}
\rput(0,3.5){$\vdots$}
\rput(-.2,4.5){\scriptsize $j=N\!-\!1$}
\rput(-.2,5.5){\scriptsize $j=N$}
\end{pspicture}
\hspace{4pt}
\begin{pspicture}[shift=-3](-.25,-.25)(2,6)
\psframe[linewidth=0pt,fillstyle=solid,fillcolor=yellow!40!white](0,0)(2,6)
\psarc[linecolor=black,linewidth=.5pt,fillstyle=solid,fillcolor=white](1,5.5){.1}{0}{360}
\psarc[linecolor=black,linewidth=.5pt,fillstyle=solid,fillcolor=white](1,3.5){.1}{0}{360}
\psarc[linecolor= red,linewidth=.5pt,fillstyle=solid,fillcolor=red](1,4.5){.1}{0}{360}
\psarc[linecolor=blue,linewidth=.5pt,fillstyle=solid,fillcolor=blue](1,1.5){.1}{0}{360}
\psarc[linecolor= red,linewidth=.5pt,fillstyle=solid,fillcolor= red](1,2.5){.1}{0}{360}
\psarc[linecolor=red,linewidth=.5pt,fillstyle=solid,fillcolor=red](1,0.5){.1}{0}{360}
\end{pspicture}
\hspace{.2cm}\ \ 
$\leftrightarrow$ \hspace{.2cm} 
$q^{\frac{1}{4}+\frac{3}{4}+\frac{5}{4}+\frac{9}{4}}
 \;=\;q^{\frac{9}{2}}$
\label{onetwo}
\end{center}
\caption{\label{singlediag}The pattern of zeros of $\La_n(u)$ 
is encoded by a single column diagram. The column corresponds 
to the 1-strings in the lower-half $u$-plane. Positions occupied by a 1-string 
are indicated by a solid red or blue circle for $j$ odd or even, respectively. Unoccupied positions are indicated by an open circle. 
The 1-string energies are given by $E_j=\half(j-\half)$. For the eigenvalue depicted, $N=6$, $\sigma=-2$, $E=\frac{9}{2}$ 
and the associated monomial is $q^E=q^{\frac{9}{2}}$.}
\end{figure}
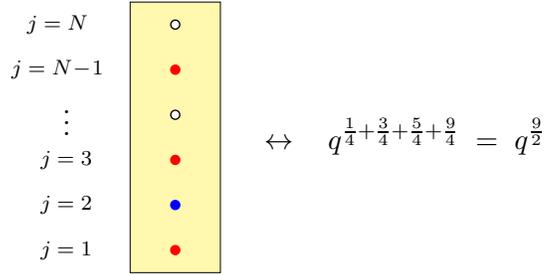

Following the description of the $\oZ_4$ sectors of critical dense polymers on the cylinder~\cite{PRV2010}, the patterns of zeros 
of $\La_n(u)$ are conveniently encoded by single column diagrams as shown in Figure~\ref{singlediag}. 
The single column corresponds to 
the 1-strings in the lower-half $u$-plane. Positions $j=1,2,\ldots,N$ occupied by a 1-string are indicated by a solid red or blue circle for 
$j$ odd or even, respectively. The unoccupied positions are indicated by an open circle. 
The number of 1-strings $m_j=0,1$ plus the number of 2-strings $n_j=0,1$ at any given position is one
\be
 m_j+n_j=1,\qquad j=1,2,\ldots,N
\ee
Each single column diagram $\{m_1,m_2,\ldots,m_N\}$ is associated with a monomial 
\be
 q^E=q^{\sum_{j=1}^N m_j E_j}
\ee

The energy $E$ of a finite excitation can be incremented by one unit either by inserting a pair 
of 1-strings at positions $j=1$ and $j=2$ or by incrementing the position $j$ of a single 1-string by 2 
units. It follows that the {\em excess}
\be
 \sigma:=m_{\text{even}}-m_{\text{odd}}
  =\sum_{k=1}^{\floor{n/2}}m_{2k}-\sum_{k=1}^{\floor{(n+1)/2}}m_{2k-1}
\label{excess}
\ee
given by the number of (blue) 1-strings at even positions $j$ minus the number of (red) 1-strings at odd positions $j$, is a {\em quantum 
number}. The single column diagrams with a given quantum number $\sigma$ are generated combinatorially by starting with the 
minimum energy configuration of 1-strings. The minimum energy configurations, for given $\sigma$, are shown in Figure~\ref{EsigmaZ4}. 
The energy of such a minimum energy configuration is 
\be
 E_\text{min}=\tfrac{1}{2}\sigma(\sigma+\tfrac{1}{2})
\label{Emin}
\ee

\begin{figure}[htbp]
\setlength{\unitlength}{.8pt}
\psset{unit=.8pt}
\begin{center}
\begin{pspicture}(-50,-15)(280,100)
\thicklines
\multirput(0,0)(40,0){8}{\psframe[linewidth=0pt,fillstyle=solid,
  fillcolor=yellow!40!white](-12,0)(12,90)}
\multiput(0,0)(40,0){8}{\line(0,10){90}}
\multirput(0,0)(40,0){8}{\psline[linewidth=1pt](-14,90)(14,90)}
\multiput(0,0)(40,0){8}{\multiput(0,0)(0,15){7}{\psarc[linecolor=black,linewidth=.5pt,fillstyle=solid,
  fillcolor=white](0,0){3}{0}{360}}}
\multiput(0,0)(0,30){3}{\psarc[linecolor=blue,linewidth=.5pt,fillstyle=solid,
  fillcolor=blue](0,15){3}{0}{360}}
\multiput(40,0)(0,30){2}{\psarc[linecolor=blue,linewidth=.5pt,fillstyle=solid,
  fillcolor=blue](0,15){3}{0}{360}}
\psarc[linecolor=blue,linewidth=.5pt,fillstyle=solid,fillcolor=blue](80,15){3}{0}{360}
\psarc[linecolor=red,linewidth=.5pt,fillstyle=solid,fillcolor=red](160,0){3}{0}{360}
\multiput(200,0)(0,30){2}{\psarc[linecolor=red,linewidth=.5pt,fillstyle=solid,
  fillcolor=red](0,0){3}{0}{360}}
\multiput(240,0)(0,30){3}{\psarc[linecolor=red,linewidth=.5pt,fillstyle=solid,
  fillcolor=red](0,0){3}{0}{360}}
\multiput(280,0)(0,30){4}{\psarc[linecolor=red,linewidth=.5pt,fillstyle=solid,
  fillcolor=red](0,0){3}{0}{360}}
\rput[B](-40,-32){$\sigma$}
\rput[B](0,-32){$3$}
\rput[B](40,-32){$2$}
\rput[B](80,-32){$1$}
\rput[B](120,-32){$0$}
\rput[B](160,-32){$-1$}
\rput[B](200,-32){$-2$}
\rput[B](240,-32){$-3$}
\rput[B](280,-32){$-4$}
\rput[B](-40,-3){\scriptsize $j=1$}
\rput[B](-40,12){\scriptsize $j=2$}
\rput[B](-40,27){\scriptsize $j=3$}
\rput[B](-40,42){\scriptsize $j=4$}
\rput[B](-40,57){\scriptsize $j=5$}
\rput[B](-40,72){\scriptsize $j=6$}
\rput[B](-40,87){\scriptsize $j=7$}
\end{pspicture}
\end{center}
\smallskip
\caption{Minimal energy configurations of the
single column diagrams. The minimal energy is $E_\text{min}=\frac{1}{2}\sigma(\sigma+\frac{1}{2})$.
The quantum number $\sigma$ is given by the excess of blue (even~$j$) over 
red (odd~$j$) 1-strings.  At each empty position $j$, there is a 2-string. 
\label{EsigmaZ4}}
\end{figure}
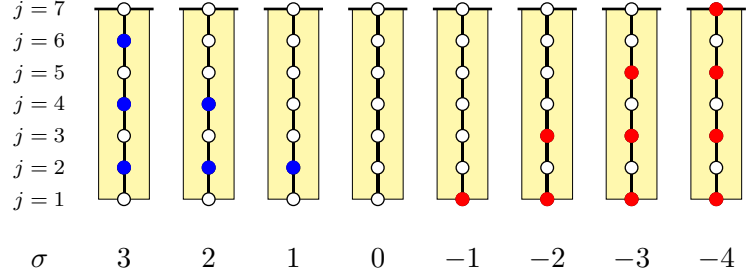

\begin{figure}[htb]
\setlength{\unitlength}{.8pt}
\psset{unit=.8pt}
\begin{center}
\begin{pspicture}(-50,-20)(400,340)
\thicklines
\multirput(0,0)(40,0){11}{\psframe[linewidth=0pt,fillstyle=solid,
  fillcolor=yellow!40!white](-12,0)(12,90)}
\multirput(0,0)(40,0){11}{\psline[linewidth=1pt](-14,90)(14,90)}
\multirput(80,120)(40,0){7}{\psframe[linewidth=0pt,fillstyle=solid,
  fillcolor=yellow!40!white](-12,0)(12,90)}
\multirput(80,120)(40,0){7}{\psline[linewidth=1pt](-14,90)(14,90)}
\multirput(160,240)(40,0){3}{\psframe[linewidth=0pt,fillstyle=solid,
  fillcolor=yellow!40!white](-12,0)(12,90)}
\multirput(160,240)(40,0){3}{\psline[linewidth=1pt](-14,90)(14,90)}
\multiput(0,0)(40,0){11}{\multiput(0,0)(0,15){7}{\psarc[linecolor=black,linewidth=.5pt,fillstyle=solid,
  fillcolor=white](0,0){3}{0}{360}}}
\multiput(80,120)(40,0){7}{\multiput(0,0)(0,15){7}{\psarc[linecolor=black,
  linewidth=.5pt,fillstyle=solid,fillcolor=white](0,0){3}{0}{360}}}
\multiput(160,240)(40,0){3}{\multiput(0,0)(0,15){7}{\psarc[linecolor=black,
  linewidth=.5pt,fillstyle=solid,fillcolor=white](0,0){3}{0}{360}}}
\psarc[linecolor=red,linewidth=.5pt,fillstyle=solid,fillcolor=red](0,0){3}{0}{360}
\psarc[linecolor=red,linewidth=.5pt,fillstyle=solid,fillcolor=red](0,30){3}{0}{360}
\psarc[linecolor=red,linewidth=.5pt,fillstyle=solid,fillcolor=red](40,0){3}{0}{360}
\psarc[linecolor=red,linewidth=.5pt,fillstyle=solid,fillcolor=red](40,60){3}{0}{360}
\psarc[linecolor=red,linewidth=.5pt,fillstyle=solid,fillcolor=red](80,30){3}{0}{360}
\psarc[linecolor=red,linewidth=.5pt,fillstyle=solid,fillcolor=red](80,60){3}{0}{360}
\psarc[linecolor=red,linewidth=.5pt,fillstyle=solid,fillcolor=red](80,120){3}{0}{360}
\psarc[linecolor=red,linewidth=.5pt,fillstyle=solid,fillcolor=red](80,210){3}{0}{360}
\psarc[linecolor=red,linewidth=.5pt,fillstyle=solid,fillcolor=red](120,0){3}{0}{360}
\psarc[linecolor=blue,linewidth=.5pt,fillstyle=solid,fillcolor=blue](120,15){3}{0}{360}
\psarc[linecolor=red,linewidth=.5pt,fillstyle=solid,fillcolor=red](120,30){3}{0}{360}
\psarc[linecolor=red,linewidth=.5pt,fillstyle=solid,fillcolor=red](120,60){3}{0}{360}
\psarc[linecolor=red,linewidth=.5pt,fillstyle=solid,fillcolor=red](120,150){3}{0}{360}
\psarc[linecolor=red,linewidth=.5pt,fillstyle=solid,fillcolor=red](120,210){3}{0}{360}
\psarc[linecolor=red,linewidth=.5pt,fillstyle=solid,fillcolor=red](160,0){3}{0}{360}
\psarc[linecolor=blue,linewidth=.5pt,fillstyle=solid,fillcolor=blue](160,45){3}{0}{360}
\psarc[linecolor=red,linewidth=.5pt,fillstyle=solid,fillcolor=red](160,30){3}{0}{360}
\psarc[linecolor=red,linewidth=.5pt,fillstyle=solid,fillcolor=red](160,60){3}{0}{360}
\psarc[linecolor=red,linewidth=.5pt,fillstyle=solid,fillcolor=red](160,120){3}{0}{360}
\psarc[linecolor=blue,linewidth=.5pt,fillstyle=solid,fillcolor=blue](160,135){3}{0}{360}
\psarc[linecolor=red,linewidth=.5pt,fillstyle=solid,fillcolor=red](160,150){3}{0}{360}
\psarc[linecolor=red,linewidth=.5pt,fillstyle=solid,fillcolor=red](160,210){3}{0}{360}
\psarc[linecolor=red,linewidth=.5pt,fillstyle=solid,fillcolor=red](160,300){3}{0}{360}
\psarc[linecolor=red,linewidth=.5pt,fillstyle=solid,fillcolor=red](160,330){3}{0}{360}
\psarc[linecolor=red,linewidth=.5pt,fillstyle=solid,fillcolor=red](200,0){3}{0}{360}
\psarc[linecolor=red,linewidth=.5pt,fillstyle=solid,fillcolor=red](200,30){3}{0}{360}
\psarc[linecolor=blue,linewidth=.5pt,fillstyle=solid,fillcolor=blue](200,75){3}{0}{360}
\psarc[linecolor=red,linewidth=.5pt,fillstyle=solid,fillcolor=red](200,60){3}{0}{360}
\psarc[linecolor=red,linewidth=.5pt,fillstyle=solid,fillcolor=red](200,120){3}{0}{360}
\psarc[linecolor=blue,linewidth=.5pt,fillstyle=solid,fillcolor=blue](200,165){3}{0}{360}
\psarc[linecolor=red,linewidth=.5pt,fillstyle=solid,fillcolor=red](200,150){3}{0}{360}
\psarc[linecolor=red,linewidth=.5pt,fillstyle=solid,fillcolor=red](200,210){3}{0}{360}
\psarc[linecolor=red,linewidth=.5pt,fillstyle=solid,fillcolor=red](200,240){3}{0}{360}
\psarc[linecolor=blue,linewidth=.5pt,fillstyle=solid,fillcolor=blue](200,255){3}{0}{360}
\psarc[linecolor=red,linewidth=.5pt,fillstyle=solid,fillcolor=red](200,300){3}{0}{360}
\psarc[linecolor=red,linewidth=.5pt,fillstyle=solid,fillcolor=red](200,330){3}{0}{360}
\psarc[linecolor=red,linewidth=.5pt,fillstyle=solid,fillcolor=red](240,0){3}{0}{360}
\psarc[linecolor=red,linewidth=.5pt,fillstyle=solid,fillcolor=red](240,30){3}{0}{360}
\psarc[linecolor=blue,linewidth=.5pt,fillstyle=solid,fillcolor=blue](240,75){3}{0}{360}
\psarc[linecolor=red,linewidth=.5pt,fillstyle=solid,fillcolor=red](240,90){3}{0}{360}
\psarc[linecolor=red,linewidth=.5pt,fillstyle=solid,fillcolor=red](240,150){3}{0}{360}
\psarc[linecolor=blue,linewidth=.5pt,fillstyle=solid,fillcolor=blue](240,135){3}{0}{360}
\psarc[linecolor=red,linewidth=.5pt,fillstyle=solid,fillcolor=red](240,180){3}{0}{360}
\psarc[linecolor=red,linewidth=.5pt,fillstyle=solid,fillcolor=red](240,210){3}{0}{360}
\psarc[linecolor=red,linewidth=.5pt,fillstyle=solid,fillcolor=red](240,240){3}{0}{360}
\psarc[linecolor=blue,linewidth=.5pt,fillstyle=solid,fillcolor=blue](240,285){3}{0}{360}
\psarc[linecolor=red,linewidth=.5pt,fillstyle=solid,fillcolor=red](240,300){3}{0}{360}
\psarc[linecolor=red,linewidth=.5pt,fillstyle=solid,fillcolor=red](240,330){3}{0}{360}
\psarc[linecolor=red,linewidth=.5pt,fillstyle=solid,fillcolor=red](280,30){3}{0}{360}
\psarc[linecolor=blue,linewidth=.5pt,fillstyle=solid,fillcolor=blue](280,45){3}{0}{360}
\psarc[linecolor=red,linewidth=.5pt,fillstyle=solid,fillcolor=red](280,60){3}{0}{360}
\psarc[linecolor=red,linewidth=.5pt,fillstyle=solid,fillcolor=red](280,90){3}{0}{360}
\psarc[linecolor=red,linewidth=.5pt,fillstyle=solid,fillcolor=red](280,120){3}{0}{360}
\psarc[linecolor=red,linewidth=.5pt,fillstyle=solid,fillcolor=red](280,180){3}{0}{360}
\psarc[linecolor=blue,linewidth=.5pt,fillstyle=solid,fillcolor=blue](280,195){3}{0}{360}
\psarc[linecolor=red,linewidth=.5pt,fillstyle=solid,fillcolor=red](280,210){3}{0}{360}
\psarc[linecolor=red,linewidth=.5pt,fillstyle=solid,fillcolor=red](320,0){3}{0}{360}
\psarc[linecolor=blue,linewidth=.5pt,fillstyle=solid,fillcolor=blue](320,15){3}{0}{360}
\psarc[linecolor=red,linewidth=.5pt,fillstyle=solid,fillcolor=red](320,30){3}{0}{360}
\psarc[linecolor=blue,linewidth=.5pt,fillstyle=solid,fillcolor=blue](320,45){3}{0}{360}
\psarc[linecolor=red,linewidth=.5pt,fillstyle=solid,fillcolor=red](320,60){3}{0}{360}
\psarc[linecolor=red,linewidth=.5pt,fillstyle=solid,fillcolor=red](320,90){3}{0}{360}
\psarc[linecolor=red,linewidth=.5pt,fillstyle=solid,fillcolor=red](320,150){3}{0}{360}
\psarc[linecolor=red,linewidth=.5pt,fillstyle=solid,fillcolor=red](320,180){3}{0}{360}
\psarc[linecolor=blue,linewidth=.5pt,fillstyle=solid,fillcolor=blue](320,195){3}{0}{360}
\psarc[linecolor=red,linewidth=.5pt,fillstyle=solid,fillcolor=red](320,210){3}{0}{360}
\psarc[linecolor=red,linewidth=.5pt,fillstyle=solid,fillcolor=red](360,0){3}{0}{360}
\psarc[linecolor=blue,linewidth=.5pt,fillstyle=solid,fillcolor=blue](360,15){3}{0}{360}
\psarc[linecolor=red,linewidth=.5pt,fillstyle=solid,fillcolor=red](360,30){3}{0}{360}
\psarc[linecolor=blue,linewidth=.5pt,fillstyle=solid,fillcolor=blue](360,75){3}{0}{360}
\psarc[linecolor=red,linewidth=.5pt,fillstyle=solid,fillcolor=red](360,60){3}{0}{360}
\psarc[linecolor=red,linewidth=.5pt,fillstyle=solid,fillcolor=red](360,90){3}{0}{360}
\psarc[linecolor=red,linewidth=.5pt,fillstyle=solid,fillcolor=red](400,0){3}{0}{360}
\psarc[linecolor=blue,linewidth=.5pt,fillstyle=solid,fillcolor=blue](400,45){3}{0}{360}
\psarc[linecolor=red,linewidth=.5pt,fillstyle=solid,fillcolor=red](400,30){3}{0}{360}
\psarc[linecolor=blue,linewidth=.5pt,fillstyle=solid,fillcolor=blue](400,75){3}{0}{360}
\psarc[linecolor=red,linewidth=.5pt,fillstyle=solid,fillcolor=red](400,60){3}{0}{360}
\psarc[linecolor=red,linewidth=.5pt,fillstyle=solid,fillcolor=red](400,90){3}{0}{360}
\rput[B](-43,-30){$\qbin75q\;=$}
\multiput(15,-30)(40,0){10}{$+$}
\rput[B](0,-30){$1$}
\rput[B](40,-30){$q$}
\rput[B](80,-30){$2q^2$}
\rput[B](120,-30){$2q^3$}
\rput[B](160,-30){$3q^4$}
\rput[B](200,-30){$3q^5$}
\rput[B](240,-30){$3q^6$}
\rput[B](280,-30){$2q^7$}
\rput[B](320,-30){$2q^8$}
\rput[B](360,-30){$q^9$}
\rput[B](400,-30){$q^{10}$}
\rput[B](-45,-3){\scriptsize $j=1$}
\rput[B](-45,12){\scriptsize $j=2$}
\rput[B](-45,27){\scriptsize $j=3$}
\rput[B](-45,42){\scriptsize $j=4$}
\rput[B](-45,57){\scriptsize $j=5$}
\rput[B](-45,72){\scriptsize $j=6$}
\rput[B](-45,87){\scriptsize $j=7$}
\end{pspicture}
\end{center}
\smallskip
\caption{For $n=7$, $\sigma=-2$, $m=\lfloor n/2\rfloor-\sigma=5$, the figure shows the combinatorial enumeration by 
single column diagrams of the $q$-binomial $\sqbin nmq=\sqbin 75q
=q^{-3/2} \sum q^{\sum_j m_j E_j}$. The excess of blue (even $j$) over red (odd $j$) 1-strings 
is given by the quantum number $\sigma=-2$. The excitation energy 
of a 1-string at position $j$ is $E_j=\half(j-\half)$. The lowest energy configuration has energy 
$E_\text{min}=\tfrac{1}{4}+\tfrac{5}{4}=\tfrac{3}{2}=\frac{1}{2}\sigma(\sigma+\frac{1}{2})$. At each empty position $j$, there is a 2-string.
Excitation increments (of energy 1) are generated by either inserting two 1-strings at positions $j=1$ and $j=2$ 
or promoting a 1-string at position $j$ to position $j+2$.  Notice that 
$\sqbin nmq=\sqbin n{n-m}q$ as \mbox{$q$-polynomials}, but they have different combinatorial 
interpretations because they have different quantum numbers $\sigma$. 
\label{binomZ4}}
\end{figure}

For modest system sizes $N$, extensive numerics were carried out in Mathematica~\cite{Wolfram} to obtain the finite-size eigenvalue 
spectra of the transfer matrices $\rho_d^{(N,w)}(\db(u))$ for the various sectors with $d$ defects and a boundary seam of width $w$. 
The exact conformal eigenenergies are read off from the patterns of 1-strings and 2-strings in the complex $u$-plane by summing 
the energies $E_j=\frac{1}{2}(j-\frac{1}{2})$ of the elementary 1-string excitations. The quantum number $\sigma$ is read off from 
the excess of even (blue) 1-strings over odd (red) 1-strings.  
From this analysis it is found that, in a sector with given $N$ and $\sigma$, the conformal spectrum or finitized partition function 
is described by a single normalized $q$-binomial of the form
\be
 q^{\frac{c}{24}+\frac{3}{32}}\chit^{(N)}_\sigma(q)
  =q^{\frac{1}{2}\sigma(\sigma+\frac{1}{2})}\qbin Nmq=q^{\frac{1}{2}\sigma(\sigma+\frac{1}{2})}
   \qbin N{\floor{\tfrac{N}{2}}-\sigma}q= 
  \sum_{\genfrac{}{}{0pt}{}{\text{$\sigma$-single}}{\text{columns}}} q^{\sum_j m_j E_j},\qquad \sigma=\floor{\tfrac{N}{2}}-m
\label{qbinBlocks}
\ee
The sum is over all single column diagrams, as in Figure~\ref{binomZ4}, associated with the fixed quantum number $\sigma$. 
In particular, this analysis yields expressions, in terms of the quantum numbers $d$ and $w$, for the lowest energy eigenvalues in 
each sector. Equating these expressions for the lowest energies to (\ref{Emin}) gives the quadratic equation
\be
\tfrac{1}{8}\big[(d+\tfrac{1}{2}-2r)^2-\tfrac{1}{4}\big]=\tfrac{1}{2}\sigma(\sigma+\tfrac{1}{2})=\Delta_{r,s-\frac{1}{2}}+\tfrac{3}{32}
\label{quadraticEq}
\ee
where we have implicitly imposed {\it selection rules} according to the empirical identifications
\be
s=d+1,\qquad r=\begin{cases}
(-1)^{d+w}\lceil \frac{w}{2}\rceil,&\mbox{$N$ even}\\[2pt]
-(-1)^{d+w}\lceil \frac{w}{2}\rceil,&\mbox{$N$ odd}
\end{cases} 
\label{selection}
\ee
This is the usual identification between $s$ and $d$~\cite{PR2007}, but we have allowed $r\in\oZ$ to take negative values. 
Solving the quadratic equation (\ref{quadraticEq})  gives a unique result for the integer $\sigma$
\be
\sigma=\begin{cases}
\frac{d}{2}-r=-\frac{1}{2}(2r-s+1),&\mbox{$d$ even ($s$ odd)}\\[2pt] 
r-\frac{d+1}{2}=\frac{1}{2}(2r-s),&\mbox{$d$ odd ($s$ even)}\end{cases}\qquad\quad
d=\begin{cases}
2(r+\sigma),&\mbox{$d$ even}\\[2pt]
2(r-\sigma)-1,&\mbox{$d$ odd}\end{cases}
\label{sigmaSoln}
\ee
On a finite lattice, the ranges of the quantum numbers $d$ and $\sigma$ are given by
\be
 0\le d\le N+w,\qquad -\lfloor \tfrac{N+1}{2}\rfloor\le\sigma\le \lfloor\tfrac{N}{2}\rfloor
\label{ranges}
\ee
where each allowed value of the quantum number $\sigma$ occurs exactly once but, depending on $N$ and $w>0$, 
some sectors given by certain disallowed values of $d$ are empty. This occurs precisely when $\sigma=\sigma(d,w)$, 
given by (\ref{sigmaSoln}), is outside of the 
allowed range (\ref{ranges}). For the twist boundary condition with $r=w=0$, the quantum number $\sigma$ of the groundstate with $d$ 
defects simplifies to
\be
\sigma=(-1)^d\big\lceil\tfrac{d}{2}\big\rceil=\begin{cases}\tfrac{d}{2},&\mbox{$d$ even}\\ -\tfrac{d+1}{2},&\mbox{$d$ odd}\end{cases},
  \qquad 
  d=\begin{cases}2\sigma,&\sigma\ge 0\\ -(2\sigma+1),&\sigma<0\end{cases}
\ee

Using the $q$-binomial building blocks (\ref{qbinBlocks}), the empirical selection rules (\ref{selection}) and summing over all of the 
sectors (labelled by $\sigma$ and weighted by $z^{-\sigma}$) 
gives the generating function for the finitized conformal partition functions
\begin{align}
Z^{(N)}(q,z)&=\sum_{\sigma=-\floor{\frac{N+1}{2}}}^{\floor{\frac{N}{2}}}\!z^{-\sigma}\, \chit^{(N)}_\sigma(q)
=q^{-\frac{c}{24}-\frac{3}{32}}\sum_{\sigma=-\lfloor \frac{N+1}{2}\rfloor}^{\lfloor \frac{N}{2}\rfloor} 
q^{\frac{1}{2}\sigma(\sigma+\frac{1}{2})} z^{-\sigma}\qbin{N}{\lfloor\frac{N}{2}\rfloor-\sigma}q
\nonumber\\
&=q^{-\frac{c}{24}-\frac{3}{32}}\prod_{k=1}^\floor{\frac{N}{2}} \big(1+q^{k-\frac{1}{4}} z^{-1}\big)
\prod_{k=1}^\floor{\frac{N+1}{2}}\big(1+q^{k-\frac{3}{4}}z\big)
\label{ZNqz}
\end{align}
This partition function is independent of $d$ and $w$ and, as discussed in Section~\ref{Sec:CFT}, coincides with the corresponding
character of $\oZ_4$ fermions. Varying $w$ acts to shuffle the building blocks $\chit^{(N)}_\sigma(q)$ among the contributions from
different defect numbers. This reshuffling is made manifest by rewriting the partition function as
\be
 Z^{(N)}(q,z)=q^{-\frac{c}{24}-\frac{3}{32}}
  \sum_{d=0}^{N+w} q^{\frac{1}{8}[(d+\frac{1}{2}-2(-1)^{N-d-w}\lceil\frac{w}{2}\rceil)^2-\frac{1}{4}]} 
   z^{(-1)^{N+w}\lceil\frac{w}{2}\rceil-(-1)^d\lceil\frac{d}{2}\rceil} \qbin{N}{\lfloor\frac{N-d}{2}\rfloor+(-1)^{N-d-w}\lceil\frac{w}{2}\rceil}q
\ee
As already indicated, despite the appearance of $w$, the partition function is independent of $w$.
We also note that
\be
 Z^{(N)}(1,1)=2^N
\label{ZN11}
\ee
which reflects the counting of Robin link states in (\ref{dimVNw}).
This is in accord with the empirical observation that each allowed eigenvalue appears exactly once in these partition functions.

\section{Conformal field theory}
\label{Sec:CFT}

\subsection{Conformal data}
\label{Sec:ConfData}

In the discussion of the lattice model above, $w$ denotes the width of the boundary seam and $d$ the number of defects.
For $\xi=-\frac{\lambda}{2}=-\frac{\pi}{4}$ as in (\ref{xi}), it was found that, in the continuum scaling limit, the Robin representations
give rise to Virasoro Verma characters of conformal weights given by
\be
 \Delta=\Delta_{r,s-\frac{1}{2}},\qquad r=(-1)^{N-d-w}\big\lceil\tfrac{w}{2}\big\rceil,\qquad s=d+1
\ee
where the Kac formula (\ref{confWts}) for critical dense polymers ${\cal LM}(1,2)$ with central charge $c=-2$ is given by
\be
 \Delta_{r,s}=\Delta_{r,s}^{1,2}=\frac{(2r-s)^2-1}{8}
\ee
The corresponding Kac table for $\D_{r,s-\frac{1}{2}}$ with $r,s\in\oZ$ is given in Figure~\ref{Kac} where we have included 
$s\in-\oN_0$ to facilitate the description of the fusion rules in Section~\ref{Sec:Fus}.
Here we note that each conformal weight in the Kac table in Figure~\ref{Kac} appears exactly once in either of the two sets
\be
 \big\{\Delta_{0,s-\frac{1}{2}};\; s\in\oN\big\},\qquad \big\{\Delta_{r,\frac{1}{2}};\; r\in\oZ\big\}
\label{columnrow}
\ee
corresponding respectively to the shaded central half-column or central row.
In the following, we will primarily work with the weights labeled as in the central row where
\be
 \Delta_{r,\frac{1}{2}}=-\frac{3}{32}+\frac{r(2r-1)}{4},\qquad r\in\oZ
\ee
For Robin boundary conditions where $r\in\oZ$ and $s\in\oN$, we thus have
\be
 \Delta_{r,s-\frac{1}{2}}=\Delta_{-\sigma,\frac{1}{2}}=\begin{cases} \Delta_{r-\frac{s-1}{2},\frac{1}{2}},\qquad &s\;\mathrm{odd}\\[.2cm]
  \Delta_{-r+\frac{s}{2},\frac{1}{2}},\qquad &s\;\mathrm{even}  \end{cases}
\label{Drow}
\ee
\begin{figure}
\begin{center}
\begin{pspicture}(-6,-6.5)(7,8)
\psframe[linewidth=0pt,fillstyle=solid,fillcolor=lightlightblue](-6,-7)(7,8)
\psframe[linewidth=0pt,fillstyle=solid,fillcolor=midblue](0,0)(1,8)
\psframe[linewidth=0pt,fillstyle=solid,fillcolor=midblue](-6,0)(7,1)
\psgrid[gridlabels=0pt,subgriddiv=1]
\multiput(-5.75,-5.6)(0,1){13}{$\cdots$}
\multiput(-4.55,7.35)(1,0){11}{$\vdots$}
\multiput(-4.55,-6.65)(1,0){11}{$\vdots$}
\multiput(6.25,-5.6)(0,1){13}{$\cdots$}
\rput(-5.65,7.65){.}\rput(-5.5,7.5){.}\rput(-5.35,7.35){.}
\rput(6.35,7.35){.}\rput(6.5,7.5){.}\rput(6.65,7.65){.}
\rput(6.35,-6.35){.}\rput(6.5,-6.5){.}\rput(6.65,-6.65){.}
\rput(-5.35,-6.35){.}\rput(-5.5,-6.5){.}\rput(-5.65,-6.65){.}
\rput(-4.5,6.5){$\frac {1085}{32}$}\rput(-3.5,6.5){$\frac {837}{32}$}\rput(-2.5,6.5){$\frac {621}{32}$}\rput(-1.5,6.5){$\frac {437}{32}$}
\rput(-0.5,6.5){$\frac {285}{32}$}
\rput(.5,6.5){$\frac {165}{32}$}\rput(1.5,6.5){$\frac {77}{32}$}\rput(2.5,6.5){$\frac{21}{32}$}\rput(3.5,6.5){$-\frac{3}{32}$}
\rput(4.5,6.5){$\frac{5}{32}$}\rput(5.5,6.5){$\frac{45}{32}$}
\rput(-4.5,5.5){$\frac {957}{32}$}\rput(-3.5,5.5){$\frac {725}{32}$}\rput(-2.5,5.5){$\frac {525}{32}$}\rput(-1.5,5.5){$\frac {357}{32}$}
\rput(-0.5,5.5){$\frac {221}{32}$}
\rput(.5,5.5){$\frac {117}{32}$}\rput(1.5,5.5){$\frac {45}{32}$}\rput(2.5,5.5){$\frac{5}{32}$}\rput(3.5,5.5){$-\frac{3}{32}$}
\rput(4.5,5.5){$\frac{21}{32}$}\rput(5.5,5.5){$\frac{77}{32}$}
\rput(-4.5,4.5){$\frac {837}{32}$}\rput(-3.5,4.5){$\frac {621}{32}$}\rput(-2.5,4.5){$\frac {437}{32}$}\rput(-1.5,4.5){$\frac {285}{32}$}
\rput(-0.5,4.5){$\frac {165}{32}$}
\rput(.5,4.5){$\frac {77}{32}$}\rput(1.5,4.5){$\frac {21}{32}$}\rput(2.5,4.5){$-\frac{3}{32}$}\rput(3.5,4.5){$\frac{5}{32}$}
\rput(4.5,4.5){$\frac{45}{32}$}\rput(5.5,4.5){$\frac{117}{32}$}
\rput(-4.5,3.5){$\frac {725}{32}$}\rput(-3.5,3.5){$\frac {525}{32}$}\rput(-2.5,3.5){$\frac {357}{32}$}\rput(-1.5,3.5){$\frac {221}{32}$}
\rput(-0.5,3.5){$\frac {117}{32}$}
\rput(.5,3.5){$\frac {45}{32}$}\rput(1.5,3.5){$\frac 5{32}$}\rput(2.5,3.5){$-\frac{3}{32}$}\rput(3.5,3.5){$\frac{21}{32}$}
\rput(4.5,3.5){$\frac{77}{32}$}\rput(5.5,3.5){$\frac{165}{32}$}
\rput(-4.5,2.5){$\frac {621}{32}$}\rput(-3.5,2.5){$\frac {437}{32}$}\rput(-2.5,2.5){$\frac {285}{32}$}\rput(-1.5,2.5){$\frac {165}{32}$}
\rput(-0.5,2.5){$\frac {77}{32}$}
\rput(.5,2.5){$\frac {21}{32}$}\rput(1.5,2.5){$-\frac 3{32}$}\rput(2.5,2.5){$\frac{5}{32}$}\rput(3.5,2.5){$\frac{45}{32}$}
\rput(4.5,2.5){$\frac{117}{32}$}\rput(5.5,2.5){$\frac{221}{32}$}
\rput(-4.5,1.5){$\frac {525}{32}$}\rput(-3.5,1.5){$\frac {357}{32}$}\rput(-2.5,1.5){$\frac {221}{32}$}\rput(-1.5,1.5){$\frac {117}{32}$}
\rput(-0.5,1.5){$\frac {45}{32}$}
\rput(.5,1.5){$\frac 5{32}$}\rput(1.5,1.5){$-\frac 3{32}$}\rput(2.5,1.5){$\frac{21}{32}$}\rput(3.5,1.5){$\frac{77}{32}$}
\rput(4.5,1.5){$\frac{165}{32}$}\rput(5.5,1.5){$\frac{285}{32}$}
\rput(-4.5,0.5){$\frac {437}{32}$}\rput(-3.5,0.5){$\frac {285}{32}$}\rput(-2.5,0.5){$\frac {165}{32}$}\rput(-1.5,0.5){$\frac {77}{32}$}
\rput(-0.5,0.5){$\frac {21}{32}$}
\rput(.5,.5){$-\frac 3{32}$}\rput(1.5,.5){$\frac 5{32}$}\rput(2.5,.5){$\frac{45}{32}$}\rput(3.5,.5){$\frac{117}{32}$}
\rput(4.5,.5){$\frac{221}{32}$}\rput(5.5,.5){$\frac{357}{32}$}
\rput(-4.5,-0.5){$\frac {357}{32}$}\rput(-3.5,-0.5){$\frac {221}{32}$}\rput(-2.5,-0.5){$\frac {117}{32}$}
\rput(-1.5,-0.5){$\frac {45}{32}$}
\rput(-0.5,-0.5){$\frac 5{32}$}\rput(0.5,-0.5){$-\frac 3{32}$}\rput(1.5,-0.5){$\frac{21}{32}$}\rput(2.5,-0.5){$\frac{77}{32}$}
\rput(3.5,-0.5){$\frac{165}{32}$}\rput(4.5,-0.5){$\frac{285}{32}$}\rput(5.5,-0.5){$\frac{437}{32}$}
\rput(-4.5,-1.5){$\frac {285}{32}$}\rput(-3.5,-1.5){$\frac {165}{32}$}\rput(-2.5,-1.5){$\frac {77}{32}$}
\rput(-1.5,-1.5){$\frac {21}{32}$}
\rput(-0.5,-1.5){$-\frac 3{32}$}\rput(0.5,-1.5){$\frac 5{32}$}\rput(1.5,-1.5){$\frac{45}{32}$}\rput(2.5,-1.5){$\frac{117}{32}$}
\rput(3.5,-1.5){$\frac{221}{32}$}\rput(4.5,-1.5){$\frac{357}{32}$}\rput(5.5,-1.5){$\frac{525}{32}$}
\rput(-4.5,-2.5){$\frac {221}{32}$}\rput(-3.5,-2.5){$\frac {117}{32}$}
\rput(-2.5,-2.5){$\frac {45}{32}$}
\rput(-1.5,-2.5){$\frac 5{32}$}\rput(-0.5,-2.5){$-\frac 3{32}$}\rput(0.5,-2.5){$\frac{21}{32}$}\rput(1.5,-2.5){$\frac{77}{32}$}
\rput(2.5,-2.5){$\frac{165}{32}$}\rput(3.5,-2.5){$\frac{285}{32}$}\rput(4.5,-2.5){$\frac{437}{32}$}\rput(5.5,-2.5){$\frac{621}{32}$}
\rput(-4.5,-3.5){$\frac {165}{32}$}\rput(-3.5,-3.5){$\frac {77}{32}$}
\rput(-2.5,-3.5){$\frac {21}{32}$}
\rput(-1.5,-3.5){$-\frac 3{32}$}\rput(-0.5,-3.5){$\frac 5{32}$}\rput(0.5,-3.5){$\frac{45}{32}$}\rput(1.5,-3.5){$\frac{117}{32}$}
\rput(2.5,-3.5){$\frac{221}{32}$}\rput(3.5,-3.5){$\frac{357}{32}$}\rput(4.5,-3.5){$\frac{525}{32}$}\rput(5.5,-3.5){$\frac{725}{32}$}
\rput(-4.5,-4.5){$\frac {117}{32}$}\rput(-3.5,-4.5){$\frac {45}{32}$}
\rput(-2.5,-4.5){$\frac 5{32}$}\rput(-1.5,-4.5){$-\frac 3{32}$}\rput(-0.5,-4.5){$\frac{21}{32}$}\rput(0.5,-4.5){$\frac{77}{32}$}
\rput(1.5,-4.5){$\frac{165}{32}$}\rput(2.5,-4.5){$\frac{285}{32}$}\rput(3.5,-4.5){$\frac{437}{32}$}\rput(4.5,-4.5){$\frac{621}{32}$}
\rput(5.5,-4.5){$\frac{837}{32}$}
\rput(-4.5,-5.5){$\frac {77}{32}$}
\rput(-3.5,-5.5){$\frac {21}{32}$}
\rput(-2.5,-5.5){$-\frac 3{32}$}\rput(-1.5,-5.5){$\frac 5{32}$}\rput(-0.5,-5.5){$\frac{45}{32}$}\rput(0.5,-5.5){$\frac{117}{32}$}
\rput(1.5,-5.5){$\frac{221}{32}$}\rput(2.5,-5.5){$\frac{357}{32}$}\rput(3.5,-5.5){$\frac{525}{32}$}\rput(4.5,-5.5){$\frac{725}{32}$}
\rput(5.5,-5.5){$\frac{957}{32}$}
{\color{blue}
\rput(-4.5,-7.5){$-5$}
\rput(-3.5,-7.5){$-4$}
\rput(-2.5,-7.5){$-3$}
\rput(-1.5,-7.5){$-2$}
\rput(-0.5,-7.5){$-1$}
\rput(.5,-7.5){$0$}
\rput(1.5,-7.5){$1$}
\rput(2.5,-7.5){$2$}
\rput(3.5,-7.5){$3$}
\rput(4.5,-7.5){$4$}
\rput(5.5,-7.5){$5$}
\rput(6.5,-7.5){$r$}
\rput(-6.5,-5.5){$-5$}
\rput(-6.5,-4.5){$-4$}
\rput(-6.5,-3.5){$-3$}
\rput(-6.5,-2.5){$-2$}
\rput(-6.5,-1.5){$-1$}
\rput(-6.5,-0.5){$0$}
\rput(-6.5,.5){$1$}
\rput(-6.5,1.5){$2$}
\rput(-6.5,2.5){$3$}
\rput(-6.5,3.5){$4$}
\rput(-6.5,4.5){$5$}
\rput(-6.5,5.5){$6$}
\rput(-6.5,6.5){$7$}
\rput(-6.5,7.5){$s$}}
\end{pspicture}
\vspace{1cm}
\caption{Kac table of conformal weights $\Delta_{r,s-\frac{1}{2}}$ for critical dense polymers ${\cal LM}(1,2)$. 
The structure of the table encodes the $sl(2)$ fusion (\ref{r's'}) with the fundamental Kac modules $(2,1)$ and $(1,2)$. 
The physical boundary conditions corresponding to $w, d\ge 0$ are given by $s\ge 1$.}
\label{Kac}
\end{center}
\end{figure}

We denote by $V(\Delta)$ the (highest-weight) Virasoro Verma module of conformal weight $\Delta$,
while the corresponding irreducible highest-weight Virasoro module is denoted by $\Vc(\Delta)$.
Two irreducible highest-weight modules of the same conformal weight are isomorphic and can be identified.
A Verma module, whose conformal weight is not in the infinitely extended {\em integer} Kac table, is irreducible
and cannot appear as a proper subquotient of an indecomposable module.
A Verma module $V(\Delta_{r,s-\frac{1}{2}})$ with $\Delta_{r,s-\frac{1}{2}}$ in Figure~\ref{Kac} is of this type, implying that
\be
 V(\Delta_{r,s-\frac{1}{2}})=\Vc(\Delta_{r,s-\frac{1}{2}}),\qquad r,s\in\oZ
\label{VV}
\ee
The Verma modules $V(\Delta_{r,s-\frac{1}{2}})$ and $V(\Delta_{r',s'-\frac{1}{2}})$ at (integer) positions $(r,s)$ and $(r',s')$ are 
therefore isomorphic when the conformal weights coincide, that is, when $|4r-2s+1|=|4r'-2s'+1|$. 
Explicitly, this occurs if $2r-s=2r'-s'$ or $2r-s+2r'-s'=-1$.

\subsection{$\oZ_4$ fermions and Virasoro modules}

We consider the spin-1 chiral fermion system $\eta(z)$ and $\xi(z)$ satisfying the standard operator product expansion
\be
 \eta(z)\xi(w)=\frac{1}{z-w}
\ee
The corresponding energy-momentum tensor is given by
\be
 T(z)=-:\eta(z)\partial\xi(z):
\ee
and the modes of $T$ defined by
\be
 T(z)=\sum_{n\in\oZ}z^{-n-2}L_n
\ee
satisfy the Virasoro algebra 
\be
 [L_n,L_m]=(n-m)L_{n+m}+\frac{c}{12}n(n^2-1)\delta_{n+m,0}
\ee
with central charge~$c=-2$.

Here we are interested in the $\oZ_4$ sector of the fermions (see a detailed description in~\cite{Kausch2000}), which means that
the fields $\eta(z)$ and $\xi(z)$ satisfy twisted periodicity conditions with respect to adding $2\pi$ to the argument $\alpha$ of 
$z=\rho e^{i\alpha}$, that is
\be
 \eta(\rho e^{i(\alpha+2\pi)})=e^{-{\pi i\over2}}\eta(\rho e^{i\alpha}),\qquad
  \xi(\rho e^{i(\alpha+2\pi)})=e^{{\pi i\over2}}\xi(\rho e^{i\alpha}),\qquad 
  \rho,\alpha\in\oR
\label{percond}
\ee
The fields $\eta(z)$ and $\xi(z)$ should thus be considered as living on a Riemann surface with four sheets, not on the complex plane.
Under the periodicity conditions (\ref{percond}), $\eta(z)$ and $\xi(z)$ have the following mode decompositions
\be
 \eta(z)=\sum_{k\in\oZ+\frac{1}{4}}z^{-k-1}\eta_k,\qquad
 \xi(z)=\sum_{k\in\oZ-\frac{1}{4}}z^{-k}\xi_k
\ee
These modes satisfy the anti-commutation rules
\be
 \big\{\eta_{n+\frac{1}{4}},\eta_{m+\frac{1}{4}}\big\}=\big\{\xi_{n-\frac{1}{4}},\xi_{m-\frac{1}{4}}\big\}=0,\qquad
 \big\{\eta_{n+\frac{1}{4}},\xi_{m-\frac{1}{4}}\big\}=\delta_{n+m,0},\qquad n,m\in\oZ
\label{anti}
\ee
With normal ordering defined by
\be
 :\!\eta_{n+\frac{1}{4}}\xi_{m-\frac{1}{4}}\!:=\begin{cases} \eta_{n+\frac{1}{4}}\xi_{m-\frac{1}{4}},\qquad &n<m\\[.15cm]
   -\xi_{m-\frac{1}{4}}\eta_{n+\frac{1}{4}},\qquad &n\geq m \end{cases}
\ee
the Virasoro algebra generators can be written as
\be
 L_n=\sum_{m\in\oZ}\big(m-\tfrac{1}{4}\big)\!:\!\eta_{n-m+\frac{1}{4}}\xi_{m-\frac{1}{4}}\!:-\tfrac{3}{32}\delta_{n,0}
\label{Ln}
\ee

The space of states in the $\oZ_4$ sector is denoted by $V_4$ and is described as follows. 
First, we separate the set of fermionic generators into the two complementary
sets
\be
 F^+:=\big\{\eta_{k-\frac{3}{4}},\,\xi_{k-\frac{1}{4}};\;k\in\oZ_{>0}\big\},\qquad
 F^-:=\big\{\eta_{k-\frac{3}{4}},\,\xi_{k-\frac{1}{4}};\;k\in\oZ_{\leq0}\big\}
\ee
The groundstate $\ket{-\frac{3}{32}}$ in the $\oZ_4$ sector is now characterised by the annihilation and eigenvalue conditions
\be
 F^+\ket{-\tfrac{3}{32}}=0,\qquad L_0\ket{-\tfrac{3}{32}}=-\tfrac{3}{32}\ket{-\tfrac{3}{32}}
\ee
in accordance with (\ref{Ln}), and
the space $V_4$ is generated by the free action of $\{I\}\cup F^-$ on this groundstate
\be
 V_4=\mathrm{span}\big\{\ket{-\tfrac{3}{32}},\, F^-\ket{-\tfrac{3}{32}}\big\}
\ee
By construction, the space $V_{4}$ is graded by $L_0$ and thus decomposes into $L_0$ eigenspaces. 

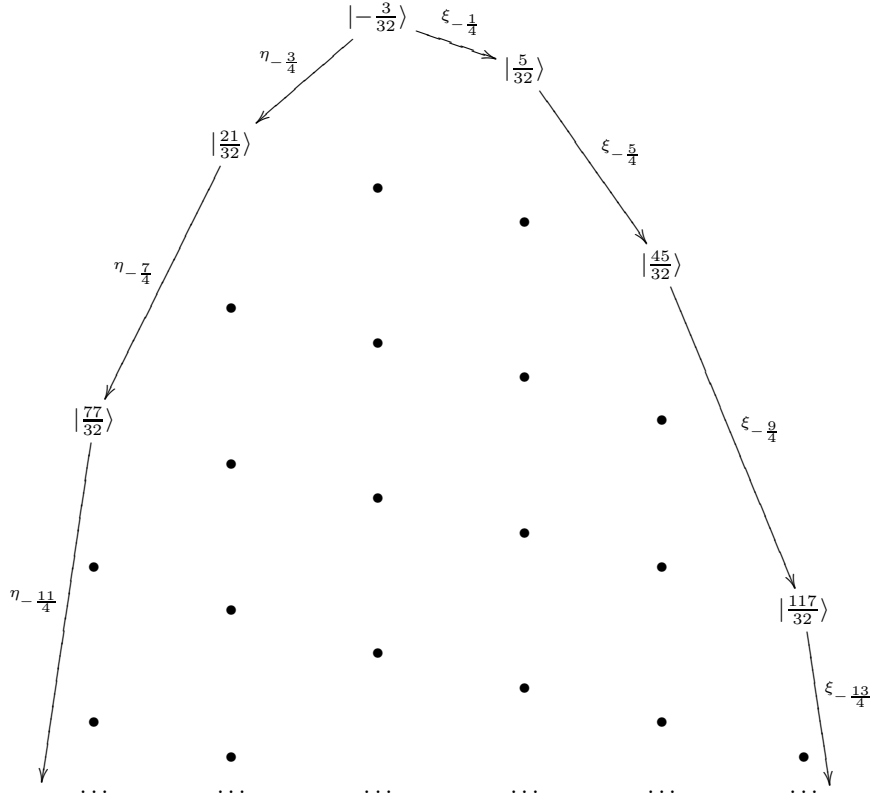
\begin{figure}[htb]
      \footnotesize \centering
    \begin{gather*}
        \mbox{}\kern-5pt \xymatrix@R=3pt@C=6pt{
         && &&&&&&\statei="x"&&&&&
          \\
          && &&&&&&&&&\stateii="y"&&
          &
          &
          &&
           \\
          &&&&&&&&&&&
          &&
          &&
          &
          &&
        \\
          &&&&&\stateiii="z"&&&&&&&&
          &&
          &
          &&
   \\
          &&&&&&&&\bullet&&&
          &&
          &&
          &
          &&
   \\
          &&&&&&&&
          &&&\bullet&
          &&&
          &
          &&
   \\
          &&&&&&&&&&&
          &&
          &\stateiv="t"
          &&
          &&
  \\
          &&&&&\bullet&&&&&&
          &&
          &&
          &
          &&
  \\
          &&&&&&&&\bullet&&&
          &&
          &&
          &
          &&
  \\
          &&&&&&&&&&&\bullet
          &&
          &&
          &
          &&
  \\
          &&\statev="u"&&&&&&&&&&&
          &\bullet
          &&
          &&
  \\
          &&&&&\bullet&&&&&&
          &&
          &&
          &
          &&
  \\
          &&&&&&&&\bullet&&&
          &&
          &&
          &
          &&
 \\
          &&&&&&&&&&&\bullet
          &&
          &&
          &
          &&
\\
          &&\bullet&&&&&&&&&
          &&
          &\bullet
          &&
          &&
\\
          &&&&&\bullet&&&&&&
          &&
          &&
          &
          &\statevi="b"&
\\
          &&&&&&&&\bullet&&&
          &&
          &&
          &
          &&
\\
          &&&&&&&&&&&\bullet
          &&
          &
          &&
          &&
\\
          &&\bullet&&&&&&&&&
          &&
          &\bullet
          &&
          &&
\\
          &&&&&\bullet&&&&&&
          &&
          &
          &&
          &\bullet&
 \\
         &\statee="a"&\stated&&&\stated&&&\stated&&&\stated
          &&&\stated
          &&
          &\stated&\statee="c"
          \ar@{->}^(.5){\xi_{-\frac{1}{4}}} "x"; "y"
           \ar@{->}^(.5){\xi_{-\frac{5}{4}}} "y"; "t"
           \ar@{->}^(.5){\xi_{-\frac{9}{4}}} "t"; "b"
          \ar@{->}^(.5){\xi_{-\frac{13}{4}}} "b"; "c"+<-12pt,1pt>
          \ar@{->}_(.5){\eta_{-\frac{3}{4}}} "x"; "z"
          \ar@{->}_(.5){\eta_{-\frac{7}{4}}} "z"; "u"
          \ar@{->}_(.5){\eta_{-\frac{11}{4}}} "u"; "a"
   }
      \end{gather*}%
      \caption[Fermions ]{\small Extremal diagram of the module $V_4$.}
      \label{fig:extremal-diagram}
    \end{figure}

Another useful grading is with respect to the zero mode of the $U(1)$ current
\be
 J(z)=-:\eta(z)\xi(z):,\qquad J(z)=\sum_{n\in\oZ}z^{-n-1}J_n
\ee
It follows from the commutation relations 
\be
 [J_0,\eta(z)]=-\eta(z),\qquad [J_0,\xi(z)]=\xi(z)
\ee 
that $\eta$ has $U(1)$ charge $-1$ and $\xi$ has charge~$1$. A $J_0$ homogeneous subspace of $V_{4}$ is thus spanned 
by the states with a
specified difference between the numbers of $\xi$ and $\eta$ modes. Using the gradings by $L_0$ and $J_0$, the space of 
states $V_{4}$ can be depicted as in Figure~\ref{fig:extremal-diagram}.
As we will discuss below, 
the space $V_4$ can be decomposed into a direct sum of irreducible Virasoro Verma modules between which fermion modes act. 
In Figure~\ref{fig:extremal-diagram}, the highest-weight states of these Virasoro modules are indicated by kets $\ket{\Delta}$
whose conformal weights are given by the $\Delta$ values.
An arrow shows the action of the highest fermion mode that does not annihilate the corresponding state. 
The diagram thus separates the grid of eigenvalues $(L_0,J_0)$ into two parts: the first one, below or on the curved 
line, occupied with at least one state at each bigrading (labeled by a $\bullet$ or a ket $\ket{\Delta}$), 
and the second one, above the curved line, with the empty space at each bigrading.

Concretely, the states on the border of the extremal diagram of $V_4$ in Figure~\ref{fig:extremal-diagram} are given 
recursively by
\be
 \ket{\Delta_{0,\frac{1}{2}}}=\ket{-\tfrac{3}{32}},\qquad
 \ket{\Delta_{-r,\frac{1}{2}}}=\eta_{\frac{1}{4}-r}\ket{\Delta_{-r+1,\frac{1}{2}}},\qquad
 \ket{\Delta_{r,\frac{1}{2}}}=\xi_{\frac{3}{4}-r}\ket{\Delta_{r-1,\frac{1}{2}}},\qquad   r\in\oN
\label{Detaxi}
\ee
yielding the ordered product expressions
\be
 \ket{\Delta_{-r,\frac{1}{2}}}=\Big(\prod_{k=-r}^{-1}\eta_{k+\frac{1}{4}}\Big)\ket{-\tfrac{3}{32}},\qquad
 \ket{\Delta_{r,\frac{1}{2}}}=\Big(\prod_{k=-r}^{-1}\xi_{k+\frac{3}{4}}\Big)\ket{-\tfrac{3}{32}},\qquad   r\in\oN
\ee
This is illustrated by
\be
 \ket{\tfrac{77}{32}}=\eta_{-\frac{7}{4}}\ket{\tfrac{21}{32}}=\eta_{-\frac{7}{4}}\eta_{-\frac{3}{4}}\ket{-\tfrac{3}{32}},\qquad
 \ket{\tfrac{45}{32}}=\xi_{-\frac{5}{4}}\ket{\tfrac{5}{32}}=\xi_{-\frac{5}{4}}\xi_{-\frac{1}{4}}\ket{-\tfrac{3}{32}}
\ee
For every $r\in\oZ$, a Virasoro module is generated from the state
$\ket{\Delta_{r,\frac{1}{2}}}$ by the free action of the non-positive Virasoro modes $L_n$, $n\leq 0$, given in (\ref{Ln}),
and since the conformal weights $\Delta_{r,\frac{1}{2}}$ do not appear in the infinitely extended {\em integer} Kac table, 
the corresponding Verma modules are irreducible (\ref{VV}).
Up to level $3$, the basis states in the Virasoro Verma module of highest weight $\Delta_{-2,\frac{1}{2}}=\frac{77}{32}$,
for example, are thus given by
\be
\begin{array}{rc}
 \mbox{level\ $0$:}\quad& \eta_{-\frac{7}{4}}\eta_{-\frac{3}{4}}\ket{-\tfrac{3}{32}} \\[.2cm]
 \mbox{level\ $1$:}\quad& \eta_{-\frac{11}{4}}\eta_{-\frac{3}{4}}\ket{-\tfrac{3}{32}}\\[.2cm]
 \mbox{level\ $2$:}\quad& \eta_{-\frac{15}{4}}\eta_{-\frac{3}{4}}\ket{-\tfrac{3}{32}},\quad
   \eta_{-\frac{11}{4}}\eta_{-\frac{7}{4}}\ket{-\tfrac{3}{32}} \\[.2cm]
 \mbox{level\ $3$:}\quad& \eta_{-\frac{19}{4}}\eta_{-\frac{3}{4}}\ket{-\tfrac{3}{32}},\quad
   \eta_{-\frac{15}{4}}\eta_{-\frac{7}{4}}\ket{-\tfrac{3}{32}},\quad
   \eta_{-\frac{11}{4}}\eta_{-\frac{7}{4}}\eta_{-\frac{3}{4}}\xi_{-\frac{1}{4}}\ket{-\tfrac{3}{32}}
\end{array}
\ee

Due to the simplicity of the anti-commutation rules (\ref{anti}), we can invert the relations (\ref{Detaxi}) and write
\be
 \ket{-\tfrac{3}{32}}=\xi_{\frac{3}{4}}\ket{\tfrac{21}{32}}=\xi_{\frac{3}{4}}\xi_{\frac{7}{4}}\ket{\tfrac{77}{32}},\qquad
 \ket{-\tfrac{3}{32}}=\eta_{\frac{1}{4}}\ket{\tfrac{5}{32}}=\eta_{\frac{1}{4}}\eta_{\frac{5}{4}}\ket{\tfrac{45}{32}}
\ee
for example. Formally, a state on the border of the extremal diagram can thus be viewed as a dense pack of $\xi$ or $\eta$ fermions 
represented by a semi-infinite product of fermion modes. As indicated, this can be done in two ways, here illustrated by
\be
 \ket{-\tfrac{3}{32}}\sim\xi_{\frac{3}{4}}\xi_{\frac{7}{4}}\xi_{\frac{11}{4}}\ldots,\quad
 \ket{\tfrac{5}{32}}\sim\xi_{-\frac{1}{4}}\xi_{\frac{3}{4}}\xi_{\frac{7}{4}}\xi_{\frac{11}{4}}\ldots,\quad
 \ket{\tfrac{21}{32}}\sim\xi_{\frac{7}{4}}\xi_{\frac{11}{4}}\xi_{\frac{15}{4}}\ldots,\quad
 \ket{\tfrac{45}{32}}\sim\xi_{-\frac{5}{4}}\xi_{-\frac{1}{4}}\xi_{\frac{3}{4}}\xi_{\frac{7}{4}}\ldots
\ee
and
\be
 \ket{-\tfrac{3}{32}}\sim\eta_{\frac{1}{4}}\eta_{\frac{5}{4}}\eta_{\frac{9}{4}}\eta_{\frac{13}{4}}\ldots,\quad
 \ket{\tfrac{5}{32}}\sim\eta_{\frac{5}{4}}\eta_{\frac{9}{4}}\eta_{\frac{13}{4}}\ldots,\quad
 \ket{\tfrac{21}{32}}\sim\eta_{-\frac{3}{4}}\eta_{\frac{1}{4}}\eta_{\frac{5}{4}}\eta_{\frac{9}{4}}\eta_{\frac{13}{4}}\ldots,\quad
 \ket{\tfrac{45}{32}}\sim\eta_{\frac{9}{4}}\eta_{\frac{13}{4}}\ldots
\ee
In general, we have the ordered products
\be
 \ket{\Delta_{r,\frac{1}{2}}}\sim\prod_{k=-r}^\infty \xi_{k+\frac{3}{4}},\qquad  
  \ket{\Delta_{r,\frac{1}{2}}}\sim\prod_{k=r}^\infty \eta_{k+\frac{1}{4}},\qquad  r\in\oZ
\label{Dirac}
\ee
In either scenario, a border state can be interpreted as a Dirac sea with a particular level of filling. With respect
to a given such Dirac sea, all states in $V_4$ are then interpreted in terms of excitations and holes.

\subsection{Characters and coinvariants}

Using the bigrading $(L_0,J_0)$ of $V_{4}$, we define the character
\be
 \chit(q,z):=\mathrm{Tr}_{V_{4}}q^{L_0-\frac{c}{24}}z^{J_0}
\label{chiqz}
\ee
This character is easily calculated and is given by
\be
 \chit(q,z)=q^{-\frac{1}{96}}\prod_{k=0}^{\infty} \Big(1+\frac{q^{k+\frac{3}{4}}}{z}\Big)\Big(1+{q^{k+\frac{1}{4}}}z\Big)
 =\sum_{r\in\oZ}z^r\chit_{r,\frac{1}{2}}(q)
\ee
where
\be
 \chit_{r,s}(q):=\mathrm{Tr}_{V(\Delta_{r,s})}q^{L_0-\frac{c}{24}}=\frac{q^{\Delta_{r,s}-\frac{c}{24}}}{\prod_{k=1}^\infty(1-q^k)}
\ee
is the character of the Virasoro Verma module $V(\Delta_{r,s})$ of conformal weight $\Delta_{r,s}$. For $s=\frac{1}{2}$, we thus have
\be
 \chit_{r,\frac{1}{2}}(q)=\frac{q^{\frac{r(2r-1)}{4}-\frac{1}{96}}}{\prod_{k=1}^\infty(1-q^k)},\qquad r\in\oZ
\ee
as the characters of the irreducible highest-weight modules generated from the states $\ket{\Delta_{r,\frac{1}{2}}}$ on the border
of the extremal diagram of $V_4$ in Figure~\ref{fig:extremal-diagram}.

Finitizations of the characters (\ref{chiqz}) can be obtained by restricting the traces to spaces of
coinvariants in $V_{4}$ with respect to certain subsets of fermionic modes. 
For each pair of nonnegative integers $P$ and $M$, we thus consider the set
\be
 C^{P,M}:=\big\{\eta_{n-\frac{3}{4}},\xi_{m-\frac{1}{4}};\,n\leq-P,\, m\leq-M\big\}
\ee
The space $C^{P,M}V_{4}\subset V_{4}$ is then defined as the linear span of all elements
of the form $x_1x_2\dots x_kv$, where $k\in\oN$, each $x_i$ is an element of $C^{P,M}$ and $v\in V_{4}$.
The space of coinvariants $V_{4}^{P,M}$ is subsequently defined as the quotient
\be
 V_{4}^{P,M}:=V_{4}\big/C^{P,M}V_{4}
\ee

To familiarize the reader with the notion of coinvariants, we digress briefly 
and calculate them explicitly for some low values of $P$ and $M$.
For $P=M=0$, we see that all states in $V_{4}$ except $\ket{-\frac{3}{32}}$
can be obtained by the action of elements from $C^{P,M}$ on vectors in $V_{4}$, so
\be
 V_{4}^{0,0}=\mathrm{span}\big\{\ket{-\tfrac{3}{32}}\big\}
\ee
It likewise follows from Figure~\ref{fig:extremal-diagram} that for $P=1$ and $M=0$, we have
\be
 V_{4}^{1,0}=\mathrm{span}\big\{\ket{-\tfrac{3}{32}},\eta_{-\frac{3}{4}}\ket{-\tfrac{3}{32}}\big\}
\ee
and that for $P=2$ and $M=1$, we have
\be
 V_{4}^{2,1}=\mathrm{span}\big\{\mathcal{A}\,\mathcal{B}\,\ket{-\tfrac{3}{32}};\,
  \mathcal{A}=I,\eta_{-\frac{3}{4}},\eta_{-\frac{7}{4}},\eta_{-\frac{7}{4}}\eta_{-\frac{3}{4}};\,\mathcal{B}=I,\xi_{-\frac{1}{4}}\big\}
\ee

As already indicated, the character of the space $V_{4}^{P,M}$ is defined as
\be
 \chit^{(P,M)}(q,z):=\mathrm{Tr}_{V_{4}^{P,M}}q^{L_0-{c\over24}}z^{J_0}
\ee
and can be viewed as a finitization of the character (\ref{chiqz}).
It is easily calculated and is given by
\be
 \chit^{(P,M)}(q,z)=q^{-\frac{1}{96}}\prod_{k=0}^{P-1} \Big(1+\frac{q^{k+\frac{3}{4}}}{z}\Big)
  \prod_{k=0}^{M-1}\Big(1+{q^{k+\frac{1}{4}}}z\Big)
 =\sum_{r=-P}^{M}\!z^r C^{(P,M)}_r(q)
\ee
where we have used
\be
 \prod_{j=0}^{n-1}(1+q^jy)=\sum_{k=0}^nq^{\frac{k(k-1)}{2}}\gauss{n}{k}_{\!q}\!y^k
\ee
and introduced
\be
 C^{(P,M)}_r(q):=q^{\frac{r(2r-1)}{4}-\frac{1}{96}}\sum_{k=\max(0,-r)}^{\min(P,M-r)}q^{k^2+r k}
  \gauss{P}{k}_{\!q}\gauss{M}{r+k}_{\!q}=q^{\frac{r(2r-1)}{4}-\frac{1}{96}}\gauss{P+M}{M-r}_{\!q}
\label{CPM}
\ee
The rewriting in (\ref{CPM}) follows from the $q$-Chu-Vandemonde identity. By construction, we have
\be
 \chit(q,z)=\lim_{P,M\to\infty}\chit^{(P,M)}(q,z)
\ee

\subsection{Interpretation of lattice observations}

Here we interpret the partition functions obtained from the lattice in terms of the characters of the $\oZ_4$ fermions.
Thus, by setting
\be
 P=\floor{\tfrac{N}{2}},\qquad M=\floor{\tfrac{N+1}{2}}
\ee
in the characterisation of coninvariants, we have
\be
 \chit^{(N)}(q,z):= \chit^{(\floor{\frac{N}{2}},\floor{\frac{N+1}{2}})}(q,z)=\sum_{r=-\floor{\frac{N}{2}}}^{\floor{\frac{N+1}{2}}}\!z^r C^{(N)}_r(q)
\ee
where
\be
 C_r^{(N)}(q):=C_r^{(\floor{\frac{N}{2}},\floor{\frac{N+1}{2}})}
  =q^{\frac{r(2r-1)}{4}-\frac{1}{96}}\gauss{N}{\floor{\frac{N+1}{2}}-r}_{\!q}
\ee
The character $\chit^{(N)}(q,z)$ is now recognised as the partition function (\ref{ZNqz}) for critical dense polymers with Robin boundary
conditions, while the finitized Virasoro Verma character
$C_r^{(N)}(q)$ is recognised as $\chit_\sigma^{(N)}(q)$ for $\sigma=-r$, that is
\be
 Z^{(N)}(q,z)=\chit^{(N)}(q,z),\qquad \chit_{-r}^{(N)}(q)=C_r^{(N)}(q)
\ee
This is in accordance with (\ref{sigmaSoln}) where $\sigma=\frac{d}{2}-r=-r$ for $d=0$ corresponding to the central row in the
Kac table in Figue~\ref{Kac}.
We refer to the (irreducible highest-weight) Virasoro Verma modules of the form ${\cal V}(\Delta_{r,s-\frac{1}{2}})$,
$r,s\in\oZ$, as {\em Robin modules}.

\subsection{Fusion rules}
\label{Sec:Fus}

As discussed in~\cite{Ras1012} and reviewed in Appendix~\ref{App:NGK},
the Kac fusion algebra of critical dense polymers ${\cal LM}(1,2)$ is finitely generated as
\be
 \big\langle(r,s);\;r,s\in\oN\big\rangle=\big\langle(1,1),\,(2,1),\,(1,2),\,(1,3)\big\rangle
\label{Kac12}
\ee
where $(1,1)$ is the identity element. This algebra contains the modules
\be
 \big\{(r,s);\;r,s\in\oN\big\}\cup\big\{\Rc_r;\;r\in\oN\big\}
\label{KR}
\ee
where the indecomposable rank-2 module $\Rc_r$ is the result of the simple fusion
\be
 (1,2)\otimes(r,2)=\Rc_r
\ee

Here we use a lattice implementation of fusion to determine the fusion of a Robin module  
with a Kac module of the form $(1,s)$. Because the Robin modules are irreducible, it follows from (\ref{columnrow}) that, 
in these evaluations, we can use the Robin boundary conditions whose labelling corresponds to $\Vc(\Delta_{0,s-\frac{1}{2}})$, that is
$d=s-1$ and $w=0$.
Subsequently, following Appendix~\ref{App:NGK}, we use the Nahm-Gaberdiel-Kausch (NGK) algorithm~\cite{Nahm,GK96} to
confirm the fusion rules inferred from the lattice and to determine the fusion of a Robin module with any module in the set (\ref{KR}).

The fusion product 
\be
 (1,s_1)\otimes\Vc(\Delta_{0,s_2-\frac{1}{2}})
\label{s1s2}
\ee
is implemented~\cite{Cardy1989,BPZ1998,PRZ,PR2007,GR1203} on the lattice by associating 
$(1,s_1)$ and $\Vc(\Delta_{0,s_2-\frac{1}{2}})$ with the left and right boundaries, respectively.
First, we characterise the boundary conditions by the corresponding defect numbers and write
\be
 (d_1):=(1,s_1),\qquad [d_2]:=\Vc(\Delta_{0,s_2-\frac{1}{2}})\qquad\mathrm{where}\qquad s_1=d_1+1,\qquad s_2=d_2+1
\ee
To fully accommodate the fusion, we shall assume that the system size is larger than the total number of defects, that is
\be
 N\geq d_1+d_2
\ee
The fusion product (\ref{s1s2}) can then be represented diagrammatically by
\be
 (d_1)\otimes[d_2]\;\sim
\begin{pspicture}[shift=-0.55](0,-0.65)(3,1)
\psline[linecolor=blue,linewidth=1.5pt](0.2,0.35)(0.2,-0.35)
\psline[linecolor=blue,linewidth=1.5pt](0.6,0.35)(0.6,-0.35)
\rput(1,0){...}
\psline[linecolor=blue,linewidth=1.5pt](1.4,0.35)(1.4,-0.35)
\rput(0.8,0.7){$\overbrace{\qquad\quad\ }^{d_1}$}
\rput(1.8,0){$\times$}
\psline[linecolor=blue,linewidth=1.5pt](2.2,0.35)(2.2,-0.35)
\psline[linecolor=blue,linewidth=1.5pt](2.6,0.35)(2.6,-0.35)
\rput(3,0){...}
\psline[linecolor=blue,linewidth=1.5pt](3.4,0.35)(3.4,-0.35)
\rput(2.8,0.7){$\overbrace{\qquad\quad\ }^{d_2}$}
\end{pspicture} 
\ee
Within each of the two batches of defects, a pair of defects are not allowed to be connected by the action from below of the transfer 
tangle. However, a defect from the left batch can connect 
to a defect from the right batch. Furthermore, according to the definition of a right Robin boundary condition,
a defect from the right batch is not allowed to link to the right boundary. 
On the other hand, as part of the fusion implementation, a defect from the left (Kac module) batch can be linked to the right boundary. 
In the end, the web of connections must be planar and thus not contain any crossings. 
For $d_1\geq d_2$, we thus have
\begin{align}
\begin{pspicture}[shift=-0.55](0,-0.65)(3.6,1)
\psline[linecolor=blue,linewidth=1.5pt](0.2,0.35)(0.2,-0.35)
\psline[linecolor=blue,linewidth=1.5pt](0.6,0.35)(0.6,-0.35)
\rput(1,0){...}
\psline[linecolor=blue,linewidth=1.5pt](1.4,0.35)(1.4,-0.35)
\rput(0.8,0.7){$\overbrace{\qquad\quad\ }^{d_1}$}
\rput(1.8,0){$\otimes$}
\psline[linecolor=blue,linewidth=1.5pt](2.2,0.35)(2.2,-0.35)
\psline[linecolor=blue,linewidth=1.5pt](2.6,0.35)(2.6,-0.35)
\rput(3,0){...}
\psline[linecolor=blue,linewidth=1.5pt](3.4,0.35)(3.4,-0.35)
\rput(2.8,0.7){$\overbrace{\qquad\quad\ }^{d_2}$}
\end{pspicture} 
&=
\begin{pspicture}[shift=-0.55](0,-0.65)(3.2,1)
\psline[linecolor=blue,linewidth=1.5pt](0.2,0.35)(0.2,-0.35)
\psline[linecolor=blue,linewidth=1.5pt](0.6,0.35)(0.6,-0.35)
\rput(1,0){...}
\psline[linecolor=blue,linewidth=1.5pt](1.4,0.35)(1.4,-0.35)
\psline[linecolor=blue,linewidth=1.5pt](1.8,0.35)(1.8,-0.35)
\psline[linecolor=blue,linewidth=1.5pt](2.2,0.35)(2.2,-0.35)
\rput(2.6,0){...}
\psline[linecolor=blue,linewidth=1.5pt](3,0.35)(3,-0.35)
\rput(1.6,0.7){$\overbrace{\qquad\qquad\qquad\quad\ \,}^{d_1+d_2}$}
\end{pspicture} 
+
\begin{pspicture}[shift=-0.55](0,-0.65)(3.2,1)
\psline[linecolor=blue,linewidth=1.5pt](0.2,0.35)(0.2,-0.35)
\rput(0.6,0){...}
\psline[linecolor=blue,linewidth=1.5pt](1,0.35)(1,-0.35)
\rput(0.6,0.7){$\overbrace{\qquad\ }^{d_1-1}$}
\psline[linecolor=blue,linewidth=1.5pt](1.4,0.35)(1.4,-0.15)
\psarc[linewidth=1.5pt,linecolor=blue](1.6,-0.15){.2}{180}{0}
\psline[linecolor=blue,linewidth=1.5pt](1.8,0.35)(1.8,-0.15)
\psline[linecolor=blue,linewidth=1.5pt](2.2,0.35)(2.2,-0.35)
\rput(2.6,0){...}
\psline[linecolor=blue,linewidth=1.5pt](3,0.35)(3,-0.35)
\rput(2.6,0.7){$\overbrace{\qquad\ }^{d_2-1}$}
\end{pspicture} 
+
\begin{pspicture}[shift=-0.55](0,-0.65)(4,1)
\psline[linecolor=blue,linewidth=1.5pt](0.2,0.35)(0.2,-0.35)
\rput(0.6,0){...}
\psline[linecolor=blue,linewidth=1.5pt](1,0.35)(1,-0.35)
\rput(0.6,0.7){$\overbrace{\qquad\ }^{d_1-2}$}
\psline[linecolor=blue,linewidth=1.5pt](1.4,0.35)(1.4,0.25)
\psline[linecolor=blue,linewidth=1.5pt](1.8,0.35)(1.8,0.25)
\psarc[linewidth=1.5pt,linecolor=blue](2,0.25){.2}{180}{0}
\psarc[linewidth=1.5pt,linecolor=blue](2,0.25){.6}{180}{0}
\psline[linecolor=blue,linewidth=1.5pt](2.2,0.35)(2.2,0.25)
\psline[linecolor=blue,linewidth=1.5pt](2.6,0.35)(2.6,0.25)
\psline[linecolor=blue,linewidth=1.5pt](3,0.35)(3,-0.35)
\rput(3.4,0){...}
\psline[linecolor=blue,linewidth=1.5pt](3.8,0.35)(3.8,-0.35)
\rput(3.4,0.7){$\overbrace{\qquad\ }^{d_2-2}$}
\end{pspicture} 
\nonumber\\[.2cm]
&+\ldots+
\begin{pspicture}[shift=-0.55](0,-0.65)(3,1)
\psline[linecolor=blue,linewidth=1.5pt](0.2,0.35)(0.2,-0.35)
\rput(0.6,0){...}
\psline[linecolor=blue,linewidth=1.5pt](1,0.35)(1,-0.35)
\rput(0.6,0.7){$\overbrace{\qquad\ }^{d_1-d_2}$}
\rput(1.65,0.33){...}
\psarc[linewidth=1.5pt,linecolor=blue](2.1,0.35){.2}{180}{0}
\psarc[linewidth=1.5pt,linecolor=blue](2.1,0.35){.7}{180}{0}
\rput(2.55,0.33){...}
\rput(2.55,0.7){$\overbrace{ }^{d_2}$}
\end{pspicture} 
\,{\red +}\,
\begin{pspicture}[shift=-0.55](0,-0.65)(3.6,1)
\psline[linecolor=blue,linewidth=1.5pt](0.2,0.35)(0.2,-0.35)
\rput(0.6,0){...}
\psline[linecolor=blue,linewidth=1.5pt](1,0.35)(1,-0.35)
\rput(0.6,0.7){$\overbrace{\qquad\ }^{d_1-d_2-1}$}
\psbezier[linewidth=1.5pt,linecolor=blue](1.4,0.35)(1.4,-0.65)(2,-0.55)(3.4,-0.55)
\rput(3.4,-0.55){$\bullet$}
\rput(2.05,0.33){...}
\psarc[linewidth=1.5pt,linecolor=blue](2.5,0.35){.2}{180}{0}
\psarc[linewidth=1.5pt,linecolor=blue](2.5,0.35){.7}{180}{0}
\rput(2.95,0.33){...}
\rput(2.95,0.7){$\overbrace{ }^{d_2}$}
\end{pspicture} 
+\ldots+
\begin{pspicture}[shift=-0.55](0,-0.65)(3.2,1)
\psbezier[linewidth=1.5pt,linecolor=blue](0.2,0.35)(0.2,-1.1)(2,-1)(3,-1)
\rput(0.6,0.33){...}
\psbezier[linewidth=1.5pt,linecolor=blue](1,0.35)(1,-0.65)(2,-0.55)(3,-0.55)
\rput(0.6,0.7){$\overbrace{\qquad\ }^{d_1-d_2}$}
\rput(1.65,0.33){...}
\psarc[linewidth=1.5pt,linecolor=blue](2.1,0.35){.2}{180}{0}
\psarc[linewidth=1.5pt,linecolor=blue](2.1,0.35){.7}{180}{0}
\rput(2.55,0.33){...}
\rput(2.55,0.7){$\overbrace{ }^{d_2}$}
\rput(3,-0.55){$\bullet$}
\rput(3,-0.68){.}
\rput(3,-0.765){.}
\rput(3,-0.85){.}
\rput(3,-1){$\bullet$}
\end{pspicture} 
\nonumber\\[.5cm]
&\sim
 \bigoplus_{d'=d_1-d_2,\,\mathrm{by}\,2}^{d_1+d_2}[d']\;{\red \oplus}\bigoplus_{d'=0}^{d_1-d_2-1}[d']
\end{align}
where the summation and direct summation in red separate, at matching places, the respective sums into two.
In the identification of the diagrams as modules, we assumed that an irreducible Robin module cannot appear as
a proper subquotient of an indecomposable module.
For $d_1\leq d_2$, we simply have
\begin{align}
\begin{pspicture}[shift=-0.55](0,-0.65)(3.6,1)
\psline[linecolor=blue,linewidth=1.5pt](0.2,0.35)(0.2,-0.35)
\psline[linecolor=blue,linewidth=1.5pt](0.6,0.35)(0.6,-0.35)
\rput(1,0){...}
\psline[linecolor=blue,linewidth=1.5pt](1.4,0.35)(1.4,-0.35)
\rput(0.8,0.7){$\overbrace{\qquad\quad\ }^{d_1}$}
\rput(1.8,0){$\otimes$}
\psline[linecolor=blue,linewidth=1.5pt](2.2,0.35)(2.2,-0.35)
\psline[linecolor=blue,linewidth=1.5pt](2.6,0.35)(2.6,-0.35)
\rput(3,0){...}
\psline[linecolor=blue,linewidth=1.5pt](3.4,0.35)(3.4,-0.35)
\rput(2.8,0.7){$\overbrace{\qquad\quad\ }^{d_2}$}
\end{pspicture} 
&=
\begin{pspicture}[shift=-0.55](0,-0.65)(3.2,1)
\psline[linecolor=blue,linewidth=1.5pt](0.2,0.35)(0.2,-0.35)
\psline[linecolor=blue,linewidth=1.5pt](0.6,0.35)(0.6,-0.35)
\rput(1,0){...}
\psline[linecolor=blue,linewidth=1.5pt](1.4,0.35)(1.4,-0.35)
\psline[linecolor=blue,linewidth=1.5pt](1.8,0.35)(1.8,-0.35)
\psline[linecolor=blue,linewidth=1.5pt](2.2,0.35)(2.2,-0.35)
\rput(2.6,0){...}
\psline[linecolor=blue,linewidth=1.5pt](3,0.35)(3,-0.35)
\rput(1.6,0.7){$\overbrace{\qquad\qquad\qquad\quad\ \,}^{d_1+d_2}$}
\end{pspicture} 
+
\begin{pspicture}[shift=-0.55](0,-0.65)(3.2,1)
\psline[linecolor=blue,linewidth=1.5pt](0.2,0.35)(0.2,-0.35)
\rput(0.6,0){...}
\psline[linecolor=blue,linewidth=1.5pt](1,0.35)(1,-0.35)
\rput(0.6,0.7){$\overbrace{\qquad\ }^{d_1-1}$}
\psline[linecolor=blue,linewidth=1.5pt](1.4,0.35)(1.4,-0.15)
\psarc[linewidth=1.5pt,linecolor=blue](1.6,-0.15){.2}{180}{0}
\psline[linecolor=blue,linewidth=1.5pt](1.8,0.35)(1.8,-0.15)
\psline[linecolor=blue,linewidth=1.5pt](2.2,0.35)(2.2,-0.35)
\rput(2.6,0){...}
\psline[linecolor=blue,linewidth=1.5pt](3,0.35)(3,-0.35)
\rput(2.6,0.7){$\overbrace{\qquad\ }^{d_2-1}$}
\end{pspicture} 
+\ldots+
\begin{pspicture}[shift=-0.55](0,-0.65)(3,1)
\rput(0.45,0.33){...}
\psarc[linewidth=1.5pt,linecolor=blue](0.9,0.35){.2}{180}{0}
\psarc[linewidth=1.5pt,linecolor=blue](0.9,0.35){.7}{180}{0}
\rput(1.35,0.33){...}
\rput(0.45,0.7){$\overbrace{ }^{d_1}$}
\psline[linecolor=blue,linewidth=1.5pt](2,0.35)(2,-0.35)
\rput(2.4,0){...}
\psline[linecolor=blue,linewidth=1.5pt](2.8,0.35)(2.8,-0.35)
\rput(2.4,0.7){$\overbrace{\qquad\ }^{d_2-d_1}$}
\end{pspicture} 
\nonumber\\[.1cm]
&\sim
 \bigoplus_{d'=d_2-d_1,\,\mathrm{by}\,2}^{d_1+d_2}[d']
\end{align}
Using
\be
 \Vc(\Delta_{0,s'-\frac{1}{2}})=\Vc(\Delta_{0,-s'+\frac{1}{2}})
\ee
we finally conclude that
\be
 (1,s_1)\otimes\Vc(\Delta_{0,s_2-\frac{1}{2}})=\bigoplus_{s'=s_1-s_2+1,\,\mathrm{by}\,2}^{s_1+s_2-1}\Vc(\Delta_{0,s'-\frac{1}{2}})
\ee

We now turn to the NGK algorithm and apply it to the fusion products
\be
 (2,1)\otimes\Vc(\Delta_{r,s-\frac{1}{2}}),\qquad (1,2)\otimes\Vc(\Delta_{r,s-\frac{1}{2}}),\qquad 
  (1,3)\otimes\Vc(\Delta_{r,s-\frac{1}{2}}),\qquad r,s\in\oZ
\label{Ac}
\ee 
To this end, it is recalled that the Kac modules $(2,1)$, $(1,2)$ and $(1,3)$ are  
constructed as the highest-weight quotient modules
\be
 (2,1)=V(1)/V(3)=\Vc(1),\qquad (1,2)=V(-\tfrac{1}{8})/V(\tfrac{15}{8})=\Vc(-\tfrac{1}{8}),\qquad (1,3)=V(0)/V(3)
\label{Q}
\ee
where $V(\Delta)$ denotes the Virasoro Verma module of highest weight $\Delta$.
The singular vectors from which the submodules in (\ref{Q}) are generated are given by
\be
 \ket{\lambda_{2,1}}=\big(L_{-1}^2-2L_{-2}\big)\ket{\Delta_{2,1}},\quad
 \ket{\lambda_{1,2}}=\big(L_{-1}^2-\tfrac{1}{2}L_{-2}\big)\ket{\Delta_{1,2}},\quad 
 \ket{\lambda_{1,3}}=\big(L_{-1}^3-2L_{-2}L_{-1}\big)\ket{\Delta_{1,3}}
\ee
According to Appendix~\ref{App:NGK}, the NGK algorithm applied to (\ref{Ac}) then yields the general fusion rules
\be
 (2,1)\otimes\Vc(\Delta_{r,\frac{1}{2}})=\Vc(\Delta_{r-1,\frac{1}{2}})\oplus\Vc(\Delta_{r+1,\frac{1}{2}})
\ee
\be
 (1,2)\otimes\Vc(\Delta_{r,\frac{1}{2}})=\Vc(\Delta_{r,-\frac{1}{2}})\oplus\Vc(\Delta_{r,\frac{3}{2}})
   =\Vc(\Delta_{-r,\frac{1}{2}})\oplus\Vc(\Delta_{-r+1,\frac{1}{2}})
\ee
and
\be
 (1,3)\otimes\Vc(\Delta_{r,\frac{1}{2}})=\Vc(\Delta_{r,-\frac{3}{2}})\oplus\Vc(\Delta_{r,\frac{1}{2}})\oplus\Vc(\Delta_{r,\frac{5}{2}})
  =\Vc(\Delta_{r+1,\frac{1}{2}})\oplus\Vc(\Delta_{r,\frac{1}{2}})\oplus\Vc(\Delta_{r-1,\frac{1}{2}})
\ee
here written in terms of the exhaustive set of conformal weights appearing in the central row of the Kac table, cf.\!\! (\ref{Drow}).
Using the Kac fusion algebra~\cite{Ras1012} of ${\cal LM}(1,2)$, 
associativity and the fact (or rather assumption, see Appendix~\ref{App:NGK})
that an irreducible Robin module does not appear as a proper subquotient of an indecomposable module, we subsequently find
\be
 (r',s')\otimes\Vc(\Delta_{r,s-\frac{1}{2}})=\bigoplus_{r''=r-r'+1,\,\mathrm{by}\,2}^{r+r'-1}\;\bigoplus_{s''=s-s'+1,\,\mathrm{by}\,2}^{s+s'-1}
  \Vc(\Delta_{r'',s''-\frac{1}{2}})
\label{r's'}
\ee
and
\be
 \Rc_{r'}\otimes\Vc(\Delta_{r,s-\frac{1}{2}})=\bigoplus_{r''=r-r'+1,\,\mathrm{by}\,2}^{r+r'-1}\Big(
  \Vc(\Delta_{r'',s-\frac{5}{2}})\oplus\Vc(\Delta_{r'',s-\frac{1}{2}})\oplus\Vc(\Delta_{r'',s-\frac{1}{2}})\oplus\Vc(\Delta_{r'',s+\frac{3}{2}})\Big)
\label{Rr'}
\ee
where $r',s'\in\oN$ and $r,s\in\oZ$. It is noted that the righthand side of (\ref{r's'}) 
is a direct sum of $r's'$ Robin modules, whereas the righthand side of (\ref{Rr'}) is a direct sum of $4r'$ such modules.

\section{Discussion}
\label{Sec:Discussion}

In this paper, we have considered Robin boundary conditions for the simplest class of $su(2)$ Yang-Baxter integrable loop models 
on the square lattice with loop fugacity $\beta=2\cos\lambda$. 
These loop models include the logarithmic minimal models ${\cal LM}(p,p')$ where the crossing parameter 
$\lambda=\tfrac{(p'-p)\pi}{p'}$ is specialized to a rational multiple of $\pi$. 
As in the case of ODEs and PDEs, Robin boundary conditions~\cite{Robin} are constructed as linear combinations of Neumann 
and Dirichlet boundary conditions. 
These boundary conditions thus allow the loop segments to either reflect or terminate on the boundary. Working in the framework of 
the one-boundary TL algebra~\cite{MaSa93,MaWood2000,NRG2005,Nichols2006a,Nichols2006b}, 
we constuct very general solutions to the boundary Yang-Baxter equation. 

Our main interest in this paper is to explore how Robin boundary conditions are incorporated into the CFT description of the logarithmic 
minimal models ${\cal LM}(p,p')$ in the continuum scaling limit. Since critical dense polymers ${\cal LM}(1,2)$~\cite{PR2007} with 
crossing parameter $\lambda=\tfrac{\pi}{2}$ is exactly solvable on arbitrary finite size lattices in all 
topologies~\cite{PRV2010,PRV1210,MDPR2013} and with any Yang-Baxter integrable boundary conditions, we focus our attention on 
this particular model. For Robin boundary conditions on one edge of the strip, as for other integrable boundary conditions on the strip, the 
double row transfer matrices are shown to satisfy a simple inversion identity which is the key to exact integrability.
When suitably specialized to give integrable lattice realizations of conformal boundary conditions, the Robin boundary conditions for 
dense polymers are naturally labelled by the quantum numbers $d$ and $w$ or, equivalently, by the Kac-type labels $r$ and 
$s-\tfrac{1}{2}$ with $r\in \oZ$, $s\in\oN$. 
Remarkably, unlike the usual Kac boundary conditions~\cite{PR2007,Ras1012,BGT1102,PRV1210}, the Robin boundary conditions are 
thus conjugate to operators or representations with half-integer Kac labels. Indeed, our detailed analytic treatment of the finite-size 
corrections using an Euler-Maclaurin formula, physical combinatorics and finite-size characters leads to the conformal weights
\be
 \Delta_{r,s-\frac{1}{2}}=\tfrac{1}{32}[(4r-2s+1)^2-4],\qquad r\in\oZ,\; s\in\oN
\ee
In fact, the existence of representations with half-integer Kac labels was posited~\cite{SaleurHalfInt87,Duplantier86} long ago in the 
context of polymers and percolation, see also~\cite{Rid0808,Delfino}. 
However, it is much less clear precisely how such representations appear in logarithmic CFTs and 
to which boundary conditions they are associated. In the case of critical dense polymers, we argue that our Robin boundary conditions 
are properly accounted for within the $\oZ_4$ sector of symplectic fermions~\cite{Kausch2000}. 
We also determine the fusion rules for the fusion of Robin modules with Kac modules. 

This paper opens several avenues for further work. It is clearly of interest to study, either numerically or analytically through more general 
functional equations~\cite{MDPR2014}, the conformal spectra of the other logarithmic minimal models ${\cal LM}(p,p')$ to confirm more 
generally that Robin boundary conditions lead to conformal weights with non-integer Kac labels. 
It would also be interesting to continue our analysis of fusion. To investigate the fusion of the Robin 
modules with themselves, in particular, requires moving to the two-boundary TL algebra~\cite{MNdeGB2004,deGN2009}.
Preliminary results indicate that such fusions lead to new types of representations.

\subsection*{Acknowledgments}

JR is supported by the Australian Research Council under the Future Fellowship scheme, project number FT100100774.
PAP is supported under the Melbourne University Research Grant Support Scheme (MRGSS).
IYT is supported by RFBR-grant 14-02-01171.
He also thanks the University of Melbourne and the University of Queensland, where parts of this work were done,
for their generous hospitality. The authors thank Alexi Morin-Duchesne for discussions and comments.
JR thanks David Ridout for comments on a draft of this paper.

\appendix

\section{Blob algebra}
\label{Sec:Blob}

For every $\nu\in\mathbb{C}^\ast$, the map 
\be
 f_N\to f'_N=\tfrac{1}{\nu}f_N,\qquad e_j\to e_j'=e_j,\qquad j=1,\ldots,N-1
\ee
generates an algebra isomorphism of the form
\be
 TL_N^{(1)}(\beta;\beta_1,\beta_2)\simeq TL_N^{(1)}(\beta;\tfrac{\beta_1}{\nu},\tfrac{\beta_2}{\nu})
\ee
In particular for $\nu=\beta_2\neq0$, we have
\be
 TL_N^{(1)}(\beta;\beta_1,\beta_2)\simeq\mathcal B_N(\beta,\beta')
\ee
where 
\be
 \mathcal B_N(\beta,\beta')=TL_N^{(1)}(\beta;\beta',1),\qquad\beta'=\tfrac{\beta_1}{\beta_2}
\ee
is the blob algebra of~\cite{MaSa93}. 
In other words, the one-boundary TL algebra with $\beta_2\neq0$ is isomorphic to the blob algebra.

The generators of the blob algebra ${\cal B}_N(\beta,\beta')$ have a simple loop representation in terms of tangles decorated 
with blobs where
\be
 I=\!
\begin{pspicture}[shift=-0.55](-0.1,-0.65)(2.0,0.45)
\rput(1.4,0.0){\small$...$}
\psline[linecolor=blue,linewidth=1.5pt](0.2,0.35)(0.2,-0.35)\rput(0.2,-0.55){$_1$}
\psline[linecolor=blue,linewidth=1.5pt](0.6,0.35)(0.6,-0.35)
\psline[linecolor=blue,linewidth=1.5pt](1.0,0.35)(1.0,-0.35)
\psline[linecolor=blue,linewidth=1.5pt](1.8,0.35)(1.8,-0.35)\rput(1.8,-0.55){$_N$}
\end{pspicture} 
 ,\qquad
 e_j=\!
 \begin{pspicture}[shift=-0.55](-0.1,-0.65)(3.2,0.45)
\rput(0.6,0.0){\small$...$}
\rput(2.6,0.0){\small$...$}
\psline[linecolor=blue,linewidth=1.5pt](0.2,0.35)(0.2,-0.35)\rput(0.2,-0.55){$_1$}
\psline[linecolor=blue,linewidth=1.5pt](1.0,0.35)(1.0,-0.35)
\psline[linecolor=blue,linewidth=1.5pt](2.2,0.35)(2.2,-0.35)
\psline[linecolor=blue,linewidth=1.5pt](3.0,0.35)(3.0,-0.35)\rput(3.0,-0.55){$_{N}$}
\psarc[linecolor=blue,linewidth=1.5pt](1.6,0.35){0.2}{180}{0}\rput(1.4,-0.55){$_j$}
\psarc[linecolor=blue,linewidth=1.5pt](1.6,-0.35){0.2}{0}{180}
\end{pspicture} 
 ,\qquad
 f_N=\!
\begin{pspicture}[shift=-0.55](-0.1,-0.65)(2.0,0.45)
\rput(1.4,0.0){\small$...$}
\psline[linecolor=blue,linewidth=1.5pt](0.2,0.35)(0.2,-0.35)\rput(0.2,-0.55){$_1$}
\psline[linecolor=blue,linewidth=1.5pt](0.6,0.35)(0.6,-0.35)
\psline[linecolor=blue,linewidth=1.5pt](1.0,0.35)(1.0,-0.35)
\psline[linecolor=blue,linewidth=1.5pt](1.8,0.35)(1.8,-0.35)
\psline[linecolor=blue,linewidth=1.5pt](2.2,0.35)(2.2,-0.35)
\rput(2.2,-0.55){$_{N}$}
\rput(2.2,0){$\bullet$}
\end{pspicture} 
\label{blob}
\ee
The corresponding relations in (\ref{1TLN}) involving $f_N$ of the blob algebra are then represented by the diagrammatic relations
\be
 e_{N-1}f_Ne_{N-1}=
\begin{pspicture}[shift=-1](-0.1,-1.1)(2.9,1.15)
\rput(1.4,-0.7){\small$...$}
\rput(1.4,0){\small$...$}
\rput(1.4,0.7){\small$...$}
\psline[linecolor=blue,linewidth=1.5pt](0.2,1.05)(0.2,-1.05)
\psline[linecolor=blue,linewidth=1.5pt](0.6,1.05)(0.6,-1.05)
\psline[linecolor=blue,linewidth=1.5pt](1.0,1.05)(1.0,-1.05)
\psline[linecolor=blue,linewidth=1.5pt](1.8,1.05)(1.8,-1.05)
\psline[linecolor=blue,linewidth=1.5pt](2.2,0.35)(2.2,-0.35)
\psline[linecolor=blue,linewidth=1.5pt](2.6,0.35)(2.6,-0.35)
\rput(2.6,0){$\bullet$}
\psarc[linecolor=blue,linewidth=1.5pt](2.4,-1.05){0.2}{0}{180}
\psarc[linecolor=blue,linewidth=1.5pt](2.4,0.35){0.2}{0}{180}
\psarc[linecolor=blue,linewidth=1.5pt](2.4,1.05){0.2}{180}{0}
\psarc[linecolor=blue,linewidth=1.5pt](2.4,-0.35){0.2}{180}{0}
\psline[linewidth=0.5pt,linestyle=dashed, dash=1pt 1pt](0,0.35)(2.8,0.35)
\psline[linewidth=0.5pt,linestyle=dashed, dash=1pt 1pt](0,-0.35)(2.8,-0.35)
\rput(0.2,-1.25){$_1$}
\rput(2.6,-1.25){$_{N}$}
\end{pspicture} 
\;=
\beta'\times\!\!
\begin{pspicture}[shift=-1](-0.1,-1.1)(2.9,0.45)
\rput(1.4,0){\small$...$}
\psline[linecolor=blue,linewidth=1.5pt](0.2,0.35)(0.2,-0.35)
\psline[linecolor=blue,linewidth=1.5pt](0.6,0.35)(0.6,-0.35)
\psline[linecolor=blue,linewidth=1.5pt](1.0,0.35)(1.0,-0.35)
\psline[linecolor=blue,linewidth=1.5pt](1.8,0.35)(1.8,-0.35)
\psarc[linecolor=blue,linewidth=1.5pt](2.4,-0.35){0.2}{0}{180}
\psarc[linecolor=blue,linewidth=1.5pt](2.4,0.35){0.2}{180}{0}
\rput(0.2,-0.55){$_1$}
\rput(2.6,-0.55){$_{N}$}
\end{pspicture} 
=\beta'e_{N-1}
\ee
and
\be
 f_N^2=
\begin{pspicture}[shift=-0.55](-0.1,-0.65)(2.5,1)
\rput(1.4,-0.35){\small$...$}
\rput(1.4,0.35){\small$...$}
\psline[linecolor=blue,linewidth=1.5pt](0.2,0.7)(0.2,-0.7)\rput(0.2,-0.9){$_1$}
\psline[linecolor=blue,linewidth=1.5pt](0.6,0.7)(0.6,-0.7)
\psline[linecolor=blue,linewidth=1.5pt](1.0,0.7)(1.0,-0.7)
\psline[linecolor=blue,linewidth=1.5pt](1.8,0.7)(1.8,-0.7)
\psline[linecolor=blue,linewidth=1.5pt](2.2,0.7)(2.2,-0.7)
\rput(2.2,-0.35){$\bullet$}
\rput(2.2,0.35){$\bullet$}
\psline[linewidth=0.5pt,linestyle=dashed, dash=1pt 1pt](0,0)(2.4,0)
\rput(2.2,-0.9){$_{N}$}
\end{pspicture} 
\;=
\begin{pspicture}[shift=-0.55](-0.1,-0.65)(2.3,0.45)
\rput(1.4,0.0){\small$...$}
\psline[linecolor=blue,linewidth=1.5pt](0.2,0.35)(0.2,-0.35)\rput(0.2,-0.55){$_1$}
\psline[linecolor=blue,linewidth=1.5pt](0.6,0.35)(0.6,-0.35)
\psline[linecolor=blue,linewidth=1.5pt](1.0,0.35)(1.0,-0.35)
\psline[linecolor=blue,linewidth=1.5pt](1.8,0.35)(1.8,-0.35)
\psline[linecolor=blue,linewidth=1.5pt](2.2,0.35)(2.2,-0.35)
\rput(2.2,0){$\bullet$}
\rput(2.2,-0.55){$_{N}$}
\end{pspicture} 
\;=f_N
\ee
In local terms, we thus have
\be
\begin{pspicture}[shift=-1](0,-1.1)(0.9,0.5)
\psarc[linecolor=blue,linewidth=1.5pt](0.4,0){0.4}{0}{360}
\rput(0.8,0){$\bullet$}
\end{pspicture} 
\;=\beta',\qquad
\begin{pspicture}[shift=-0.55](-0.1,-0.65)(0.4,0.9)
\psline[linecolor=blue,linewidth=1.5pt](0.2,0.7)(0.2,-0.7)
\rput(0.2,-0.35){$\bullet$}
\rput(0.2,0.35){$\bullet$}
\end{pspicture} 
\;=
\begin{pspicture}[shift=-0.55](-0.1,-0.65)(0.4,0.5)
\psline[linecolor=blue,linewidth=1.5pt](0.2,0.35)(0.2,-0.35)
\rput(0.2,0){$\bullet$}
\end{pspicture} 
\vspace{-.3cm}
\ee
meaning that blobbed loops have fugacity $\beta'$, as opposed to the usual (non-blobbed) loops which have fugacity $\beta$,
and that blobbing a strand is an idempotent process.

In the special case $\beta'=\beta$, we can represent the generators of the blob algebra
by diagrams without blobs.
In the case $\beta'=1$, which corresponds to $\beta_1=\beta_2$ in the one-boundary loop language,
we do not distinguish between the two types of boundary loops.
From the isomorphisms above, we see that we may set $\beta_1=\beta_2=1$ in this case.

Despite the simplicity of the loop representation (\ref{blob}) of the blob algebra, we find it convenient
to work with the loop representation (\ref{Ief}) of the one-boundary TL algebra $TL_N^{(1)}(\beta;\beta_1,\beta_2)$,
even for $\beta_2\neq0$.
The main reason is that to determine whether a blob diagram actually can be constructed as a
word in the blob algebra generators, one must check whether all blobs can be pulled or stretched to a virtual edge on the right in a 
non-crossing manner.
In the one-boundary loop picture, on the other hand, non-crossing of loops readily singles out the allowed diagrams.

\section{General inversion identity}
\label{App:GenInv}

In this appendix, we consider critical dense polymers ${\cal LM}(1,2)$ in which case $\lambda=\frac{\pi}{2}$ and $\beta=0$.
We initially keep $\beta_1$, $\beta_2$ and $\xi$ free to obtain the most general inversion identity possible
for the transfer tangles with Neumann and Robin boundary conditions on the left and right, respectively.
Focus here is on the transfer tangles $\Db(u)$. The inversion identities for the corresponding renormalised transfer tangles $\db(u)$
are readily obtained from the results presented in the following.
\begin{Proposition}
For $\beta=0$, the transfer tangle $\Db(u)$ defined in (\ref{Duxi}) satisfies the inversion identity
\be
 \Db(u)\Db(u+\tfrac{\pi}{2})=G_N^{(w)}(u,\xi)I
\label{DDG}
\ee
where
\be
 G_N^{(w)}(u,\xi)=-\tan^2 2u\,\eta^{(w)}(u,\xi)\eta^{(w)}(u+\tfrac{\pi}{2},\xi)\Big(A\,[\cos u]^{4N}-2B\,[\cos u\sin u]^{2N}
  +C\,[\sin u]^{4N}\Big)
\ee
with
\begin{align}
 A=&\;\Gamma(u)\big(\Gamma(u+\tfrac{\pi}{2})+\beta_1\cos2u\big)\,\frac{\eta^{(w)}(u,\xi)}{\eta^{(w)}(u+\tfrac{\pi}{2},\xi)}\nn
 B=&\;\Gamma(u)\Gamma(u+\tfrac{\pi}{2})-\tfrac{1}{2}\beta_1\beta_2\cos2u\sin2u
 \label{ABC}\\
 C=&\;\big(\Gamma(u)-\beta_1\cos2u\big)\Gamma(u+\tfrac{\pi}{2})\,\frac{\eta^{(w)}(u+\tfrac{\pi}{2},\xi)}{\eta^{(w)}(u,\xi)}\nonumber
\end{align}
\label{Prop:InvGen}
\end{Proposition}
\noindent{\scshape Proof:}
The product $\Db(u)\Db(u+\frac{\pi}{2})$ is the $(N+w)$-tangle
\be
 \Db(u)\Db(u+\tfrac{\pi}{2})=\
\psset{unit=1cm}
\begin{pspicture}[shift=-0.89](0.4,1)(10.5,4)
\facegrid{(1,0)}{(9,4)}
\pspolygon[fillstyle=solid,fillcolor=lightlightblue](9,1)(10,2)(10,0)(9,1)
\pspolygon[fillstyle=solid,fillcolor=lightlightblue](9,3)(10,4)(10,2)(9,3)
\psarc[linewidth=0.025](1,0){0.16}{0}{90}
\psarc[linewidth=0.025](1,1){0.16}{0}{90}
\psarc[linewidth=0.025](1,2){0.16}{0}{90}
\psarc[linewidth=0.025](1,3){0.16}{0}{90}
\psarc[linewidth=0.025](2,0){0.16}{0}{90}
\psarc[linewidth=0.025](2,1){0.16}{0}{90}
\psarc[linewidth=0.025](2,2){0.16}{0}{90}
\psarc[linewidth=0.025](2,3){0.16}{0}{90}
\psarc[linewidth=0.025](4,0){0.16}{0}{90}
\psarc[linewidth=0.025](4,1){0.16}{0}{90}
\psarc[linewidth=0.025](4,2){0.16}{0}{90}
\psarc[linewidth=0.025](4,3){0.16}{0}{90}
\rput(1.5,0.5){$_{u}$}
\rput(2.5,0.5){$_{u}$}
\rput(4.5,0.5){$_{u}$}
\rput(1.5,1.5){$_{\frac{\pi}{2}-u}$}
\rput(2.5,1.5){$_{\frac{\pi}{2}-u}$}
\rput(4.5,1.5){$_{\frac{\pi}{2}-u}$}
\rput(1.5,2.5){$_{u+\frac{\pi}{2}}$}
\rput(2.5,2.5){$_{u+\frac{\pi}{2}}$}
\rput(4.5,2.5){$_{u+\frac{\pi}{2}}$}
\rput(1.5,3.5){$_{-u}$}
\rput(2.5,3.5){$_{-u}$}
\rput(4.5,3.5){$_{-u}$}
\rput(3.5,0.5){$\ldots$}
\rput(3.5,1.5){$\ldots$}
\rput(3.5,2.5){$\ldots$}
\rput(3.5,3.5){$\ldots$}
\rput(5.55,0.5){$_{u-\xi_{w}}$}
\rput(7.55,0.5){$_{u-\xi_2}$}
\rput(8.55,0.5){$_{u-\xi_1}$}
\rput(5.5,1.5){$_{-\!u\!-\!\xi_{w\!-\!1}}$}
\rput(7.5,1.5){$_{-u-\xi_1}$}
\rput(8.5,1.5){$_{-u-\xi_0}$}
\rput(5.53,2.5){$_{u-\xi_{w\!-\!1}}$}
\rput(7.53,2.5){$_{u-\xi_1}$}
\rput(8.53,2.5){$_{u-\xi_0}$}
\rput(5.52,3.5){$_{-u-\xi_{w}}$}
\rput(7.52,3.5){$_{-u-\xi_2}$}
\rput(8.52,3.5){$_{-u-\xi_1}$}
\rput(6.5,0.5){$\ldots$}
\rput(6.5,1.5){$\ldots$}
\rput(6.5,2.5){$\ldots$}
\rput(6.5,3.5){$\ldots$}
\rput(9.65,1){$_{u}$}
\rput(9.6,3){$_{u+\frac{\pi}{2}}$}
\psarc[linewidth=1.5pt,linecolor=blue](1,1){.5}{90}{270}
\psarc[linewidth=1.5pt,linecolor=blue](1,3){.5}{90}{270}
\psline[linecolor=blue,linewidth=1.5pt]{-}(9,0.5)(9.5,0.5)
\psline[linecolor=blue,linewidth=1.5pt]{-}(9,1.5)(9.5,1.5)
\psline[linecolor=blue,linewidth=1.5pt]{-}(9,2.5)(9.5,2.5)
\psline[linecolor=blue,linewidth=1.5pt]{-}(9,3.5)(9.5,3.5)
\rput(3,-0.5){$\underbrace{\qquad \qquad \qquad \qquad\qquad}_N$}
\rput(7,-0.5){$\underbrace{\qquad \qquad \qquad \qquad\qquad}_w$}
\end{pspicture}
\vspace{1.6cm}
\label{DDuu}
\ee
Inserting the horizontal identity tangle on four strands
\be
\psset{unit=0.77cm}
\begin{pspicture}[shift=-0.89](0.4,1)(2,3.6)
\psline[linecolor=blue,linewidth=1.5pt]{-}(0,0.5)(1.5,0.5)
\psline[linecolor=blue,linewidth=1.5pt]{-}(0,1.5)(1.5,1.5)
\psline[linecolor=blue,linewidth=1.5pt]{-}(0,2.5)(1.5,2.5)
\psline[linecolor=blue,linewidth=1.5pt]{-}(0,3.5)(1.5,3.5)
\end{pspicture}
=\frac{1}{\cos^22u}\qquad
\begin{pspicture}[shift=-0.89](0.4,1)(2,3.6)
\pspolygon[fillstyle=solid,fillcolor=lightlightblue](0,2)(1,1)(2,2)(1,3)(0,2)
\pspolygon[fillstyle=solid,fillcolor=lightlightblue](2,2)(3,1)(4,2)(3,3)(2,2)
\psline[linecolor=blue,linewidth=1.5pt]{-}(0,0.5)(4,0.5)
\psline[linecolor=blue,linewidth=1.5pt]{-}(0,1.5)(0.5,1.5)
\psline[linecolor=blue,linewidth=1.5pt]{-}(0,2.5)(0.5,2.5)
\psline[linecolor=blue,linewidth=1.5pt]{-}(1.5,1.5)(2.5,1.5)
\psline[linecolor=blue,linewidth=1.5pt]{-}(1.5,2.5)(2.5,2.5)
\psline[linecolor=blue,linewidth=1.5pt]{-}(3.5,1.5)(4,1.5)
\psline[linecolor=blue,linewidth=1.5pt]{-}(3.5,2.5)(4,2.5)
\psline[linecolor=blue,linewidth=1.5pt]{-}(0,3.5)(4,3.5)
\psarc[linewidth=0.025](0,2){0.16}{-45}{45}
\psarc[linewidth=0.025](2,2){0.16}{-45}{45}
\rput(1,2){$2u$}
\rput(3,2){$-2u$}
\end{pspicture}
\vspace{0.3cm}
\ee
somewhere in the interior of the diagram in (\ref{DDuu}), and using the YBE to push the $2$-tangles
$
\begin{pspicture}[shift=-0.25](0.8,0.8)
\psset{unit=0.4}
\pspolygon[fillstyle=solid,fillcolor=lightlightblue](0,1)(1,0)(2,1)(1,2)(0,1)
\psarc[linewidth=0.025](0,1){0.16}{-45}{45}
\rput(1,1){$_{_{2u}}$}
\end{pspicture}
$ 
and 
$
\begin{pspicture}[shift=-0.25](0.8,0.8)
\psset{unit=0.4}
\pspolygon[fillstyle=solid,fillcolor=lightlightblue](0,1)(1,0)(2,1)(1,2)(0,1)
\psarc[linewidth=0.025](0,1){0.16}{-45}{45}
\rput(1,1){$_{_{-2u}}$}
\end{pspicture}
$ 
to the left and right, respectively, yields
\be
 \Db(u)\Db(u+\tfrac{\pi}{2})=\frac{1}{\cos^22u}
\quad
\psset{unit=1cm}
\begin{pspicture}[shift=-0.89](0,1)(5.5,4.5)
\pspolygon[fillstyle=solid,fillcolor=lightlightblue](0,2)(1,1)(2,2)(1,3)(0,2)
\pspolygon[fillstyle=solid,fillcolor=pink](2,0)(3.5,0)(3.5,4)(2,4)(2,0)
\rput(2.75,2){$N,w$}
\pspolygon[fillstyle=solid,fillcolor=lightlightblue](3.5,2)(4.5,1)(5.5,2)(4.5,3)(3.5,2)
\pspolygon[fillstyle=solid,fillcolor=lightlightblue](4.5,3)(5.5,4)(5.5,0)(4.5,1)(5.5,2)(4.5,3)
\psarc[linewidth=0.025](0,2){0.16}{-45}{45}
\psarc[linewidth=0.025](3.5,2){0.16}{-45}{45}
\psline[linecolor=blue,linewidth=1.5pt]{-}(0.5,0.5)(2,0.5)
\psline[linecolor=blue,linewidth=1.5pt]{-}(1.5,1.5)(2,1.5)
\psline[linecolor=blue,linewidth=1.5pt]{-}(1.5,2.5)(2,2.5)
\psline[linecolor=blue,linewidth=1.5pt]{-}(0.5,3.5)(2,3.5)
\psarc[linewidth=1.5pt,linecolor=blue](0.5,1){.5}{90}{270}
\psarc[linewidth=1.5pt,linecolor=blue](0.5,3){.5}{90}{270}
\rput(1,2){$2u$}
\rput(4.5,2){$-2u$}
\rput(5.15,1){$u$}
\rput(5.1,3){$_{u+\frac{\pi}{2}}$}
\psline[linecolor=blue,linewidth=1.5pt]{-}(3.5,0.5)(5,0.5)
\psline[linecolor=blue,linewidth=1.5pt]{-}(3.5,1.5)(4,1.5)
\psline[linecolor=blue,linewidth=1.5pt]{-}(3.5,2.5)(4,2.5)
\psline[linecolor=blue,linewidth=1.5pt]{-}(3.5,3.5)(5,3.5)
\end{pspicture}
\label{Dpink}
\vspace{0.6cm}
\ee
where
\be
\psset{unit=1cm}
\qquad
\begin{pspicture}[shift=-0.89](0.8,1)(1.5,4.5)
\pspolygon[fillstyle=solid,fillcolor=pink](0,0)(1.5,0)(1.5,4)(0,4)(0,0)
\rput(0.75,2){$N,w$}
\end{pspicture}
\ \ :=\
\begin{pspicture}[shift=-0.89](0.8,1)(9,4.5)
\facegrid{(1,0)}{(9,4)}
\psarc[linewidth=0.025](1,0){0.16}{0}{90}
\psarc[linewidth=0.025](1,1){0.16}{0}{90}
\psarc[linewidth=0.025](1,2){0.16}{0}{90}
\psarc[linewidth=0.025](1,3){0.16}{0}{90}
\psarc[linewidth=0.025](2,0){0.16}{0}{90}
\psarc[linewidth=0.025](2,1){0.16}{0}{90}
\psarc[linewidth=0.025](2,2){0.16}{0}{90}
\psarc[linewidth=0.025](2,3){0.16}{0}{90}
\psarc[linewidth=0.025](4,0){0.16}{0}{90}
\psarc[linewidth=0.025](4,1){0.16}{0}{90}
\psarc[linewidth=0.025](4,2){0.16}{0}{90}
\psarc[linewidth=0.025](4,3){0.16}{0}{90}
\rput(1.5,0.5){$_{u}$}
\rput(2.5,0.5){$_{u}$}
\rput(4.5,0.5){$_{u}$}
\rput(1.5,2.5){$_{\frac{\pi}{2}-u}$}
\rput(2.5,2.5){$_{\frac{\pi}{2}-u}$}
\rput(4.5,2.5){$_{\frac{\pi}{2}-u}$}
\rput(1.5,1.5){$_{u+\frac{\pi}{2}}$}
\rput(2.5,1.5){$_{u+\frac{\pi}{2}}$}
\rput(4.5,1.5){$_{u+\frac{\pi}{2}}$}
\rput(1.5,3.5){$_{-u}$}
\rput(2.5,3.5){$_{-u}$}
\rput(4.5,3.5){$_{-u}$}
\rput(3.5,0.5){$\ldots$}
\rput(3.5,1.5){$\ldots$}
\rput(3.5,2.5){$\ldots$}
\rput(3.5,3.5){$\ldots$}
\rput(5.55,0.5){$_{u-\xi_{w}}$}
\rput(7.55,0.5){$_{u-\xi_2}$}
\rput(8.55,0.5){$_{u-\xi_1}$}
\rput(5.5,2.5){$_{-\!u\!-\!\xi_{w\!-\!1}}$}
\rput(7.5,2.5){$_{-u-\xi_1}$}
\rput(8.5,2.5){$_{-u-\xi_0}$}
\rput(5.53,1.5){$_{u-\xi_{w\!-\!1}}$}
\rput(7.53,1.5){$_{u-\xi_1}$}
\rput(8.53,1.5){$_{u-\xi_0}$}
\rput(5.52,3.5){$_{-u-\xi_{w}}$}
\rput(7.52,3.5){$_{-u-\xi_2}$}
\rput(8.52,3.5){$_{-u-\xi_1}$}
\rput(6.5,0.5){$\ldots$}
\rput(6.5,1.5){$\ldots$}
\rput(6.5,2.5){$\ldots$}
\rput(6.5,3.5){$\ldots$}
\end{pspicture}
\vspace{1.2cm}
\label{pink}
\ee
Decomposing the two $2$-tangles in (\ref{Dpink}) subsequently yields
\be
 \Db(u)\Db(u+\tfrac{\pi}{2})=-\frac{\cos2u\sin2u}{\cos^22u}\,\mathfrak{D}_1+\frac{\cos^22u}{\cos^22u}\,\mathfrak{D}_2
   -\frac{\sin^22u}{\cos^22u}\,\mathfrak{D}_3+\frac{\sin2u\cos2u}{\cos^22u}\,\mathfrak{D}_4
\ee
where
\be
 \mathfrak{D}_1:=
\psset{unit=0.8cm}
\begin{pspicture}[shift=-0.89](0,1)(4.2,4.2)
\pspolygon[fillstyle=solid,fillcolor=pink](1,0)(2.5,0)(2.5,4)(1,4)(1,0)
\psarc[linewidth=1.5pt,linecolor=blue](1,1){.5}{90}{270}
\psarc[linewidth=1.5pt,linecolor=blue](1,3){.5}{90}{270}
\rput(1.75,2){$N,w$}
\psline[linecolor=blue,linewidth=1.5pt]{-}(2.5,0.5)(3.5,0.5)
\psline[linecolor=blue,linewidth=1.5pt]{-}(2.5,3.5)(3.5,3.5)
\psarc[linewidth=1.5pt,linecolor=blue](2.5,2){.5}{-90}{90}
\pspolygon[fillstyle=solid,fillcolor=lightlightblue](4,0)(4,4)(3,3)(4,2)(3,1)(4,0)
\rput(3.65,1){$u$}
\rput(3.6,3){$_{u+\frac{\pi}{2}}$}
\psline[linecolor=blue,linewidth=1.5pt]{-}(3.5,1.5)(3.5,2.5)
\end{pspicture}
=\,g_N^{(w)}(u,\xi)I\ \
\begin{pspicture}[shift=-0.89](0,1)(1.7,4.2)
\psarc[linewidth=1.5pt,linecolor=blue](0.5,1){.5}{180}{315}
\psarc[linewidth=1.5pt,linecolor=blue](0.5,3){.5}{45}{180}
\psline[linecolor=blue,linewidth=1.5pt]{-}(0,1)(0,3)
\pspolygon[fillstyle=solid,fillcolor=lightlightblue](1.5,0)(1.5,4)(0.5,3)(1.5,2)(0.5,1)(1.5,0)
\psline[linecolor=blue,linewidth=1.5pt]{-}(1,1.5)(1,2.5)
\rput(1.15,1){$u$}
\rput(1.1,3){$_{u+\frac{\pi}{2}}$}
\end{pspicture}
=-\beta_1^2\cos2u\sin2u\,g_N^{(w)}(u,\xi)I
\label{D1}
\vspace{1cm}
\ee
\be
 \mathfrak{D}_2:=
\psset{unit=0.8cm}
\begin{pspicture}[shift=-0.89](0,1)(4.2,4.2)
\pspolygon[fillstyle=solid,fillcolor=pink](1,0)(2.5,0)(2.5,4)(1,4)(1,0)
\psarc[linewidth=1.5pt,linecolor=blue](1,1){.5}{90}{270}
\psarc[linewidth=1.5pt,linecolor=blue](1,3){.5}{90}{270}
\rput(1.75,2){$N,w$}
\psline[linecolor=blue,linewidth=1.5pt]{-}(2.5,0.5)(3.5,0.5)
\psline[linecolor=blue,linewidth=1.5pt]{-}(2.5,1.5)(3.5,1.5)
\psline[linecolor=blue,linewidth=1.5pt]{-}(2.5,2.5)(3.5,2.5)
\psline[linecolor=blue,linewidth=1.5pt]{-}(2.5,3.5)(3.5,3.5)
\pspolygon[fillstyle=solid,fillcolor=lightlightblue](4,0)(4,4)(3,3)(4,2)(3,1)(4,0)
\rput(3.65,1){$u$}
\rput(3.6,3){$_{u+\frac{\pi}{2}}$}
\end{pspicture}
=\,g_N^{(w)}(u,\xi)I\ \
\begin{pspicture}[shift=-0.89](0,1)(1.7,4.2)
\psarc[linewidth=1.5pt,linecolor=blue](0.5,1){.5}{45}{315}
\psarc[linewidth=1.5pt,linecolor=blue](0.5,3){.5}{45}{315}
\pspolygon[fillstyle=solid,fillcolor=lightlightblue](1.5,0)(1.5,4)(0.5,3)(1.5,2)(0.5,1)(1.5,0)
\rput(1.15,1){$u$}
\rput(1.1,3){$_{u+\frac{\pi}{2}}$}
\end{pspicture}
=-\beta_1^2\sin^22u\,g_N^{(w)}(u,\xi)I
\label{D2}
\vspace{1cm}
\ee
\be
 \mathfrak{D}_3:=
\psset{unit=0.8cm}
\begin{pspicture}[shift=-0.89](0,1)(4.2,4.2)
\pspolygon[fillstyle=solid,fillcolor=pink](1,0)(2.5,0)(2.5,4)(1,4)(1,0)
\psarc[linewidth=1.5pt,linecolor=blue](1,2){.5}{90}{270}
\psbezier[linewidth=1.5pt,linecolor=blue](1,0.5)(0,0.45)(0,3.55)(1,3.5)
\rput(1.75,2){$N,w$}
\psline[linecolor=blue,linewidth=1.5pt]{-}(2.5,0.5)(3.5,0.5)
\psline[linecolor=blue,linewidth=1.5pt]{-}(2.5,3.5)(3.5,3.5)
\psarc[linewidth=1.5pt,linecolor=blue](2.5,2){.5}{-90}{90}
\pspolygon[fillstyle=solid,fillcolor=lightlightblue](4,0)(4,4)(3,3)(4,2)(3,1)(4,0)
\rput(3.65,1){$u$}
\rput(3.6,3){$_{u+\frac{\pi}{2}}$}
\psline[linecolor=blue,linewidth=1.5pt]{-}(3.5,1.5)(3.5,2.5)
\end{pspicture}
=\G(u)\G(u+\tfrac{\pi}{2})
\begin{pspicture}[shift=-0.89](0,1)(3.5,4.2)
\pspolygon[fillstyle=solid,fillcolor=pink](1,0)(2.5,0)(2.5,4)(1,4)(1,0)
\psarc[linewidth=1.5pt,linecolor=blue](1,2){.5}{90}{270}
\psbezier[linewidth=1.5pt,linecolor=blue](1,0.5)(0,0.45)(0,3.55)(1,3.5)
\rput(1.75,2){$N,w$}
\psarc[linewidth=1.5pt,linecolor=blue](2.5,2){.5}{-90}{90}
\psbezier[linewidth=1.5pt,linecolor=blue](2.5,0.5)(3.5,0.45)(3.5,3.55)(2.5,3.5)
\end{pspicture}
-\beta_1\cos2u\sin2u
\psset{unit=0.8cm}
\begin{pspicture}[shift=-0.89](0,1)(3.2,4.2)
\pspolygon[fillstyle=solid,fillcolor=pink](1,0)(2.5,0)(2.5,4)(1,4)(1,0)
\psarc[linewidth=1.5pt,linecolor=blue](1,2){.5}{90}{270}
\psbezier[linewidth=1.5pt,linecolor=blue](1,0.5)(0,0.45)(0,3.55)(1,3.5)
\rput(1.75,2){$N,w$}
\psline[linecolor=blue,linewidth=1.5pt]{-}(2.5,0.5)(3,0.5)
\psline[linecolor=blue,linewidth=1.5pt]{-}(2.5,3.5)(3,3.5)
\psarc[linewidth=1.5pt,linecolor=blue](2.5,2){.5}{-90}{90}
\rput(3,0.5){$\bullet$}
\rput(3,3.5){$\bullet$}
\end{pspicture}
\label{D3}
\vspace{1cm}
\ee
and
\be
 \mathfrak{D}_4:=
\psset{unit=0.8cm}
\begin{pspicture}[shift=-0.89](0,1)(4.2,4.2)
\pspolygon[fillstyle=solid,fillcolor=pink](1,0)(2.5,0)(2.5,4)(1,4)(1,0)
\psarc[linewidth=1.5pt,linecolor=blue](1,2){.5}{90}{270}
\psbezier[linewidth=1.5pt,linecolor=blue](1,0.5)(0,0.45)(0,3.55)(1,3.5)
\rput(1.75,2){$N,w$}
\psline[linecolor=blue,linewidth=1.5pt]{-}(2.5,0.5)(3.5,0.5)
\psline[linecolor=blue,linewidth=1.5pt]{-}(2.5,1.5)(3.5,1.5)
\psline[linecolor=blue,linewidth=1.5pt]{-}(2.5,2.5)(3.5,2.5)
\psline[linecolor=blue,linewidth=1.5pt]{-}(2.5,3.5)(3.5,3.5)
\pspolygon[fillstyle=solid,fillcolor=lightlightblue](4,0)(4,4)(3,3)(4,2)(3,1)(4,0)
\rput(3.65,1){$u$}
\rput(3.6,3){$_{u+\frac{\pi}{2}}$}
\end{pspicture}
=\hat{g}_N^{(w)}(u,\xi)I-\sin^22u
\begin{pspicture}[shift=-0.89](0,1)(3.2,4.2)
\pspolygon[fillstyle=solid,fillcolor=pink](1,0)(2.5,0)(2.5,4)(1,4)(1,0)
\psarc[linewidth=1.5pt,linecolor=blue](1,2){.5}{90}{270}
\psbezier[linewidth=1.5pt,linecolor=blue](1,0.5)(0,0.45)(0,3.55)(1,3.5)
\rput(1.75,2){$N,w$}
\psline[linecolor=blue,linewidth=1.5pt]{-}(2.5,0.5)(3,0.5)
\psline[linecolor=blue,linewidth=1.5pt]{-}(2.5,1.5)(3,1.5)
\psline[linecolor=blue,linewidth=1.5pt]{-}(2.5,2.5)(3,2.5)
\psline[linecolor=blue,linewidth=1.5pt]{-}(2.5,3.5)(3,3.5)
\rput(3,0.5){$\bullet$}
\rput(3,1.5){$\bullet$}
\rput(3,2.5){$\bullet$}
\rput(3,3.5){$\bullet$}
\end{pspicture}
\vspace{0.8cm}
\label{D4}
\ee
with
\begin{align}
 g_N^{(w)}(u,\xi):=&(-1)^{N+w}[\cos u\sin u]^{2N}\prod_{j=1}^w\cos^2(u+\xi_j)\sin^2(u-\xi_j)
\nonumber\\[.1cm]
 \hat{g}_N^{(w)}(u,\xi):=&\beta_1\sin2u\,\Big(\G(u+\tfrac{\pi}{2})\,[\sin u]^{4N}\prod_{j=1}^w\sin^2(u+\xi_j)\sin^2(u-\xi_j)
\\[.1cm]
 &\qquad\qquad\qquad -\G(u)\,[\cos u]^{4N}\prod_{j=1}^w\cos^2(u+\xi_j)\cos^2(u-\xi_j)\Big)
\nonumber
\end{align}
The rewritings of the expressions (\ref{D1})-(\ref{D4}) follow by decomposing the boundary triangles and using the trigonometric identity
\be
 \G(u)s(2u+\pi)+s(2u)\G(u+\tfrac{\pi}{2})+\beta_2 s(2u)s(2u+\pi)=-\beta_1\cos^22u\sin^22u
\ee 
{}From~\cite{PRV1210}, we know that
\begin{align}
\psset{unit=.8cm}
\begin{pspicture}[shift=-0.89](0,1)(3.5,4)
\pspolygon[fillstyle=solid,fillcolor=pink](1,0)(2.5,0)(2.5,4)(1,4)(1,0)
\psarc[linewidth=1.5pt,linecolor=blue](1,2){.5}{90}{270}
\psbezier[linewidth=1.5pt,linecolor=blue](1,0.5)(0,0.45)(0,3.55)(1,3.5)
\rput(1.75,2){$N,w$}
\psarc[linewidth=1.5pt,linecolor=blue](2.5,2){.5}{-90}{90}
\psbezier[linewidth=1.5pt,linecolor=blue](2.5,0.5)(3.5,0.45)(3.5,3.55)(2.5,3.5)
\end{pspicture}
&=\eta^{(w)}(u,\xi)\eta^{(w)}(u+\tfrac{\pi}{2},\xi)
\\[-0.4cm]
&\times\Big(\frac{\eta^{(w)}(u,\xi)}{\eta^{(w)}(u+\tfrac{\pi}{2},\xi)}\,[\cos u]^{4N}-2\,[\cos u\sin u]^{2N}
  +\frac{\eta^{(w)}(u+\tfrac{\pi}{2},\xi)}{\eta^{(w)}(u,\xi)}\,[\sin u]^{4N}\Big)I
\nonumber
\end{align}
which combined with (\ref{D1})-(\ref{D4}) implies
\begin{align}
 \Db(u)\Db(u+\tfrac{\pi}{2})=&-\frac{\sin^22u}{\cos^22u}\Big([\cos u]^{4N}[\eta^{(w)}(u,\xi)]^2\G(u)\big(\G(u+\tfrac{\pi}{2})+\beta_1\cos2u\big)
\nonumber\\[.2cm]
 &\qquad\qquad\quad-2\,[\cos u\sin u]^{2N}\eta^{(w)}(u,\xi)\eta^{(w)}(u+\tfrac{\pi}{2},\xi)\G(u)\G(u+\tfrac{\pi}{2})
\nonumber\\[.2cm]
 &\qquad\qquad\quad+[\sin u]^{4N}[\eta^{(w)}(u+\tfrac{\pi}{2},\xi)]^2\big(\G(u)-\beta_1\cos2u\big)\G(u+\tfrac{\pi}{2})\Big)I
\nonumber\\[.2cm]
 &+\frac{\sin^32u}{\cos^22u}
\left(
 \beta_1
\psset{unit=0.8cm}
\begin{pspicture}[shift=-0.89](0,1)(3.2,4.2)
\pspolygon[fillstyle=solid,fillcolor=pink](1,0)(2.5,0)(2.5,4)(1,4)(1,0)
\psarc[linewidth=1.5pt,linecolor=blue](1,2){.5}{90}{270}
\psbezier[linewidth=1.5pt,linecolor=blue](1,0.5)(0,0.45)(0,3.55)(1,3.5)
\rput(1.75,2){$N,w$}
\psline[linecolor=blue,linewidth=1.5pt]{-}(2.5,0.5)(3,0.5)
\psline[linecolor=blue,linewidth=1.5pt]{-}(2.5,3.5)(3,3.5)
\psarc[linewidth=1.5pt,linecolor=blue](2.5,2){.5}{-90}{90}
\rput(3,0.5){$\bullet$}
\rput(3,3.5){$\bullet$}
\end{pspicture}
-
\begin{pspicture}[shift=-0.89](0,1)(3.2,4.2)
\pspolygon[fillstyle=solid,fillcolor=pink](1,0)(2.5,0)(2.5,4)(1,4)(1,0)
\psarc[linewidth=1.5pt,linecolor=blue](1,2){.5}{90}{270}
\psbezier[linewidth=1.5pt,linecolor=blue](1,0.5)(0,0.45)(0,3.55)(1,3.5)
\rput(1.75,2){$N,w$}
\psline[linecolor=blue,linewidth=1.5pt]{-}(2.5,0.5)(3,0.5)
\psline[linecolor=blue,linewidth=1.5pt]{-}(2.5,1.5)(3,1.5)
\psline[linecolor=blue,linewidth=1.5pt]{-}(2.5,2.5)(3,2.5)
\psline[linecolor=blue,linewidth=1.5pt]{-}(2.5,3.5)(3,3.5)
\rput(3,0.5){$\bullet$}
\rput(3,1.5){$\bullet$}
\rput(3,2.5){$\bullet$}
\rput(3,3.5){$\bullet$}
\end{pspicture}
\right)
\label{DDbeta1}
\end{align}
To verify that the $(N+w)$-tangle within parentheses in this expression is proportional to the identity tangle $I$, we first establish that
\begin{align}
\psset{unit=0.8cm}
\begin{pspicture}[shift=-0.89](0,1)(2.2,4.2)
\facegrid{(1,0)}{(2,4)}
\psarc[linewidth=1.5pt,linecolor=blue](1,2){.5}{90}{270}
\psbezier[linewidth=1.5pt,linecolor=blue](1,0.5)(0,0.45)(0,3.55)(1,3.5)
\psarc[linewidth=0.025](1,0){0.16}{0}{90}
\psarc[linewidth=0.025](1,1){0.16}{0}{90}
\psarc[linewidth=0.025](1,2){0.16}{0}{90}
\psarc[linewidth=0.025](1,3){0.16}{0}{90}
\rput(1.5,0.5){$_{\mu}$}
\rput(1.53,1.5){$_{\mu+\frac{\pi}{2}}$}
\rput(1.53,2.5){$_{\nu+\frac{\pi}{2}}$}
\rput(1.5,3.5){$_{\nu}$}
\end{pspicture}
&=\cos\nu\cos\mu\cos(\nu-\mu)
\psset{unit=0.8cm}
\left(\ \,
\begin{pspicture}[shift=-0.89](0,1)(1.2,4.2)
\pspolygon[fillstyle=solid,fillcolor=lightlightblue](0,0)(1,0)(1,4)(0,4)(0,0)
\psarc[linewidth=1.5pt,linecolor=blue](1,3){.5}{90}{270}
\psbezier[linewidth=1.5pt,linecolor=blue](0.3,4)(0.2,1.45)(0.5,1.55)(1,1.5)
\psarc[linewidth=1.5pt,linecolor=blue](1,0){.5}{90}{180}
\end{pspicture}
+\
\begin{pspicture}[shift=-0.89](0,1)(1.2,4.2)
\pspolygon[fillstyle=solid,fillcolor=lightlightblue](0,0)(1,0)(1,4)(0,4)(0,0)
\psarc[linewidth=1.5pt,linecolor=blue](1,1){.5}{90}{270}
\psbezier[linewidth=1.5pt,linecolor=blue](0.3,0)(0.2,2.55)(0.5,2.45)(1,2.5)
\psarc[linewidth=1.5pt,linecolor=blue](1,4){.5}{180}{270}
\end{pspicture}
\right)
\nonumber\\[.3cm]
&+\cos\nu\sin\mu\cos(\nu-\mu)\
\psset{unit=0.8cm}
\begin{pspicture}[shift=-0.89](0,1)(1.2,4.2)
\pspolygon[fillstyle=solid,fillcolor=lightlightblue](0,0)(1,0)(1,4)(0,4)(0,0)
\psarc[linewidth=1.5pt,linecolor=blue](1,1){.5}{90}{270}
\psarc[linewidth=1.5pt,linecolor=blue](1,3){.5}{90}{270}
\psline[linecolor=blue,linewidth=1.5pt]{-}(0.3,0)(0.3,4)
\end{pspicture}
-\tfrac{1}{4}\sin2\nu\sin2\mu\
\begin{pspicture}[shift=-0.89](0,1)(1.2,4.2)
\pspolygon[fillstyle=solid,fillcolor=lightlightblue](0,0)(1,0)(1,4)(0,4)(0,0)
\psbezier[linewidth=1.5pt,linecolor=blue](1,1.5)(0.6,1.45)(0.6,2.55)(1,2.5)
\psbezier[linewidth=1.5pt,linecolor=blue](1,0.5)(0.3,0.45)(0.3,3.55)(1,3.5)
\psline[linecolor=blue,linewidth=1.5pt]{-}(0.3,0)(0.3,4)
\end{pspicture}
\\[.3cm]\nonumber
\end{align}
We also note that
\be
\psset{unit=0.8cm}
\beta_1
\begin{pspicture}[shift=-0.89](0,1)(3.2,4.2)
\psarc[linewidth=1.5pt,linecolor=blue](1,3){.5}{90}{270}
\psbezier[linewidth=1.5pt,linecolor=blue](0.3,4)(0.2,1.45)(0.5,1.55)(1,1.5)
\psarc[linewidth=1.5pt,linecolor=blue](1,0){.5}{90}{180}
\pspolygon[fillstyle=solid,fillcolor=pink](1,0)(2.5,0)(2.5,4)(1,4)(1,0)
\rput(1.75,2){$N,w$}
\psline[linecolor=blue,linewidth=1.5pt]{-}(2.5,0.5)(3,0.5)
\psline[linecolor=blue,linewidth=1.5pt]{-}(2.5,3.5)(3,3.5)
\psarc[linewidth=1.5pt,linecolor=blue](2.5,2){.5}{-90}{90}
\rput(3,0.5){$\bullet$}
\rput(3,3.5){$\bullet$}
\end{pspicture}
\ \ -\
\begin{pspicture}[shift=-0.89](0,1)(3.2,4.2)
\psarc[linewidth=1.5pt,linecolor=blue](1,3){.5}{90}{270}
\psbezier[linewidth=1.5pt,linecolor=blue](0.3,4)(0.2,1.45)(0.5,1.55)(1,1.5)
\psarc[linewidth=1.5pt,linecolor=blue](1,0){.5}{90}{180}
\pspolygon[fillstyle=solid,fillcolor=pink](1,0)(2.5,0)(2.5,4)(1,4)(1,0)
\rput(1.75,2){$N,w$}
\psline[linecolor=blue,linewidth=1.5pt]{-}(2.5,0.5)(3,0.5)
\psline[linecolor=blue,linewidth=1.5pt]{-}(2.5,1.5)(3,1.5)
\psline[linecolor=blue,linewidth=1.5pt]{-}(2.5,2.5)(3,2.5)
\psline[linecolor=blue,linewidth=1.5pt]{-}(2.5,3.5)(3,3.5)
\rput(3,0.5){$\bullet$}
\rput(3,1.5){$\bullet$}
\rput(3,2.5){$\bullet$}
\rput(3,3.5){$\bullet$}
\end{pspicture}
\ =0,\qquad
\beta_1
\begin{pspicture}[shift=-0.89](0,1)(3.2,4.2)
\psarc[linewidth=1.5pt,linecolor=blue](1,1){.5}{90}{270}
\psbezier[linewidth=1.5pt,linecolor=blue](0.3,0)(0.2,2.55)(0.5,2.45)(1,2.5)
\psarc[linewidth=1.5pt,linecolor=blue](1,4){.5}{180}{270}
\pspolygon[fillstyle=solid,fillcolor=pink](1,0)(2.5,0)(2.5,4)(1,4)(1,0)
\rput(1.75,2){$N,w$}
\psline[linecolor=blue,linewidth=1.5pt]{-}(2.5,0.5)(3,0.5)
\psline[linecolor=blue,linewidth=1.5pt]{-}(2.5,3.5)(3,3.5)
\psarc[linewidth=1.5pt,linecolor=blue](2.5,2){.5}{-90}{90}
\rput(3,0.5){$\bullet$}
\rput(3,3.5){$\bullet$}
\end{pspicture}
\ \ -\
\begin{pspicture}[shift=-0.89](0,1)(3.2,4.2)
\psarc[linewidth=1.5pt,linecolor=blue](1,1){.5}{90}{270}
\psbezier[linewidth=1.5pt,linecolor=blue](0.3,0)(0.2,2.55)(0.5,2.45)(1,2.5)
\psarc[linewidth=1.5pt,linecolor=blue](1,4){.5}{180}{270}
\pspolygon[fillstyle=solid,fillcolor=pink](1,0)(2.5,0)(2.5,4)(1,4)(1,0)
\rput(1.75,2){$N,w$}
\psline[linecolor=blue,linewidth=1.5pt]{-}(2.5,0.5)(3,0.5)
\psline[linecolor=blue,linewidth=1.5pt]{-}(2.5,1.5)(3,1.5)
\psline[linecolor=blue,linewidth=1.5pt]{-}(2.5,2.5)(3,2.5)
\psline[linecolor=blue,linewidth=1.5pt]{-}(2.5,3.5)(3,3.5)
\rput(3,0.5){$\bullet$}
\rput(3,1.5){$\bullet$}
\rput(3,2.5){$\bullet$}
\rput(3,3.5){$\bullet$}
\end{pspicture}
\ =0
\vspace{1cm}
\ee
and
\be
\psset{unit=0.8cm}
\beta_1
\begin{pspicture}[shift=-0.89](0,1)(3.2,4.2)
\psarc[linewidth=1.5pt,linecolor=blue](1,1){.5}{90}{270}
\psarc[linewidth=1.5pt,linecolor=blue](1,3){.5}{90}{270}
\psline[linecolor=blue,linewidth=1.5pt]{-}(0.25,0)(0.25,4)
\pspolygon[fillstyle=solid,fillcolor=pink](1,0)(2.5,0)(2.5,4)(1,4)(1,0)
\rput(1.75,2){$N,w$}
\psline[linecolor=blue,linewidth=1.5pt]{-}(2.5,0.5)(3,0.5)
\psline[linecolor=blue,linewidth=1.5pt]{-}(2.5,3.5)(3,3.5)
\psarc[linewidth=1.5pt,linecolor=blue](2.5,2){.5}{-90}{90}
\rput(3,0.5){$\bullet$}
\rput(3,3.5){$\bullet$}
\end{pspicture}
\ \ -\
\begin{pspicture}[shift=-0.89](0,1)(3.2,4.2)
\psarc[linewidth=1.5pt,linecolor=blue](1,1){.5}{90}{270}
\psarc[linewidth=1.5pt,linecolor=blue](1,3){.5}{90}{270}
\psline[linecolor=blue,linewidth=1.5pt]{-}(0.25,0)(0.25,4)
\pspolygon[fillstyle=solid,fillcolor=pink](1,0)(2.5,0)(2.5,4)(1,4)(1,0)
\rput(1.75,2){$N,w$}
\psline[linecolor=blue,linewidth=1.5pt]{-}(2.5,0.5)(3,0.5)
\psline[linecolor=blue,linewidth=1.5pt]{-}(2.5,1.5)(3,1.5)
\psline[linecolor=blue,linewidth=1.5pt]{-}(2.5,2.5)(3,2.5)
\psline[linecolor=blue,linewidth=1.5pt]{-}(2.5,3.5)(3,3.5)
\rput(3,0.5){$\bullet$}
\rput(3,1.5){$\bullet$}
\rput(3,2.5){$\bullet$}
\rput(3,3.5){$\bullet$}
\end{pspicture}
\ =0
\vspace{1cm}
\ee
so that
\be
 \beta_1
\psset{unit=0.8cm}
\begin{pspicture}[shift=-0.89](0,1)(3.2,4.2)
\pspolygon[fillstyle=solid,fillcolor=pink](1,0)(2.5,0)(2.5,4)(1,4)(1,0)
\psarc[linewidth=1.5pt,linecolor=blue](1,2){.5}{90}{270}
\psbezier[linewidth=1.5pt,linecolor=blue](1,0.5)(0,0.45)(0,3.55)(1,3.5)
\rput(1.75,2){$N,w$}
\psline[linecolor=blue,linewidth=1.5pt]{-}(2.5,0.5)(3,0.5)
\psline[linecolor=blue,linewidth=1.5pt]{-}(2.5,3.5)(3,3.5)
\psarc[linewidth=1.5pt,linecolor=blue](2.5,2){.5}{-90}{90}
\rput(3,0.5){$\bullet$}
\rput(3,3.5){$\bullet$}
\end{pspicture}
-
\begin{pspicture}[shift=-0.89](0,1)(3.2,4.2)
\pspolygon[fillstyle=solid,fillcolor=pink](1,0)(2.5,0)(2.5,4)(1,4)(1,0)
\psarc[linewidth=1.5pt,linecolor=blue](1,2){.5}{90}{270}
\psbezier[linewidth=1.5pt,linecolor=blue](1,0.5)(0,0.45)(0,3.55)(1,3.5)
\rput(1.75,2){$N,w$}
\psline[linecolor=blue,linewidth=1.5pt]{-}(2.5,0.5)(3,0.5)
\psline[linecolor=blue,linewidth=1.5pt]{-}(2.5,1.5)(3,1.5)
\psline[linecolor=blue,linewidth=1.5pt]{-}(2.5,2.5)(3,2.5)
\psline[linecolor=blue,linewidth=1.5pt]{-}(2.5,3.5)(3,3.5)
\rput(3,0.5){$\bullet$}
\rput(3,1.5){$\bullet$}
\rput(3,2.5){$\bullet$}
\rput(3,3.5){$\bullet$}
\end{pspicture}
=\,[\cos u\sin u]^2
\left(
\beta_1
\begin{pspicture}[shift=-0.89](0,1)(3.2,4.2)
\pspolygon[fillstyle=solid,fillcolor=pink](1,0)(2.5,0)(2.5,4)(1,4)(1,0)
\psbezier[linewidth=1.5pt,linecolor=blue](1,1.5)(0.6,1.45)(0.6,2.55)(1,2.5)
\psbezier[linewidth=1.5pt,linecolor=blue](1,0.5)(0.3,0.45)(0.3,3.55)(1,3.5)
\psline[linecolor=blue,linewidth=1.5pt]{-}(0.25,0)(0.25,4)
\rput(1.75,2){$_{N-1,w}$}
\psline[linecolor=blue,linewidth=1.5pt]{-}(2.5,0.5)(3,0.5)
\psline[linecolor=blue,linewidth=1.5pt]{-}(2.5,3.5)(3,3.5)
\psarc[linewidth=1.5pt,linecolor=blue](2.5,2){.5}{-90}{90}
\rput(3,0.5){$\bullet$}
\rput(3,3.5){$\bullet$}
\end{pspicture}
-
\begin{pspicture}[shift=-0.89](0,1)(3.2,4.2)
\pspolygon[fillstyle=solid,fillcolor=pink](1,0)(2.5,0)(2.5,4)(1,4)(1,0)
\psbezier[linewidth=1.5pt,linecolor=blue](1,1.5)(0.6,1.45)(0.6,2.55)(1,2.5)
\psbezier[linewidth=1.5pt,linecolor=blue](1,0.5)(0.3,0.45)(0.3,3.55)(1,3.5)
\psline[linecolor=blue,linewidth=1.5pt]{-}(0.25,0)(0.25,4)
\rput(1.75,2){$_{N-1,w}$}
\psline[linecolor=blue,linewidth=1.5pt]{-}(2.5,0.5)(3,0.5)
\psline[linecolor=blue,linewidth=1.5pt]{-}(2.5,1.5)(3,1.5)
\psline[linecolor=blue,linewidth=1.5pt]{-}(2.5,2.5)(3,2.5)
\psline[linecolor=blue,linewidth=1.5pt]{-}(2.5,3.5)(3,3.5)
\rput(3,0.5){$\bullet$}
\rput(3,1.5){$\bullet$}
\rput(3,2.5){$\bullet$}
\rput(3,3.5){$\bullet$}
\end{pspicture}
\right)
\ee
Applying this repeatedly yields
\begin{align}
 \beta_1
\psset{unit=0.8cm}
\begin{pspicture}[shift=-0.89](0,1)(3.2,4.2)
\pspolygon[fillstyle=solid,fillcolor=pink](1,0)(2.5,0)(2.5,4)(1,4)(1,0)
\psarc[linewidth=1.5pt,linecolor=blue](1,2){.5}{90}{270}
\psbezier[linewidth=1.5pt,linecolor=blue](1,0.5)(0,0.45)(0,3.55)(1,3.5)
\rput(1.75,2){$N,w$}
\psline[linecolor=blue,linewidth=1.5pt]{-}(2.5,0.5)(3,0.5)
\psline[linecolor=blue,linewidth=1.5pt]{-}(2.5,3.5)(3,3.5)
\psarc[linewidth=1.5pt,linecolor=blue](2.5,2){.5}{-90}{90}
\rput(3,0.5){$\bullet$}
\rput(3,3.5){$\bullet$}
\end{pspicture}
-
\begin{pspicture}[shift=-0.89](0,1)(3.2,4.2)
\pspolygon[fillstyle=solid,fillcolor=pink](1,0)(2.5,0)(2.5,4)(1,4)(1,0)
\psarc[linewidth=1.5pt,linecolor=blue](1,2){.5}{90}{270}
\psbezier[linewidth=1.5pt,linecolor=blue](1,0.5)(0,0.45)(0,3.55)(1,3.5)
\rput(1.75,2){$N,w$}
\psline[linecolor=blue,linewidth=1.5pt]{-}(2.5,0.5)(3,0.5)
\psline[linecolor=blue,linewidth=1.5pt]{-}(2.5,1.5)(3,1.5)
\psline[linecolor=blue,linewidth=1.5pt]{-}(2.5,2.5)(3,2.5)
\psline[linecolor=blue,linewidth=1.5pt]{-}(2.5,3.5)(3,3.5)
\rput(3,0.5){$\bullet$}
\rput(3,1.5){$\bullet$}
\rput(3,2.5){$\bullet$}
\rput(3,3.5){$\bullet$}
\end{pspicture}
=-\beta_1\beta_2\,[\cos u\sin u]^{2N}\eta^{(w)}(u,\xi)\eta^{(w)}(u+\tfrac{\pi}{2},\xi)I
\label{beta1NN}
\\[.2cm] \nonumber
\end{align}
The identity (\ref{DDG}) now follows by combining (\ref{DDbeta1}) with (\ref{beta1NN}).
\hfill $\square$
\medskip

\noindent
For $\lambda=\frac{\pi}{2}$, the function $\eta^{(w)}(u,\xi)$ satisfies
\be
 \eta^{(w)}(u,\xi)=\prod_{j=1}^w\cos(u+\xi_j)\cos(u-\xi_j),\qquad
 \eta^{(w)}(u+\tfrac{\pi}{2},\xi)=\prod_{j=1}^w\sin(u+\xi_j)\sin(u-\xi_j)
\ee
and
\be
 \frac{\eta^{(w)}(u+\tfrac{\pi}{2},\xi)}{\eta^{(w)}(u,\xi)}\,=\,\begin{cases}1,\quad &w\;\mbox{even}\\[.2cm]
   \cot(u+\xi)\cot(u-\xi),\quad &w\;\mbox{odd}  \end{cases}
\ee
For $w>0$, we can thus simplify the expressions for $A$ and $C$ in (\ref{ABC}) as
\be
 A=\Big(\big(\beta_1^2+\beta_2^2\big)\cos^2\!u-\big(\beta_1\sin\xi+\beta_2\cos\xi\big)^2\Big)\times\begin{cases}
  \cos(u+\xi)\cos(u-\xi),\quad &w\;\mbox{even}\\[.2cm]  \sin(u+\xi)\sin(u-\xi),\quad &w\;\mbox{odd}\end{cases}
\ee
\be
 C=\Big(\big(\beta_1^2+\beta_2^2\big)\sin^2\!u-\big(\beta_1\sin\xi+\beta_2\cos\xi\big)^2\Big)\times\begin{cases}
  \sin(u+\xi)\sin(u-\xi),\quad &w\;\mbox{even}\\[.2cm]  \cos(u+\xi)\cos(u-\xi),\quad &w\;\mbox{odd}\end{cases}
\ee
while
\be
 B=-\tfrac{1}{8}\big(\beta_1^2-\beta_2^2\big)\big(\!\cos4u-\cos4\xi\big)-\tfrac{1}{4}\beta_1\beta_2\sin4\xi
\ee
where we recall the trigonometric identities
\be
 \cos(u+\xi)\cos(u-\xi)=\cos^2\!u-\sin^2\!\xi,\qquad \sin(u+\xi)\sin(u-\xi)=\sin^2\!u-\sin^2\!\xi
\ee

These expressions for $A$, $B$ and $C$ are all homogenous of degree $2$ in the boundary loop fugacities $\beta_1$ and $\beta_2$.
It follows that, for $\beta_1=\beta_2$, the righthand side of the inversion identity in Proposition~\ref{Prop:InvGen} is 
proportional to $\beta_1^2$. The explicit form is easily obtained, but not given here.
Instead, using (\ref{minuspi4}) and
\be
 \cos(u+\xi)\cos(u-\xi)\big|_{\xi=-\frac{\pi}{4}}=-\sin(u+\xi)\sin(u-\xi)\big|_{\xi=-\frac{\pi}{4}}=\tfrac{1}{2}\cos2u
\ee
we obtain the following corollaries.
\begin{Corollary}
For $\xi=-\frac{\pi}{4}$, the inversion identity in Proposition~\ref{Prop:InvGen} is given by
\begin{align}
 \Db(u)\Db(u+\tfrac{\pi}{2})=&-\frac{[\sin2u]^2\,[\cos2u]^{2w-1}}{4^{w+1}}
  \Big(2\big(\beta_1^2+\beta_2^2\big)\big(\!\cos^{4N+2}\!u-\sin^{4N+2}\!u\big)\nn
   &-(\beta_1-\beta_2)^2\big(\!\cos^{4N}\!u-\sin^{4N}\!u\big)
   +2(-1)^w\big(\beta_1^2-\beta_2^2\big)\cos2u[\cos u\sin u]^{2N}\Big)I
\end{align}
\end{Corollary}
\begin{Corollary}
For $\beta_1=\beta_2$ and $\xi=-\frac{\pi}{4}$, the inversion identity in Proposition~\ref{Prop:InvGen} is given by
\be
  \Db(u)\Db(u+\tfrac{\pi}{2})=-\frac{\beta_1^2\,[\sin2u]^2\,[\cos2u]^{2w-1}}{4^w}
  \big(\!\cos^{4N+2}\!u-\sin^{4N+2}\!u\big)I
\label{Inv11pi4}
\ee
\end{Corollary}
For $\beta_1=\beta_2=1$, the inversion identity in (\ref{Inv11pi4}) 
is readily seen to yield the inversion identity (\ref{ddI}) for the renormalised transfer tangle $\db(u)$.
\begin{Corollary}
For $\beta_1=0$ and $\xi=-\frac{\pi}{4}$, the inversion identity in Proposition~\ref{Prop:InvGen} is given by
\be
 \Db(u)\Db(u+\tfrac{\pi}{2})=-\frac{\beta_2^2\,[\sin2u]^2\,[\cos2u]^{2w}}{4^{w+1}}
  \big(\!\cos^{2N}\!u-(-1)^w\sin^{2N}\!u\big)^2I
\label{b10}
\ee
\label{Cor:b10}
\end{Corollary}
\begin{Corollary}
For $\beta_2=0$ and $\xi=-\frac{\pi}{4}$, the inversion identity in Proposition~\ref{Prop:InvGen} is given by
\be
 \Db(u)\Db(u+\tfrac{\pi}{2})=-\frac{\beta_1^2\,[\sin2u]^2\,[\cos2u]^{2w}}{4^{w+1}}
  \big(\!\cos^{2N}\!u+(-1)^w\sin^{2N}\!u\big)^2I
\label{b20}
\ee
\end{Corollary}
The Corollary~\ref{Cor:b10} generalises to the following proposition.
\begin{Proposition}
For $\beta_1=0$, the inversion identity in Proposition~\ref{Prop:InvGen} is given by
\be
 \Db(u)\Db(u+\tfrac{\pi}{2})=-\tan^2 2u\,\Gamma(u)\Gamma(u+\tfrac{\pi}{2})
   \Big([\cos u]^{2N}\eta^{(w)}(u,\xi)-[\sin u]^{2N}\eta^{(w)}(u+\tfrac{\pi}{2},\xi)\Big)^{\!2}I
\label{b10gen}
\ee
\end{Proposition}
Up to the normalisations of the Neumann parts by $\Gamma(u)$ and $\Gamma(u+\frac{\pi}{2})$ in the construction of the 
Robin twist boundary conditions (\ref{twist}), this inversion identity is recognised as the inversion identity in the case with 
Neumann boundary conditions on {\em both} sides of the strip~\cite{PR2007,PRV1210}. For $w>0$, the inversion identity 
(\ref{b10gen}) can be written as
\be
  \Db(u)\Db(u+\tfrac{\pi}{2})=-\tfrac{1}{4}\beta_2^2\tan^2 2u\,\big(\!\cos^2 2u-\cos^2 2\xi\big)
   \Big([\cos u]^{2N}\eta^{(w)}(u,\xi)-[\sin u]^{2N}\eta^{(w)}(u+\tfrac{\pi}{2},\xi)\Big)^{\!2}I
\ee
from which one recovers (\ref{b10}) by setting $\xi=-\frac{\pi}{4}$. Simplifications of the general inversion identity
in Proposition~\ref{Prop:InvGen} can of course be worked out in many other cases, such as for $\beta_1\sin\xi+\beta_2\cos\xi=0$,
but we will not do that here.

\section{Fusion and the NGK algorithm}
\label{App:NGK}

Here we consider the general logarithmic minimal model ${\cal LM}(p,p')$. The associated Kac fusion algebra
\be
 \big\langle(r,s);\;r,s\in\oN\big\rangle
\label{KacAlg}
\ee
is the algebra generated by repeated fusion of the Kac modules
$(r,s)$, $r,s\in\oN$. Higher-rank modules are generated by this fusion procedure, and
the algebra contains an infinite family of indecomposable rank-2 modules and, for $p>1$, an additional infinite family of 
indecomposable rank-3 modules. Modules of rank higher than $3$ have not been observed.

The fundamental fusion algebra~\cite{RP0707}
\be
 \big\langle(1,1),\,(2,1),\,(1,2)\big\rangle\subset\big\langle(r,s);\;r,s\in\oN\big\rangle
\ee
is the subalgebra of (\ref{KacAlg}) generated by repeated fusion of the fundamental Kac modules $(2,1)$ and $(1,2)$.
The full Kac fusion algebra (\ref{KacAlg}) has only been worked out explicitly for $p=1$ in which case~\cite{Ras1012}
\be
 \big\langle(r,s);\;r,s\in\oN\big\rangle=\big\langle(1,1),\,(2,1),\,(1,2),\,(1,p'+1)\big\rangle\qquad\quad (p=1)
\label{p1}
\ee

A Virasoro Verma module whose highest weight $\Delta$ is {\em not} in the infinitely extended {\em integer} Kac table associated
with ${\cal LM}(p,p')$
\be
 \Delta\notin\big\{\Delta_{r,s};\;r,s\in\oN\big\},\qquad \Delta_{r,s}=\frac{(rp'-sp)^2-(p'-p)^2}{4pp'}
\label{Dnot}
\ee
is {\em irreducible} and, excluding trivial self-extensions (which we believe do not arise in our fusion prescription), 
{\em cannot} appear as a proper subquotient of an indecomposable module. 
It is noted, though, that a conformal weight parameterised as $\Delta_{\frac{k}{p'},\frac{k'}{p}}$ {\em is} in the integer
Kac table for all $k,k'\in\oZ$, despite the fractional Kac labels.
The conformal weight $\Delta_{r,s-\frac{1}{2}}$, $r,s\in\oZ$, of a Robin module in ${\cal LM}(1,2)$ is recognised as being outside of the
corresponding integer Kac table.

We now parameterise the conformal weight $\Delta$ not in the integer Kac table as $\Delta=\Delta_{\rho,\sigma}$.
Since the corresponding highest-weight Verma module is irreducible, it contains no proper submodules. 
As we are considering the Virasoro algebra only, so-called spurious states
should therefore not arise in the application of the Nahm-Gaberdiel-Kausch (NGK) algorithm~\cite{Nahm,GK96}
to the fusion product $(r,s)\otimes\Vc(\Delta_{\rho,\sigma})$ where $(r,s)$ is a Kac module and $\Vc(\Delta_{\rho,\sigma})$ is the 
irreducible highest-weight module of conformal weight $\Delta_{\rho,\sigma}$.
Here we only consider fusions involving Kac modules of the form $(r,1)$ or $(1,s)$ which are 
highest-weight modules constructed as the quotients
\be
 (r,1)=V(\Delta_{r,1})/V(\Delta_{r,1}+r),\qquad (1,s)=V(\Delta_{1,s})/V(\Delta_{1,s}+s),\qquad r,s\in\oN
\ee
where $V(h)$ is the highest-weight Verma module of conformal weight $h$.
In the case of the fundamental Kac module $(2,1)$ or $(1,2)$, the corresponding submodule is generated from the singular vector
\be
 \ket{\lambda_{r,s}}=\Big(L_{-1}^2-\tfrac{2}{3}(1+2\Delta_{r,s})L_{-2}\Big)\ket{\Delta_{r,s}},\qquad (r,s)=(2,1),\,(1,2)
\ee
whereas the similar submodule in the case of $(3,1)$ or $(1,3)$ is generated from the singular vector
\be
 \ket{\lambda_{r,s}}=\Big(L_{-1}^3-2(1+\Delta_{r,s})L_{-2}L_{-1}+\Delta_{r,s}(1+\Delta_{r,s})L_{-3}\Big)\ket{\Delta_{r,s}},\qquad 
   (r,s)=(3,1),\,(1,3)
\ee

In the NGK algorithm, the decomposition of a fusion product like $(r,s)\otimes\Vc(\Delta)$, with $\Delta$ as in (\ref{Dnot}), 
relies on the analysis of the 
action of certain co-products $\Delta^\ell(L_n)$ of the Virasoro modes on a finite-dimensional state space
and is carried out to Nahm level $\ell\in\oN_0$. Since there are no spurious states in this case, 
this can be done in terms of $rs$-dimensional matrices.
For $(r,s)=(2,1),(1,2)$ and in the basis 
$\{\ket{\Delta_{r,s}}\times\ket{\Delta},\,L_{-1}\ket{\Delta_{r,s}}\times\ket{\Delta}\}$, 
the co-product of $L_0$ on $(r,s)\otimes\Vc(\Delta)$ restricted to Nahm level $0$ is thus given by
\be
 \Delta^0(L_0)=\begin{pmatrix} \Delta_{r,s}+\Delta&\tfrac{2}{3}(1+2\Delta_{r,s})\Delta \\[.15cm] 
    1&\tfrac{1}{3}(1-\Delta_{r,s})+\Delta\end{pmatrix},\qquad (r,s)=(2,1),\,(1,2)
\ee
Writing $\Delta=\Delta_{\rho,\sigma}$, the sets of eigenvalues are given by
\be
 \big\{\Delta_{\rho-1,\sigma},\;\Delta_{\rho+1,\sigma};\;(r,s)=(2,1)\big\},\qquad 
  \big\{\Delta_{\rho,\sigma-1},\;\Delta_{\rho,\sigma+1};\;(r,s)=(1,2)\big\}
\ee
Since neither of these conformal weights is in the integer Kac table, we conclude that
\be
 (2,1)\otimes\Vc(\Delta_{\rho,\sigma})=\Vc(\Delta_{\rho-1,\sigma})\oplus\Vc(\Delta_{\rho+1,\sigma}),\qquad
 (1,2)\otimes\Vc(\Delta_{\rho,\sigma})=\Vc(\Delta_{\rho,\sigma-1})\oplus\Vc(\Delta_{\rho,\sigma+1}) 
\ee

Likewise for $(r,s)=(3,1),(1,3)$, in the basis 
$\{\ket{\Delta_{r,s}}\times\ket{\Delta},\,L_{-1}\ket{\Delta_{r,s}}\times\ket{\Delta},\,L_{-1}^2\ket{\Delta_{r,s}}\times\ket{\Delta}\}$, 
the co-product of $L_0$ on $(r,s)\otimes\Vc(\Delta)$ restricted to Nahm level $0$ is given by
\be
 \Delta^0(L_0)=\begin{pmatrix} \Delta_{r,s}+\Delta&0&2\Delta_{r,s}(1+\Delta_{r,s})\Delta \\[.15cm] 
    1&\Delta_{r,s}+\Delta+1&(1+\Delta_{r,s})(2\Delta-\Delta_{r,s})\\[.15cm]
    0&1&\Delta-\Delta_{r,s}\end{pmatrix},\qquad (r,s)=(3,1),\,(1,3)
\ee
Again writing $\Delta=\Delta_{\rho,\sigma}$, the sets of eigenvalues are given by
\be
 \big\{\Delta_{\rho-2,\sigma},\;\Delta_{\rho,\sigma},\;\Delta_{\rho+2,\sigma};\;(r,s)=(3,1)\big\},\qquad 
  \big\{\Delta_{\rho,\sigma-2},\;\Delta_{\rho,\sigma},\;\Delta_{\rho,\sigma+2};\;(r,s)=(1,3)\big\}
\ee
Since neither of these conformal weights is in the integer Kac table, we conclude that
\begin{align}
 (3,1)\otimes\Vc(\Delta_{\rho,\sigma})
   &=\Vc(\Delta_{\rho-2,\sigma})\oplus\Vc(\Delta_{\rho,\sigma})\oplus\Vc(\Delta_{\rho+2,\sigma})\nn
 (1,3)\otimes\Vc(\Delta_{\rho,\sigma})&=\Vc(\Delta_{\rho,\sigma-2})\oplus\Vc(\Delta_{\rho,\sigma})\oplus\Vc(\Delta_{\rho,\sigma+2}) 
\label{13}
\end{align}

General fusion rules between a module in the fundamental fusion algebra and $\Vc(\Delta)$, 
with $\Delta$ as in (\ref{Dnot}), can now be inferred from detailed knowledge of the fusion algebra, the requirement of associativity 
and the assumption that $\Vc(\Delta)$ cannot appear as a proper subquotient of an indecomposable module.
In the case of critical dense polymers ${\cal LM}(1,2)$, the explicit evaluation of $(1,3)\otimes\Vc(\Delta_{\rho,\sigma})$
in (\ref{13}) and the observation that the Kac fusion algebra is finitely generated as in (\ref{p1}) imply that 
we can infer the decomposition of ${\cal A}\otimes\Vc(\Delta_{\rho,\sigma})$ for every module ${\cal A}$ in the Kac fusion algebra
of ${\cal LM}(1,2)$. The result for $\Delta_{\rho,\sigma}=\Delta_{r,s-\frac{1}{2}}$ with $r,s\in\oZ$ is presented in Section~\ref{Sec:Fus}.


\end{document}